\documentclass[twocolumn,nofootinbib]{aastex63}
\usepackage{natbib}
\usepackage[utf8]{inputenc}
\usepackage{amsmath,units}
\usepackage{booktabs}
\usepackage{makecell} 
\usepackage{color}
\usepackage{graphicx}
\usepackage{acronym}   
\usepackage{xspace}
\usepackage{enumitem} 
\usepackage{calc}
\usepackage[caption=false]{subfig}
\usepackage{import}
\usepackage{amsmath}	
\usepackage{bbm} 
\usepackage{mathtools} 
\definecolor{limegreen}{rgb}{0.64, 0.86, 0.215}

\definecolor{forestgreen}{rgb}{0.0, 0.27, 0.13}

\definecolor{ochre}{rgb}{0.8, 0.47, 0.13}

\definecolor{indigo}{rgb}{0.18,0.03,0.33}

\definecolor{azure}{rgb}{0.0,0.5,1.0}

\definecolor{phlox}{rgb}{0.87, 0.0, 1.0}

\definecolor{thistle}{rgb}{0.87, 0.80, 0.87}

\newcommand{\vCOMPAS}{\ensuremath{\rm{v}02.21.00}\xspace} 
\newcommand{\docCOMPAS}{\url{https://compas.science/docs/}}

\newcommand{\monei}{\ensuremath{M_{1,\rm{i}}}\xspace}
\newcommand{\mtwoi}{\ensuremath{M_{2,\rm{i}}}\xspace}
\newcommand{\monef}{\ensuremath{M_{1}}\xspace}
\newcommand{\mtwof}{\ensuremath{M_{2}}\xspace}
\newcommand{\ai}{\ensuremath{a_{\rm{i}}}\xspace}
\newcommand{\qi}{\ensuremath{q_{\rm{i}}}\xspace}
\newcommand{\Zi}{\ensuremath{Z_{\rm{i}}}\xspace}
\newcommand{\vk}{\ensuremath{v_{\rm{k}}}\xspace}
\newcommand{\thetak}{\ensuremath{{\theta}_{\rm{k}}}\xspace}

\newcommand{\ei}{\ensuremath{{e}_{\rm{i}}}\xspace}

\newcommand{\minus}{\scalebox{0.75}[1.0]{$-$}}

\newcommand{\Rsun}{\ensuremath{\,\rm{R}_{\odot}}\xspace}

\newcommand{\kms}{\ensuremath{\,\rm{km}\,\rm{s}^{-1}}\xspace}
\newcommand{\Msun}{\ensuremath{\,\rm{M}_{\odot}}\xspace}
\newcommand{\Zsun}{\ensuremath{\,\rm{Z}_{\odot}}\xspace}
\newcommand{\Lsun}{\ensuremath{\,\rm{L}_{\odot}}\xspace}

\newcommand{\AU}{\ensuremath{\,\mathrm{au}}\xspace}

\newcommand{\Gyr}{\ensuremath{\,\mathrm{Gyr}}\xspace}

\newcommand{\yearmin}{\ensuremath{\,\rm{yr}^{-1}}\xspace}
\newcommand{\MpcminThree}{\ensuremath{\,\rm{Mpc}^{-3}}\xspace}

\newcommand{\MSFR}{\ensuremath{{M}_{\rm{SFR}}}\xspace}
\newcommand{\tdelay}{\ensuremath{{t}_{\rm{delay}}}\xspace}
\newcommand{\tDCO}{\ensuremath{{t}_{\rm{DCO}}}\xspace}
\newcommand{\ts}{\ensuremath{{t}_{\rm{s}}}\xspace}
\newcommand{\tevolve}{\ensuremath{{t}_{\rm{evolve}}}\xspace}
\newcommand{\tform}{\ensuremath{{t}_{\rm{form}}}\xspace}
\newcommand{\tmerger}{\ensuremath{{t}_{\rm{m}}}\xspace}
\newcommand{\tinspiral}{\ensuremath{{t}_{\rm{inspiral}}}\xspace}

\newcommand{\tdet}{\ensuremath{{t}_{\rm{det}}}\xspace}
\newcommand{\Nform}{\ensuremath{{N}_{\rm{form}}}\xspace}
\newcommand{\Ndet}{\ensuremath{{N}_{\rm{det}}}\xspace}
\newcommand{\Nmerger}{\ensuremath{{N}_{\rm{merger}}}\xspace}

\newcommand{\Pdet}{\ensuremath{{P}_{\rm{det}}}\xspace}

\newcommand{\Vc}{\ensuremath{{V}_{\rm{c}}}\xspace}

\newcommand{\SFRD}{\text{SFRD}\ensuremath{(Z_{\rm{i}},z)}\xspace}

\newcommand*\diff{\mathop{}\!\mathrm{d}}
      \newcommand\rate{\mathcal{R}} 
\newcommand\COMPAS{{\sc{COMPAS}}\xspace}  

\acrodef{OOP}{object-oriented programming}
\acrodef{ZAMS}{zero-age main sequence}
\acrodef{MS}{main sequence}
\acrodef{HG}{Hertzsprung gap}
\acrodef{HR}{Hertzsprung-Russell}
\acrodef{NS}{neutron star}
\acrodefplural{NS}[NSs]{neutron stars}
\acrodef{BH}{black hole}
\acrodefplural{BH}[BHs]{black holes}
\acrodef{CO}{compact object}
\acrodefplural{CO}[COs]{compact objects}
\acrodef{CHE}{chemically homogeneous evolution}
\acrodef{CHeB}{core helium burning}
\acrodef{AGB}{asymptotic giant branch}
\acrodef{EAGB}{early asymptotic giant branch}
\acrodef{HeMS}{helium main sequence}
\acrodef{HeGB}{helium giant branch}

\acrodef{GW}{gravitational wave}
\acrodef{BBH}{binary black hole}
\acrodefplural{BBH}[BBHs]{binary black holes}
\acrodef{DNS}{double neutron star}
\acrodefplural{DNS}[DNSs]{double neutron stars}
\acrodef{DCO}{double compact object}
\acrodefplural{DCO}[DCOs]{double compact objects}
\acrodef{BH--NS}{black hole-neutron star}
\acrodef{WD}{white dwarf}
\acrodefplural{WD}[WDs]{white dwarfs}
\acrodef{HeWD}{helium white dwarf}
\acrodefplural{HeWD}[HeWDs]{helium white dwarfs}
\acrodef{COWD}{carbon-oxygen white dwarf}
\acrodefplural{COWD}[COWDs]{carbon-oxygen white dwarf}
\acrodef{ONeWD}{oxygen-neon white dwarf}
\acrodefplural{ONeWD}[ONeWD]{oxygen-neon white dwarf}

\acrodef{GRB}{gamma--ray burst}
\acrodef{RLOF}{Roche-lobe overflow}
\acrodef{CE}{common envelope}
\acrodef{SN}{supernova}
\acrodefplural{SN}[SNe]{supernovae}
\acrodef{ECSN}{electron-capture supernova}
\acrodefplural{ECSN}[ECSNe]{electron-capture supernovae}
\acrodef{USSN}{ultra-stripped supernova}
\acrodefplural{USSN}[USSNe]{ultra-stripped supernovae}
\acrodef{CCSN}{core-collapse supernova}
\acrodefplural{CCSN}[CCSNe]{core-collapse supernovae}
\acrodef{PISN}{pair-instability supernova}
\acrodefplural{PISN}[PISNe]{pair-instability supernovae}
\acrodef{LRN}{luminous red nova}
\acrodefplural{LRN}[LRNe]{luminous red novae}
\acrodef{LBV}{luminous blue variable}
\acrodefplural{LBV}[LBVs]{luminous blue variables}
\acrodef{WR}{Wolf-Rayet}

\acrodef{SNR}{signal-to-noise ratio}

\acrodef{COMPAS}{
Compact Object Mergers: Population Astrophysics and Statistics}
\acrodef{BPS}{binary population synthesis} 
\acrodef{SSE}{single star evolution} 
\acrodef{BSE}{binary star evolution} 
\acrodef{LVC}{LIGO-Virgo Collaboration}
\acrodef{LVK}{LIGO-Virgo-KAGRA Collaboration}
\acrodef{LISA}{Laser Interferometer Space Antenna}
\acrodef{IMF}{initial mass function}

\acrodef{GSMF}{galaxy mass function, the number density of galaxies per logarithmic mass bin}
\acrodef{MZR}{mass-metallicity relation}
\acrodef{SFRD}{star formation rate density}

\acrodef{CSV}{comma separated values}
\acrodef{HDF5}{hierarchical data format, version 5}
\acrodef{TSV}{tab separated values}
\acrodef{TXT}{plain text}

\begin{document}

\title{Rapid stellar and binary population synthesis with COMPAS}

\shorttitle{COMPAS code paper}
\shortauthors{Team COMPAS}

\author{Team COMPAS}
\affiliation{The public~\COMPAS code is a product of work by the entire \COMPAS collaboration over many years; we therefore kindly request that, in recognition of this team effort, the paper is cited as ``Team \COMPAS: J.~Riley et al.''}
\author{Jeff Riley}
\affiliation{School of Physics and Astronomy,
Monash University, Clayton, Victoria 3800, Australia}
\affiliation{OzGrav, Australian Research Council Centre of Excellence for Gravitational Wave Discovery, Australia}
\author{Poojan Agrawal}
\affiliation{Centre for Astrophysics and Supercomputing, Swinburne University of Technology, Hawthorn, VIC 3122, Australia}
\affiliation{OzGrav, Australian Research Council Centre of Excellence for Gravitational Wave Discovery, Australia}
\author{Jim W. Barrett}
\affiliation{Institute of Gravitational Wave Astronomy and School of Physics and Astronomy, University of Birmingham, Birmingham, B15 2TT, United Kingdom}
\author{Kristan N. K. Boyett}
\affiliation{Department of Physics, University of Oxford, Denys Wilkinson Building, Keble Road, Oxford OX1 3RH, UK}
\author{Floor S. Broekgaarden}
\affiliation{Center for Astrophysics \textbar{} Harvard $\&$ Smithsonian,
60 Garden St., Cambridge, MA 02138, USA}
\author{Debatri Chattopadhyay}
\affiliation{Centre for Astrophysics and Supercomputing, Swinburne University of Technology, Hawthorn, VIC 3122, Australia}
\affiliation{OzGrav, Australian Research Council Centre of Excellence for Gravitational Wave Discovery, Australia}
\author{Sebastian M. Gaebel}
\affiliation{Max Planck Institute for Gravitational Physics (Albert Einstein Institute), Callinstrasse 38, D-30167 Hannover, Germany}
\author{Fabian Gittins}
\affiliation{Mathematical Sciences and STAG Research Centre, University of Southampton, Southampton SO17 1BJ, UK}
\author{Ryosuke Hirai}
\affiliation{School of Physics and Astronomy,
Monash University, Clayton, Victoria 3800, Australia}
\affiliation{OzGrav, Australian Research Council Centre of Excellence for Gravitational Wave Discovery, Australia}
\author{George Howitt}
\affiliation{School of Physics, University of Melbourne, Parkville, Victoria, 3010, Australia}
\affiliation{OzGrav, Australian Research Council Centre of Excellence for Gravitational Wave Discovery, Australia}
\author{Stephen Justham}
\affiliation{Anton Pannekoek Institute of Astronomy and GRAPPA, Science Park 904, University of Amsterdam, 1098XH Amsterdam, The Netherlands}
\affiliation{School of Astronomy \& Space Science, University of the Chinese Academy of Sciences, Beijing 100012, China}
\affiliation{Max Planck Institute for Astrophysics, Karl-Schwarzschild-Str. 1, 85748 Garching, Germany}
\author{Lokesh Khandelwal}
\affiliation{Anton Pannekoek Institute of Astronomy and GRAPPA, Science Park 904, University of Amsterdam, 1098XH Amsterdam, The Netherlands}
\author{Floris Kummer}
\affiliation{Anton Pannekoek Institute of Astronomy and GRAPPA, Science Park 904, University of Amsterdam, 1098XH Amsterdam, The Netherlands}
\author{Mike Y. M. Lau}
\affiliation{School of Physics and Astronomy,
Monash University, Clayton, Victoria 3800, Australia}
\affiliation{OzGrav, Australian Research Council Centre of Excellence for Gravitational Wave Discovery, Australia}
\author{Ilya Mandel}
\affiliation{School of Physics and Astronomy,
Monash University, Clayton, Victoria 3800, Australia}
\affiliation{OzGrav, Australian Research Council Centre of Excellence for Gravitational Wave Discovery, Australia}
\affiliation{Institute of Gravitational Wave Astronomy and School of Physics and Astronomy, University of Birmingham, Birmingham, B15 2TT, United Kingdom}
\author{Selma E. de Mink}
\affiliation{Max Planck Institute for Astrophysics, Karl-Schwarzschild-Str. 1, 85748 Garching, Germany}
\affiliation{Anton Pannekoek Institute of Astronomy and GRAPPA, Science Park 904, University of Amsterdam, 1098XH Amsterdam, The Netherlands}
\affiliation{Center for Astrophysics \textbar{} Harvard $\&$ Smithsonian,
60 Garden St., Cambridge, MA 02138, USA}
\author{Coenraad Neijssel}
\affiliation{Institute of Gravitational Wave Astronomy and School of Physics and Astronomy, University of Birmingham, Birmingham, B15 2TT, United Kingdom}
\affiliation{OzGrav, Australian Research Council Centre of Excellence for Gravitational Wave Discovery, Australia}
\author{Tim Riley}
\affiliation{School of Physics and Astronomy,
Monash University, Clayton, Victoria 3800, Australia}
\affiliation{OzGrav, Australian Research Council Centre of Excellence for Gravitational Wave Discovery, Australia}
\author{Lieke van Son}
\affiliation{Center for Astrophysics \textbar{} Harvard $\&$ Smithsonian,
60 Garden St., Cambridge, MA 02138, USA}
\affiliation{Anton Pannekoek Institute of Astronomy and GRAPPA, Science Park 904, University of Amsterdam, 1098XH Amsterdam, The Netherlands}
\affiliation{Max Planck Institute for Astrophysics, Karl-Schwarzschild-Str. 1, 85748 Garching, Germany}
\author{Simon Stevenson}
\affiliation{Centre for Astrophysics and Supercomputing, Swinburne University of Technology, Hawthorn, VIC 3122, Australia}
\affiliation{OzGrav, Australian Research Council Centre of Excellence for Gravitational Wave Discovery, Australia}
\author{Alejandro Vigna-G\'omez}
\affiliation{DARK, Niels Bohr Institute, University of Copenhagen, Jagtvej 128, 2200, Copenhagen, Denmark}
\affiliation{Niels Bohr International Academy, The Niels Bohr Institute, Blegdamsvej 17, 2100 Copenhagen, Denmark}
\author{Serena Vinciguerra}
\affiliation{Anton Pannekoek Institute of Astronomy and GRAPPA, Science Park 904, University of Amsterdam, 1098XH Amsterdam, The Netherlands}
\author{Tom Wagg}
\affiliation{Department of Astronomy, University of Washington, Seattle, WA, 98195}
\affiliation{Center for Astrophysics \textbar{} Harvard $\&$ Smithsonian,
60 Garden St., Cambridge, MA 02138, USA}
\affiliation{Max Planck Institute for Astrophysics, Karl-Schwarzschild-Str. 1, 85748 Garching, Germany}
\author{Reinhold Willcox}
\affiliation{School of Physics and Astronomy, Monash University, Clayton, Victoria 3800, Australia}
\affiliation{OzGrav, Australian Research Council Centre of Excellence for Gravitational Wave Discovery, Australia}
\email{jeff.riley@monash.edu, floor.broekgaarden@cfa.harvard.edu, ilya.mandel@monash.edu, spstevenson@swin.edu.au }
\date{\today}


\begin{abstract}
 Compact Object Mergers: Population Astrophysics and Statistics (\href{https://compas.science}{COMPAS}; \url{https://compas.science}) is a public rapid binary population synthesis code.  COMPAS generates populations of isolated stellar binaries under a set of parameterized assumptions in order to allow comparisons against observational data sets, such as those coming from gravitational-wave observations of merging compact remnants.  It includes a number of tools for population processing in addition to the core binary evolution components.  COMPAS is publicly available via the github repository \url{https://github.com/TeamCOMPAS/COMPAS/}, and is designed to allow for flexible modifications as evolutionary models improve. This paper describes the methodology and implementation of COMPAS.  It is a living document that will be updated as new features are added to COMPAS; the current document describes COMPAS \vCOMPAS. 
\end{abstract}

\vspace{2.48cm} 

\tableofcontents
\newpage 


\section{Introduction}
\label{sec:intro}

The majority of massive stars are born in a stellar binary- or multiple-star system with other stellar companions \citep[e.g.,][]{chini2012spectroscopic,Sana:2012Sci,Sana:2014,kobulnicky2014toward,almeida2017tarantula,Moe:2017ApJS}.  
The subsequent evolution of massive stellar binaries plays a critical role in many fields of astronomy.
Massive binaries are thought to play key roles in the reionization of the universe \citep[e.g.,][]{Conroy:2012,Ma:2016,Eldridge:2017,Rosdahl:2018, Goetberg:2019,Goetberg:2020}, nucleosynthesis \citep[e.g.,][]{Dray:2003, 2006A&A...460..565I,woosley2007nucleosynthesis,Langer:2012}, and the diversity of observed supernovae \citep[e.g.,][]{Podsiadlowski:1992ApJ,Eldridge:2013,Eldridge:2018,Eldrdige:2019,2015MNRAS.451.2123T, 2017MNRAS.466.2085M, Yoon:2017, Yoon:2019, Zapartas:2019,Zapartas:2021}.  
Some massive binaries will evolve into systems containing one or two compact objects, which can be observed as X-ray binaries \citep[e.g.,][]{RemillardMcClintock:2006}, double neutron stars \citep[e.g.,][]{Tauris:2017omb}, short gamma-ray bursts  \citep[e.g.,][]{WoosleyBloom:2006,Berger:2014}, and gravitational-wave transients \citep[e.g.][]{mandel2018merging,2019PhRvX...9c1040A,MandelBroekgaarden:2021}.  

Many physical processes in the evolution of a binary system are uncertain.  
The uncertainties in stellar wind mass loss, mass transfer, common envelope physics, supernova remnant masses, and natal kicks, among others, can be constrained by comparing the observed populations listed above against theoretical predictions under a range of assumptions.
The essence of \ac{BPS} simulations is to enable modeling of large populations by combining prescriptions for \ac{SSE} and \ac{BSE}, modelling the evolution of stars from \ac{ZAMS} until their final states.

In this paper we present the publicly available \ac{BPS} suite Compact  Object  Mergers:    Population  Astrophysics  and  Statistics (\COMPAS). 
The core of \COMPAS is a \ac{BPS} code that models the evolution of a population of binary stars using a set of simplified prescriptions or recipes for stellar and binary evolution.
By doing so, \COMPAS can compute the full evolution of a typical binary system in around $10\,$ms on a modern laptop, and compute the evolution of a million binaries in a few CPU hours.  
The general approach  is similar to other \ac{BPS} codes including the Scenario Machine \citep{1996A&A...310..489L,1996ApJ...466..234L,2009ARep...53..915L}, IBiS \citep{1996MNRAS.280.1035T}, SeBa \citep{1996A&A...309..179P,1998A&A...332..173P,2001A&A...365..491N,2012A&A...546A..70T},  BSE \citep{Hurley:2002rf}, StarTrack \citep{Belczynski:2001uc,Belczynski:2008,2020A&A...636A.104B},  binary$\_$c \citep{2004MNRAS.350..407I,2006A&A...460..565I,2009A&A...508.1359I},  MOBSE \citep{2018MNRAS.480.2011G,2018MNRAS.474.2959G}  and COSMIC \citep{2019arXiv191100903B}.

In addition to the core \ac{BPS} code, \COMPAS also provides several other publicly available tools, including rapid single stellar evolution, postprocessing tools to study the evolution of populations over cosmic time \citep{2019MNRAS.490.3740N}, postprocessing scripts to model the detectability of \ac{DCO} [comprising \ac{DNS}, \ac{BBH}, and \ac{BH--NS}] mergers by ground-based gravitational-wave observatories \citep[e.g.,][]{Barrett:2017fcw}, a statistical sampling framework to optimize the computational cost of \ac{BPS} \citep{Broekgaarden:2019qnw}, and models for specific evolutionary phases such as X-ray binaries and pulsars \citep[e.g.,][]{Chattopadhyay:2019xye,Vinciguerra:2020}.

\COMPAS was developed with a primary focus on the study of compact object mergers that serve as sources of gravitational waves. 
It has been used extensively to investigate the properties of compact binaries containing neutron stars and black holes.  
\citet{Stevenson2017FormationEvolution} studied the formation history of the first three \acp{BBH} detected via gravitational waves. 
\citet{Barrett:2017fcw} explored how future gravitational-wave observations will allow us to determine the physics of massive binary evolution. 
\citet{Stevenson:2019rcw} included the impact of (pulsational) pair-instability supernovae.  
\citet{2019MNRAS.490.3740N} investigated the consequences of uncertain metallicity-specific star formation history on the rate and properties of \ac{DCO} mergers. 
\citet{Lau:2019} predicted the number of \acp{DNS} detectable with LISA and the inference these observations will enable.  
\citet{Bavera:2019} used \COMPAS to predict the spin distribution of merging \acp{BBH}.  
\citet{vanSon:2020zbk} investigated the robustness of the predicted pair-instability mass gap to uncertainties in the accretion efficiency for black holes. 
\citet{Riley:2020} studied chemically homogeneous evolution as a pathway to \ac{BBH} formation.  
\citet{Mandel:2020} considered the consequences of an alternative, stochastic recipe for compact remnant masses.  
\citet{VignaGomez:2021} explored sequential \ac{BBH} mergers during triple evolution.  
\citet{Broekgaarden:2021} and \citet{Broekgaarden:2021hlu} focused on the formation of black hole -- neutron star binaries.
\citet{vanSon:2021} investigated the relative contributions of dynamically stable and unstable mass transfer to \ac{BBH} mergers over cosmic history.   \citet{Broekgaarden:2021dco} investigated the relative impact from uncertain metallicity-specific star formation history to that from uncertain stellar evolution on the rate and properties of \ac{DCO} mergers. 

A number of other consequences of massive binary evolution have been explored with \COMPAS.   \citet{2018MNRAS.481.4009V} used observations of Galactic \acp{DNS} to constrain evolutionary physics.   \citet{Chattopadhyay:2019xye} used the same population to constrain birth distributions of pulsar spin periods and magnetic field strengths, as well as magnetic field decay scales. \citet{Schroder:2019xqq} investigated optical counterparts to \ac{CE} events which fail to eject the envelope and result in a merger.  \citet{2020MNRAS.492.3229H} explored luminous red novae: red optical transients associated with \ac{CE} events.  \citet{Vigna-Gomez:2020bgo} cataloged the \ac{CE} events en route to \ac{DNS} formation.  \citet{MandelMueller:2020} took advantage of \COMPAS \ac{SSE} modules in formulating a stochastic compact remnant mass prescription and momentum-conserving natal kick prescription.
\citet{Vinciguerra:2020} studied the population of Be X-ray binaries to establish constraints for mass accretion efficiency.   \citet{Chattopadhyay:2020lff} explored the prospects for observing neutron star--black hole binaries in future radio pulsar surveys.  \citet{MillerJones:2021} and \citet{Neijssel:2020CygX1} constrained the massive stellar wind mass loss with observations of the \ac{BH} high-mass X-ray binary Cygnus X-1.  \citet{Willcox:2021kbg} compared \COMPAS models to observed pulsar velocities in order to study neutron star natal kicks.

\COMPAS has also played an important role in developing more efficient sampling, inference, and model emulation techniques.  \citet{Barrett:2017tug} attempted to generate computationally efficient surrogate models of binary population synthesis through Gaussian process emulation.   \citet{TaylorGerosa:2018} used \COMPAS public data as an example in their hierarchical inference study.  \citet{Broekgaarden:2019qnw} applied importance sampling to enable computationally efficient \ac{DCO} simulations.  \citet{Lin:2021} developed new tools for classification and emulation based on (local) Gaussian process models and the COMPAS suite.

In this paper we introduce the \COMPAS code and discuss its methodology and implementation.  The paper is organized as follows.  Section \ref{sec:COMPAS} presents the main features of \COMPAS. 
Section~\ref{sec:single_star} discusses the implementation of single stellar evolution in \COMPAS.
Section~\ref{sec:binary_evolution} discusses binary stellar evolution prescriptions.
In Section~\ref{sec:populations} we describe how \COMPAS can be used to model populations of binaries.
Section~\ref{sec:postprocessing} describes the postprocessing tools available within \COMPAS. Section~\ref{sec:results} illustrates a few applications of \COMPAS.  We conclude in Section \ref{sec:conclusions}.

\section{\COMPAS}\label{sec:COMPAS}

\subsection{Overview}\label{subsec:COMPAS_intro}

\COMPAS\footnote{\url{compas.science}. Code available at \url{https://github.com/TeamCOMPAS/COMPAS}} is an open-source integrated suite of software tools combining a robust, rapid, and flexible population synthesis application for both single-star evolution and binary star evolution, with tools for deployment on a range of platforms, including high-performance computing platforms, and a set of Python postprocessing analysis and plotting scripts.

The heart of the \COMPAS suite is the \ac{SSE} and \ac{BSE} simulation code, developed in the C++ programming language. C++ is a cross platform, \ac{OOP} language that gives a clear structure to programs and allows code to be easily understood and maintained - it is one of the world's most popular programming languages, and has an established history of being used to create high-performance applications. The \COMPAS C++ code is a modular, object-oriented code, designed to be easily understood and extended.

Results produced with \COMPAS are publicly available\footnote{ \url{https://zenodo.org/communities/compas/}}. 

\subsection{Mode: Single Star versus Binary Star}\label{subsec:COMPAS_modes}

\COMPAS operates in either one of two selectable modes: \ac{SSE} or \ac{BSE}. As the names suggest, in \ac{SSE} mode \COMPAS evolves single stars, and in \ac{BSE} mode \COMPAS evolves binary stars.

The \ac{SSE} algorithm covers all evolution phases from the Zero-age Main Sequence (\ac{ZAMS}) up to and including the remnant stages.  
The allowed range of \ac{ZAMS} masses,  $M\in[0.1,150]$\Msun, extends the range of \citet{Hurley:2000pk} models by extrapolation; the allowed range of metallicities $Z\in[10^{\minus{4}},0.03]$ follows these models (see Section~\ref{sec:single_star} for a full description).

In \ac{BSE} mode, the \ac{SSE} code provides the stellar attributes (e.g., luminosity, radius, temperature, etc.) for each of the component stars as they evolve. \COMPAS evolves binary stars until a \ac{DCO} is formed, the component stars merge, or, optionally, the binary is disrupted.

In either mode (\ac{SSE} or \ac{BSE}), users can specify a maximum evolution time and/or maximum number of evolutionary steps, after which evolution is halted.

\subsection{Architecture}\label{subsec:COMPAS_architecture}

\begin{figure*}
    \begin{center}
	    \includegraphics[scale=0.85]{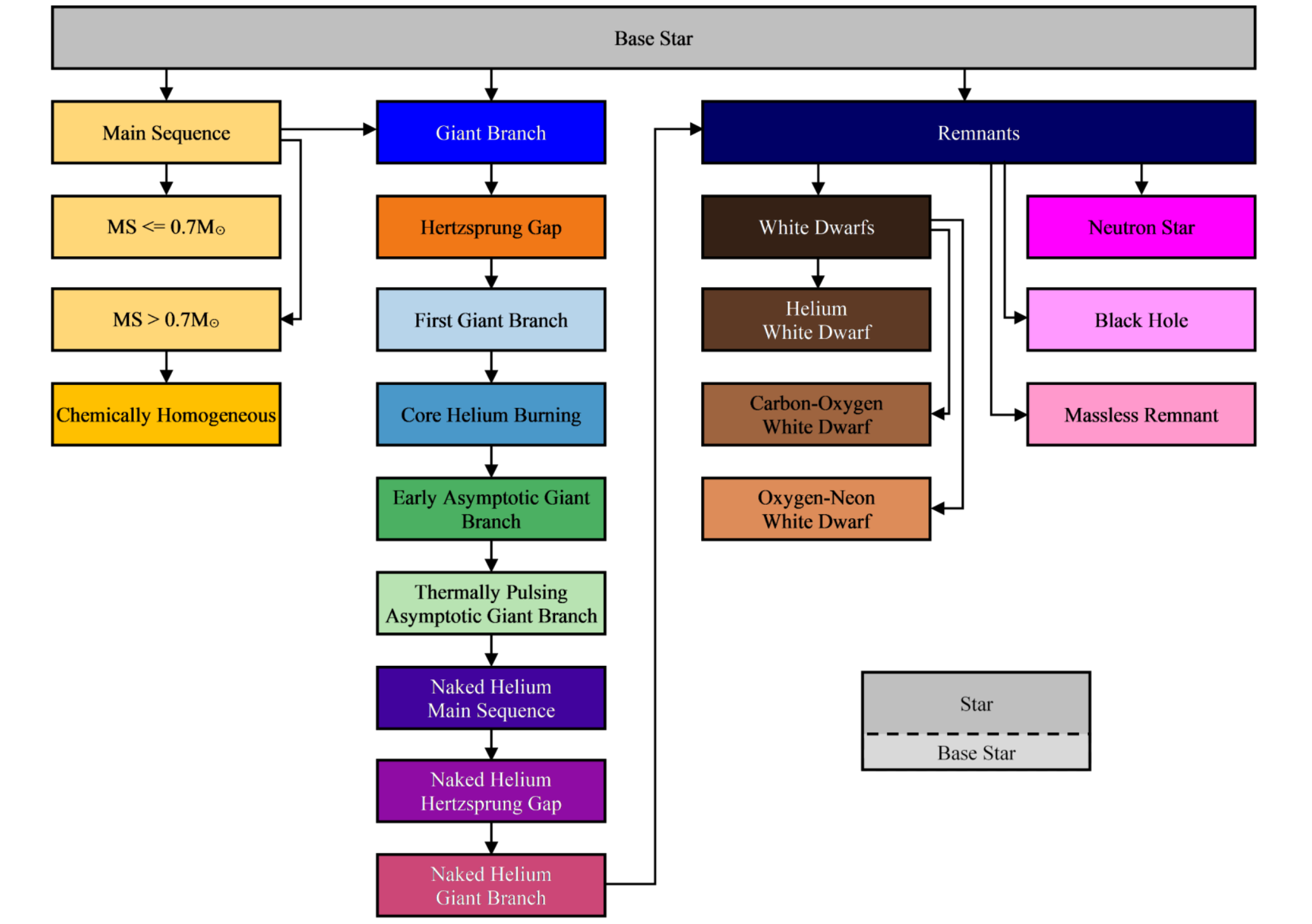}
    \end{center}
    \vspace{-2.00mm}
    \caption{SSE class and container diagram (arrows indicate inheritance).
    }
    \label{fig:SSE_ClassDiagram}
\end{figure*}

The \ac{SSE} and \ac{BSE} simulation code at the heart of \COMPAS is written in C++ using \ac{OOP} concepts. The architecture is based on stellar type, with each stellar type being described by a separate C++ class. Figure~\ref{fig:SSE_ClassDiagram} shows the \ac{SSE} class and container diagram, where the arrows indicate inheritance (the \ac{OOP} mechanism used to base one class upon another: the inheriting class inherits the implementation of the inherited class).  The \COMPAS C++ code is implemented using multiple inheritance, and all stellar classes also inherit directly from the \textit{BaseStar} class (arrows not shown in Figure~\ref{fig:SSE_ClassDiagram} for clarity). Each of the stellar classes encapsulates data structures and algorithms specific to the evolutionary phase corresponding to the class.\footnote{The class names shown in Figure~\ref{fig:SSE_ClassDiagram} do not match the class names used in the \COMPAS C++ code, which are abbreviated into single words or acronyms.}

The \textit{Star} class shown in Figure~\ref{fig:SSE_ClassDiagram} is a container class for the stellar classes. An instance of the \textit{Star} class is a single star being evolved by \COMPAS, and contains an object that is created as a \textit{BaseStar} object, and evolves, over time, through various \ac{SSE} classes shown in Figure~\ref{fig:SSE_ClassDiagram}.

\begin{figure*}
    \begin{center}
	    \includegraphics[scale=0.08]{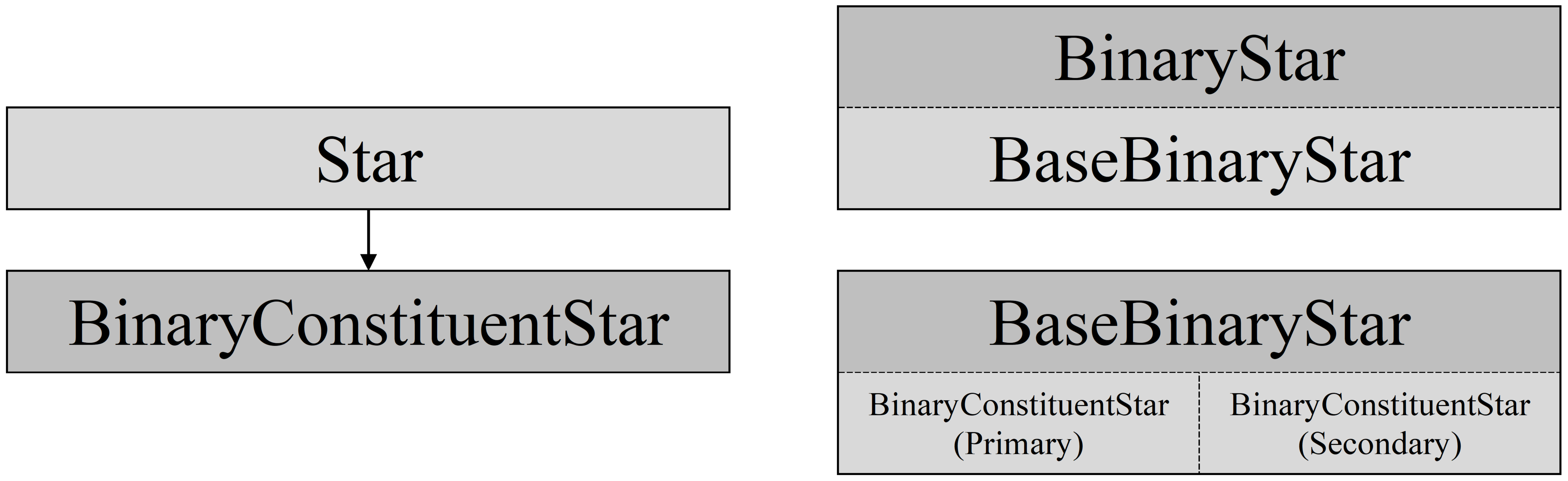}
    \end{center}
    \vspace{-2.00mm}
    \caption{BSE class and container diagram.}
    \label{fig:BSE_ClassDiagram}
\end{figure*}

Figure~\ref{fig:BSE_ClassDiagram} shows the \ac{BSE} class and container diagram. The main class for binary star evolution is the \textit{BinaryStar} class. The \textit{BinaryStar} class is a wrapper, containing a \textit{BaseBinaryStar} class object, and abstracts away the details of the binary star and the evolution.
The \textit{BaseBinaryStar} class is a container class for the objects that represent the component stars of a binary system. An instance of the \textit{BaseBinaryStar} class is a binary system being evolved by COMPAS, and contains a \textit{BinaryConstituentStar} class object for each of the component stars (i.e.\ the primary and secondary stars), as well as data structures and algorithms specific to the evolution of a binary system.
The \textit{BinaryConstituentStar} class inherits from the \ac{SSE} \textit{Star} class, so objects instantiated from the \textit{BinaryConstituentStar} class inherit the characteristics of the \ac{SSE} \textit{Star} class, particularly the stellar evolution model. The \textit{BinaryConstituentStar} class defines additional data structures and algorithms (to the data structures and algorithms provided by the \ac{SSE} classes) required to support the evolution of a binary system component star.

\subsection{Evolutionary Models}\label{subsec:COMPAS_evolution}

\subsubsection{SSE Model}\label{subsubsec:COMPAS_SSE_evolution}

The \ac{SSE} model implemented in \COMPAS (see Section~\ref{sec:single_star} for more details) follows \citet{Hurley:2000pk}, using their analytical fits to the models of \citet{Pols:1998MNRAS}. 
After the creation of the star according to the initial conditions (specified or sampled), the evolution of a single star proceeds by integrating the attributes of the star over its lifetime, and stops when the star evolves to a remnant, or the maximum time, or maximum number of time steps is reached.

\begin{figure}
    \begin{center}
	    \includegraphics[scale=0.1]{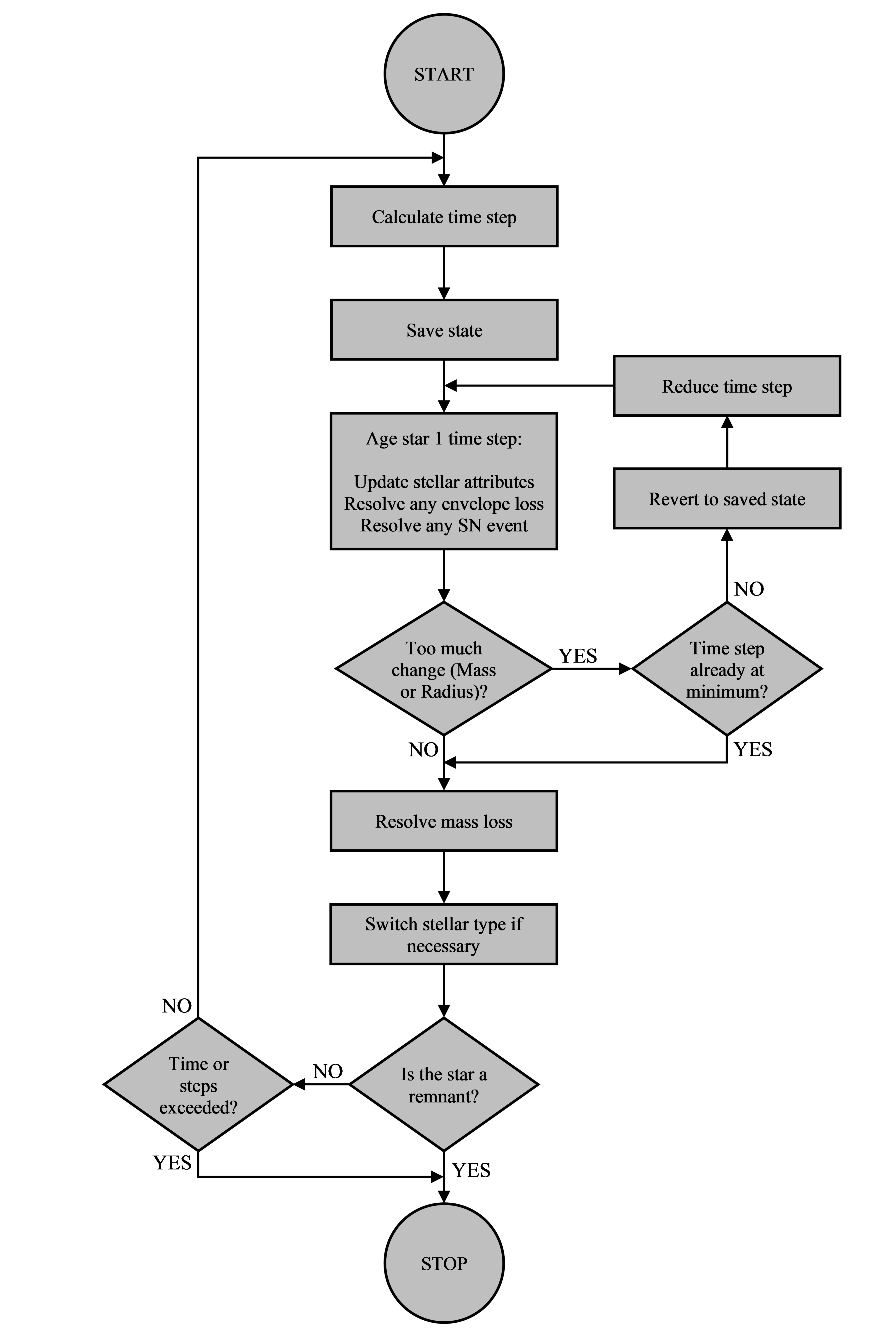}
    \end{center}
    \vspace{-2.00mm}
    \caption{High-level SSE evolution.}
    \label{fig:SSE_FlowChart}
\end{figure}

Figure~\ref{fig:SSE_FlowChart} shows a high-level overview of the code flow for \ac{SSE}.

\subsubsection{BSE Model}\label{subsubsec:COMPAS_BSE_evolution}

The binary evolution model implemented in \COMPAS is broadly similar to the BSE population synthesis application \citep{Hurley:2002rf}, and other population synthesis applications derived from it, such as binary\_c \citep{2004MNRAS.350..407I,2006A&A...460..565I,2009A&A...508.1359I,de2013rotation} and StarTrack \citep{Belczynski:2001uc,Belczynski:2008}.

After the creation of the binary system according to the initial conditions (specified or sampled), the evolution of a binary system proceeds by integrating the attributes of the system over its lifetime and stops if the component stars merge, when the system forms a \ac{DCO}, is disrupted, or the maximum time or maximum number of time steps is reached.

\begin{figure}
    \begin{center}
	    \includegraphics[scale=0.08]{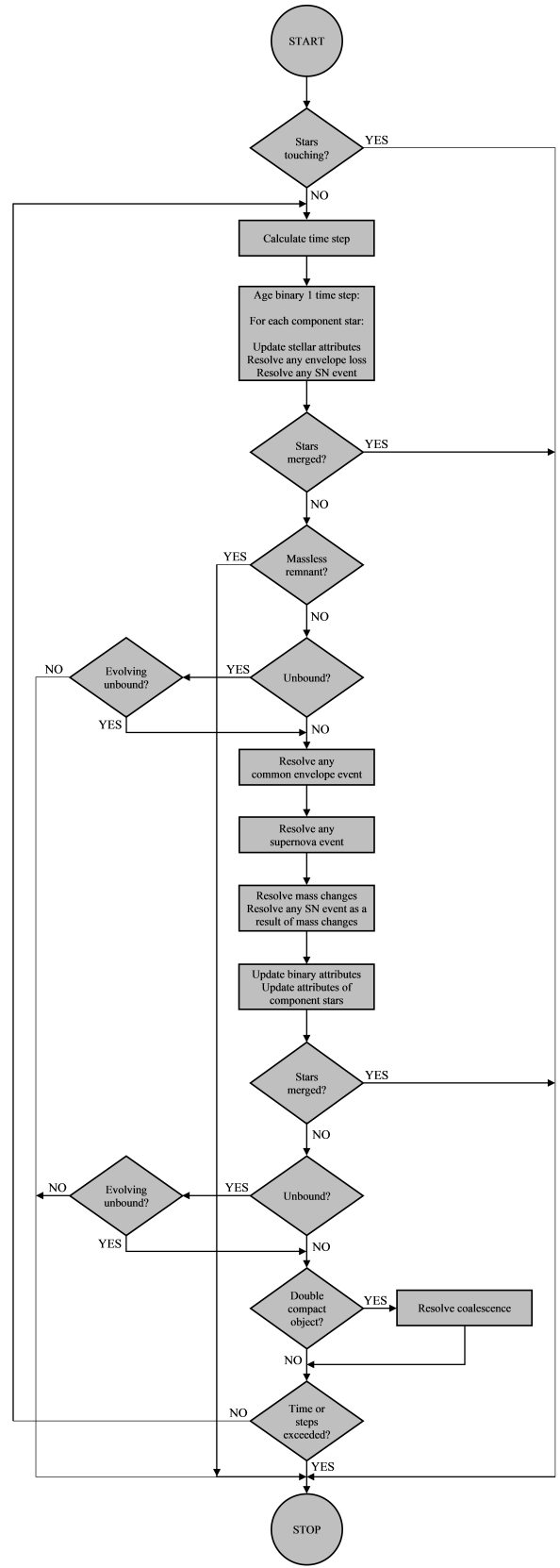}
    \end{center}
    \vspace{-2.00mm}
    \caption{High-level BSE evolution.}
    \label{fig:BSE_FlowChart}
\end{figure}

Figure~\ref{fig:BSE_FlowChart} shows a high-level overview of the code flow for \ac{BSE}.

\subsection{Time Stepping}\label{subsec:COMPAS_timestepping}

The initial estimate of the time step used in \ac{SSE} follows \citet{Hurley:2000pk}, where the time step varies depending upon the evolutionary phase of the star (see Section~\ref{sec:single_star} for a full list of the stellar types).  \COMPAS then checks whether the time step produces excessive change, defined as either
\begin{enumerate}[label={(\alph*)}]
    \item mass loss greater than 1\%, or
    \item radial change greater than 10\%
\end{enumerate}

\noindent{over the time step, and limits the time step accordingly.}

For nuclear timescale evolution, we limit the time step to a minimum of 100 yr, and we impose an overall minimum time step of 100 s,  including for dynamical timescale evolution (see Section \ref{subsec:timescales}). 

For the \ac{BSE} time step, \COMPAS uses the minimum of the binary constituent stars' \ac{SSE} time steps: this allows the constituent stars to evolve using time steps that do not produce excessive change. Changes in binary properties are not separately considered when calculating the time step since large changes in binary properties would be accompanied by similarly large changes in constituent star properties.

\COMPAS provides a mechanism for the user to scale the calculated time step by a positive scaling factor. Scaling is performed prior to limiting of the time step.

\subsection{Input and Configuration}\label{subsec:COMPAS_input}

\COMPAS provides wide-ranging functionality and affords users much flexibility in determining how the synthesis and evolution of stars (single or binary) is conducted. Users configure \COMPAS's functionality and provide initial conditions via the use of program options and input grid files.  The full list and description of program options and grid files can be found in the \COMPAS online documentation, {\docCOMPAS}.

\subsection{Default Model}\label{subsec:default_model}

Investigating massive star and binary evolution requires modeling many complicated astrophysical processes. 
In \ac{BPS} codes, this is often done through the use of simple analytic prescriptions, which are calibrated either to theoretical predictions or to results from observations (see Sections \ref{sec:single_star} and \ref{sec:binary_evolution} for more details in the context of \ac{SSE} and \ac{BSE}). 
In this paper, we refer to our default set of modeling assumptions, including which prescriptions are used, as our \emph{Default} (or `default') model.  This model is summarized in Table \ref{tab:COMPAS}. 
This model has been calibrated to match a range of observations (see Section \ref{sec:intro} for a brief overview). 
There are, however, large uncertainties in the prescriptions used. \COMPAS is flexible, and in many cases, we provide additional options to allow users to easily vary their choices from our defaults.

\begin{table*}
\caption{Initial Values and Default Settings for Binary Population Synthesis Simulation with {\sc{COMPAS}}.  Note: \COMPAS users who wish to provide a similar table accompanying their publication can find a template at \url{https://github.com/FloorBroekgaarden/templateForTableBPSsettings}. }
\label{tab:COMPAS}
\centering
\resizebox{\textwidth}{!}{%
\begin{tabular}{lll}
\hline  \hline
Description and Name                                 														& Value/Range                       & Note/Setting   \\ \hline  \hline
\multicolumn{3}{c}{Initial conditions}                                                                      \\ \hline
Initial mass \monei                               															& $[5, 150]$\Msun    & \citet{2001MNRAS.322..231K} IMF  $\propto  {\monei}^{-\alpha_{\rm{IMF}}}$  with $\alpha_{\rm{IMF}} = 2.3$ for stars in this mass range	  \\
Initial mass ratio $\qi = \mtwoi / \monei $           												& $[0.01, 1]$                          &       We assume a flat mass ratio distribution  $p(\qi) \propto  1$ with \mtwoi $\geq 0.1\Msun$   \\
Initial semi-major axis \ai                                            											& $[0.01, 1000]$\AU & Distributed flat-in-log $p(\ai) \propto 1 / {\ai}$ \\   
Initial metallicity \Zi                                           											& $[0.0001, 0.03]$                 & Distributed flat-in-log $p(\Zi) \propto 1/\Zi$      \\
Initial orbital eccentricity \ei                                 							 				& 0                                & All binaries are assumed to be circular at birth  \\
%
\hline
\multicolumn{3}{c}{Fiducial parameter settings}                                                            \\ \hline
Chemically homogeneous evolution & Enabled & Following \citet{Riley:2020}, ``pessimistic'' version checking for threshold throughout evolution (\S\ref{subsec:rotation}).\\
Stellar winds  for hydrogen rich stars                                   																&      \citet{Belczynski:2010ApJ}    &   Based on {\citet{2000A&A...362..295V,2001A&A...369..574V}}, including  LBV wind mass loss with $f_{\rm{LBV}} = 1.5$.   \\
Stellar winds for hydrogen-poor helium stars &  \citet{Belczynski:2010ApJ} & Based on   {\citet{1998A&A...335.1003H}} and  {\citealt{2005A&A...442..587V}}.  \\

%
Mass transfer stability criteria & $\zeta$-prescription & Based on \citet[][]{2018MNRAS.481.4009V} and references therein     \\ 
 Mass transfer accretion rate & Thermal timescale & Limited by thermal timescale for stars  \citet[][]{2018MNRAS.481.4009V,Vinciguerra:2020} \\ 
 & Eddington-limited  & Accretion rate is Eddington-limited for compact objects  \\
Non-conservative mass loss & Isotropic re-emission &  {\citet[][]{1975MmSAI..46..217M,1991PhR...203....1B}} \\ 
& &  {\citet{Soberman:1997mq,2006csxs.book..623T}} \\
 Case BB mass transfer stability                                														& Always stable         &       Based on  \citet{2015MNRAS.451.2123T,Tauris:2017omb,2018MNRAS.481.4009V}    (\S\ref{subsec:binary_mass_transfer})      \\ 
 Circularisation at the onset of RLOF                               														& On        &     Instantly circularised to periapsis (\S\ref{subsec:binary_mass_transfer})       \\ 
 %
%
CE prescription & $\alpha-\lambda$ & Based on  \citet{1984ApJ...277..355W,1990ApJ...358..189D}  \\
 CE efficiency $\alpha$-parameter                     												& 1.0                               &      (\S\ref{subsubsec:binary_common_envelope})        \\
CE $\lambda$-parameter                               													& $\lambda_{\rm{Nanjing}}$                             &        Based on \citet{2010ApJ...716..114X,2010ApJ...722.1985X} and  \citet{2012ApJ...759...52D}       \\
 Hertzsprung gap (HG) donor in {CE}                       														& Pessimistic                       &  Defined in \citet{2012ApJ...759...52D}:  HG donors do not survive a {CE}  phase        \\
%
%
{SN} natal-kick magnitude for white dwarfs & 0 & We assume WDs do not receive natal kicks (\S\ref{subsubsec:white-dwarf-formation})\\
{SN} natal-kick magnitude \vk  for \ac{NS}                       									& $[0, \infty)$\kms & Drawn from Maxwellian distribution    with standard deviation $\sigma_{\rm{rms}}^{\rm{1D}}$          \\
{SN} natal-kick magnitude for \ac{BH}                       									& $[0, \infty)$\kms & Reduced relative to NS kicks by the fallback fraction \citep{Fryer:2012ApJ}, see Section \ref{subsec:kicks}      \\
 {SN} natal-kick polar angle $\thetak$          											& $[0, \pi]$                        & $p(\thetak) = \sin(\thetak)/2$ \\
 {SN} natal-kick azimuthal angle $\phi_k$                           					  	& $[0, 2\pi]$                        & Uniform $p(\phi) = 1/ (2 \pi)$   \\
 {SN} mean anomaly of the orbit                    											&     $[0, 2\pi]$                             & Uniformly distributed  \\
 Core-collapse  {SN} remnant mass prescription          									     &  Delayed                     &  From \citet{Fryer:2012ApJ}, which  has no lower {BH} mass gap  \\%
 USSN  remnant mass prescription          									     &  Delayed                     &  From \citet{Fryer:2012ApJ}   \\%
ECSN  remnant mass prescription                        												&                                 $m_{\rm{f}} = 1.26\Msun$ &      Based on Equation~8 in \citet{1996ApJ...457..834T}          \\
 Core-collapse  {SN}  velocity dispersion $\sigma_{\rm{rms}}^{\rm{1D}}$ 			& 265\kms           & 1D rms value based on              \citet{2005MNRAS.360..974H}                          \\
 USSN  and ECSN  velocity dispersion $\sigma_{\rm{rms}}^{\rm{1D}}$ 							 	& 30\kms             &            1D rms value based on e.g.,    \citet{2002ApJ...571L..37P}, \citet{2004ApJ...612.1044P}    \\
 PISN/PPISN remnant mass prescription               											& \citet{Marchant:2019}                    &       As implemented in \citet{Stevenson:2019rcw}      \\
 Maximum NS mass                                      & $\rm{max}_{\rm{NS}} = 2.5$\Msun &   Mass division between NS and BH   (\S\ref{subsubsec:supernovae})        \\
Tides and rotation & & No tides and/or rotation except chemically homogeneous evolution\\
\hline
\multicolumn{3}{c}{Simulation settings}                                                                     \\ \hline
%
Binary fraction                                      & $f_{\rm{bin}} = 1$ &       Corrected factor to be consistent with, e.g., {\citet[][]{2017IAUS..329..110S}}        \\
Solar metallicity \Zsun                             & \Zsun = 0.0142 & Based on {\citet{Asplund}} \\
\hline \hline
\end{tabular}%
}
\end{table*}

\subsubsection{Program Options}\label{subsubsec:COMPAS_options}

\COMPAS provides a rich set of configuration parameters via program options, allowing users to vary many parameters that define the initial attributes and/or affect the evolution of single and binary stars being evolved. 
Furthermore, \COMPAS allows some parameters to be specified as ranges or sets of values via the program options, allowing users to specify a grid of parameter values on the command line. 
Combining command-line program options, particularly ranges and sets, with a grid file allows users great flexibility in specifying more complex combinations of parameter values.

\subsubsection{Grid files}
\label{subsubsec:COMPAS_grid_files}

A grid file allows users to specify, in plain text, initial values and physics assumptions for multiple systems for both \ac{SSE} and \ac{BSE}. 
Each line of a grid file is used by \COMPAS to set the initial values of an individual single star (\ac{SSE}) or an individual binary system (\ac{BSE}), and the physics assumptions to be used to evolve the star or system.

\subsection{Output}
\label{subsec:COMPAS_output}

\COMPAS provides real-time status information during the evolution of systems. Detailed and summary information about the star or system being evolved is written to log files as the evolution proceeds.

A number of \COMPAS log files may be produced depending upon the simulation type (\ac{SSE} or \ac{BSE}) and user specifications. These log files record, for each star or system being evolved:

\begin{itemize}
    \item {summary information at the completion of evolution,}
    \item {detailed information at each time step,}
    \item {detailed information at the time of each stellar type switch,}
    \item {summary information for all \ac{SN} events,}
    \item {summary information for all \ac{CE} events during \ac{BSE},}
    \item {detailed information for all \ac{RLOF} events during \ac{BSE},}
    \item {summary information for all \acp{DCO} formed during \ac{BSE}, and}
    \item {detailed pulsar evolution information.}
\end{itemize}

\COMPAS log files are created and written as \ac{HDF5} files,\footnote{\url{https://www.hdfgroup.org/}}, \ac{CSV} files, \ac{TSV} files, or plain text files, as specified by the user.

The \COMPAS software suite includes a Python postprocessing script to combine all \COMPAS output \ac{HDF5}, \ac{CSV}, or \ac{TSV} files into a single \ac{HDF5} file, which is especially useful if a single large experiment is spread over several virtual machines.

\section{Single Stellar Evolution}
\label{sec:single_star}

As stars evolve, they experience nuclear fusion while balancing gravity with pressure and radiating away excess energy.  Consequently, their composition, radius, temperature, and luminosity all change, they may lose mass in stellar winds, and sufficiently massive stars may explode in supernovae at the end of their lives.

COMPAS currently relies on rapid algorithms that provide estimates for how various fundamental stellar properties, such as their radii and luminosities, change as a star evolves through different evolutionary phases. 
The algorithms that capture how stars evolve---the SSE library---are at the core of the \COMPAS code. These routines govern the evolution of single stars, as the name suggests, but they are also used to capture how a star evolves under the external influence of a binary companion.

In this section we begin by giving an overview of the evolutionary algorithms implemented in COMPAS that govern the evolution of the main properties of stars (Section~\ref{subsec:evol_alogrith}). We then discuss some of the key evolutionary stages that stars evolve through (Section~\ref{subsec:evolutionary_stages}), and highlight several important evolutionary timescales in stellar evolution (Section~\ref{subsec:timescales}). 
We proceed to describe the effects of rapid rotation that are included in COMPAS (Section~\ref{subsec:rotation}), and describe the prescriptions available in COMPAS for incorporating mass loss in stellar winds (Section~\ref{subsec:winds}). We end this section by discussing the properties of stellar remnants: \acp{WD}, \acp{NS}, and \acp{BH}.

\subsection {Evolutionary Algorithms}
\label{subsec:evol_alogrith}

COMPAS computes the properties of a star (such as luminosity, radius, or core mass) as functions of a star's mass, metallicity, and age using analytic formulae, fit to detailed stellar models. 
The \ac{ZAMS} radius $R_\mathrm{ZAMS}$ and luminosity $L_\mathrm{ZAMS}$ are calculated as functions of mass and metallicity using the analytic formulae from \citet{1996MNRAS.281..257T}. 
During the evolution we use the formulae from \citet{Hurley:2000pk}, developed to match the detailed stellar models of \citet{Pols:1998MNRAS} (see also  \citealt{1989ApJ...347..998E} and \citealt{1997MNRAS.291..732T} for the basis of this approach).

The \citet{Pols:1998MNRAS} stellar models are for nonrotating stars and span \ac{ZAMS} masses between 0.1\Msun and 50\Msun. 
The original models do not include any mass loss. We incorporate mass loss following \citet{Hurley:2000pk} as described in Section~\ref{subsec:winds}.
Since the \citet{Hurley:2000pk} formulae are polynomials in $M_\mathrm{ZAMS}$, they can easily be extrapolated to higher masses by  evaluating them outside of this range. We find that the fits extrapolate smoothly to at least 150\Msun. We note that this approach is not ideal, but significantly improving upon this is far from trivial. The evolution of high-mass stars is still very uncertain \citep{Maeder:2000review,Langer:2012,Agrawal:2020,Bowman:2020,Belczynski:2021}. This is particularly true for the later, faster evolutionary phases, where observations are scarce, and for internal properties such as core masses, which we cannot probe directly. Although grids of detailed models exist, they vary widely in their predictions.
Given the limitations in the accuracy of massive star models and the absence of robust rapid prescriptions that clearly provide a significant improvement, we opt for simply extrapolating the fits in the present version of \COMPAS. 

The models are applicable for metallicities between $Z=10^{-4}$ and $Z=0.03$. We \emph{do not} extrapolate in metallicity, as we find that some of the fitting formulae are not well behaved outside of this range. We discuss the implications for population studies in Section~\ref{sec:cosmic_history}.

\subsection{Evolutionary Stages}
\label{subsec:evolutionary_stages}

To introduce some of the evolutionary stages captured by the \ac{SSE} formulae, we briefly summarize the main evolutionary stages of single stars.  Our extremely brief sketch of stellar evolution generally follows the characteristic behavior of massive stars, and should not be taken as fully general.

We follow the lives of stars from the \ac{ZAMS} (we do not include pre-main sequence evolution in COMPAS). Stars spend most of their lives on the main sequence, fusing hydrogen in their cores. Following \citet{Hurley:2000pk}, we distinguish between low-mass ($M_\mathrm{ZAMS} \leq 0.7$\Msun) main sequence stars (stellar type 0) which are expected to be fully convective, and more massive main sequence stars ($M_\mathrm{ZAMS} > 0.7$\Msun, stellar type 1). During main sequence evolution, most stars increase their luminosity and decrease their effective temperature $T_\mathrm{eff}$.
Following the main sequence, sufficiently massive stars experience a rapid thermal timescale (see Section~\ref{subsec:timescales}) phase of expansion, evolving to lower effective temperatures at near constant luminosity. This phase of evolution is sometimes known as the \ac{HG}, as the short timescale leads to a paucity of stars observed in this phase. 
Stars then begin a longer-lived phase of \ac{CHeB}, evolving onto and up the giant branch. We note that at low metallicity,  massive stars may not reach the giant branch before beginning \ac{CHeB}. 
Helium-shell-burning stars evolve along the \ac{AGB}. 
We choose to use the name \ac{EAGB} for H-rich massive stars with a C/O core, in addition to its usual meaning.
Stars that lose their outer hydrogen envelopes, either through stellar winds or binary mass transfer, become \ac{HeMS} stars, which then evolve analogously to hydrogen-rich stars through to the \ac{HeGB}.  
Finally, most stars end their lives as a stellar remnant, either a \ac{WD}, \ac{NS} or \ac{BH} depending on their initial mass, while some stars leave behind no remnant. 

\begin{figure*}[htb]
    \centering
    \includegraphics[width=2\columnwidth]{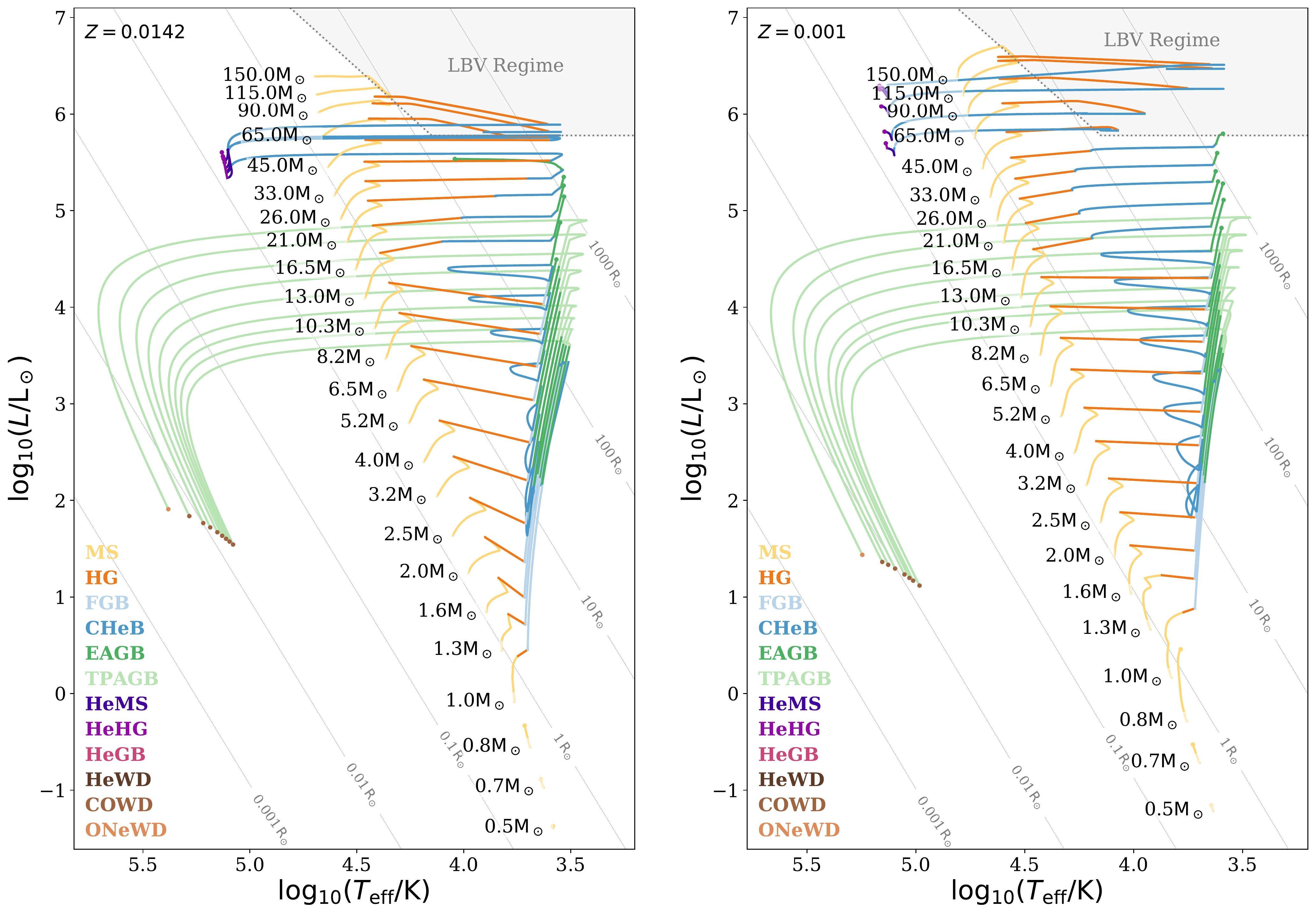}
    \caption{Evolutionary tracks in the Hertzsprung--Russell diagrams for single stars with \ac{ZAMS} masses between 0.5 and 150\Msun at solar ($Z = 0.0142$) and low ($Z = 0.001$) metallicity using \COMPAS default settings, limited to a maximum evolution time of 14 Gyr. The shaded region (bounded by the Humphreys--Davidson limit and a minimum luminosity) illustrates the regime in which luminous blue-variable (LBV) mass loss dominates. Diagonal contours show lines of constant radii.}
    \label{fig:hr_diagram_single}
\end{figure*}

We show the evolution of single stars of particular masses in the mass range 0.5--150\Msun at solar ($Z = 0.0142$, \citealt{Asplund}) and low ($Z = 0.001$) metallicity in the Hertzsprung--Russell diagram according to COMPAS in Figure~\ref{fig:hr_diagram_single}, with the different evolutionary phases identified for each track. 

\begin{table}[t]
\caption{Stellar phases and abbreviations used in COMPAS. Types 0--15 are from \citet{Hurley:2000pk}, whilst CHE stars (type 16) are an addition in COMPAS \citep{Riley:2020}.}
\label{table:SSE_stellar_phases}
\centering
\begin{tabular}{@{}lll@{}}
\toprule
Stellar phase & Abbreviation & Number \\ \midrule
Main sequence, $M < 0.7$ M$_\odot$ & MS & 0 \\
Main sequence, $M > 0.7$ M$_\odot$ & MS & 1 \\
Hertzsprung gap & HG & 2 \\
First giant branch & FGB & 3 \\
Core helium burning & CHeB & 4 \\
Early asymptotic giant branch\footnote{We apply the term EAGB to all H-rich stars with an inert C/O core that are primarily powered by He shell fusion, regardless of mass or location in the Hertzsprung-Russell diagram.} & EAGB & 5 \\
Thermally pulsing \\ \quad asymptotic giant branch & TPAGB & 6 \\
Helium main sequence & HeMS & 7 \\
Helium hertzsprung gap & HeHG & 8 \\
Helium giant branch & HeGB & 9 \\
Helium white dwarf & HeWD & 10 \\
Carbon-oxygen white dwarf & COWD & 11 \\
Oxygen-Neon white dwarf & ONeWD & 12 \\
Neutron star & NS & 13 \\
Black hole & BH & 14 \\
Massless remnant & MR & 15 \\
Chemically homogeneously evolving & CHE & 16 \\
... & NONE\footnote{Stellar type 19 (NONE) can sometimes appear as a temporary placeholder in COMPAS outputs before the stellar type is set.} & 19 \\
\bottomrule
\end{tabular}
\end{table}

A complete list of stellar phases is given in Table~\ref{table:SSE_stellar_phases}, following \citet{Hurley:2000pk}. See Figure \ref{fig:SSE_ClassDiagram} for a schematic of the corresponding classes in the COMPAS code.  

\begin{figure*}
\centering
\includegraphics[width=\textwidth]{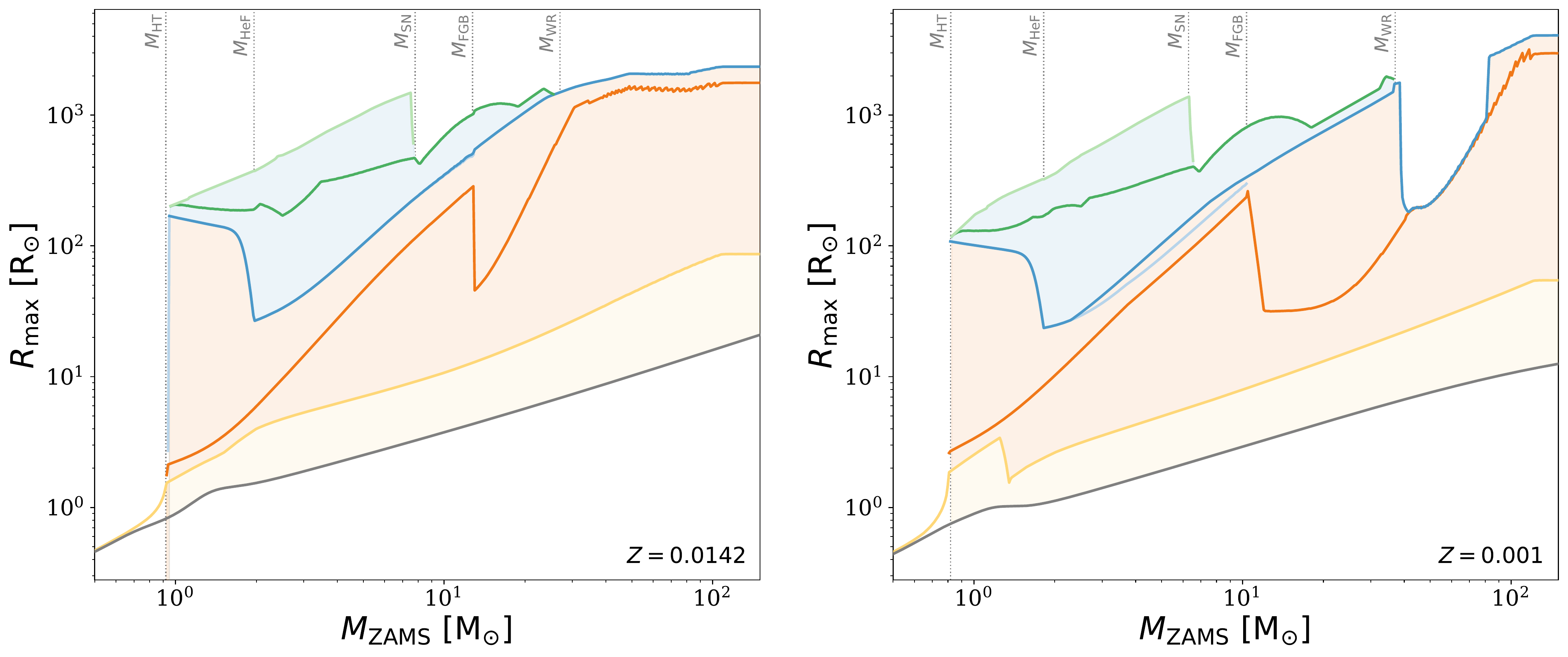}
\caption{Maximum radial extent of stars with \ac{ZAMS} masses between 0.5 and 150\Msun during different phases of stellar evolution, from the main sequence to the helium giant branch, using default settings. Solid lines show the maximum extent of stars during a given evolutionary phase, whose colors have the same meaning as in Figure \ref{fig:hr_diagram_single}.  We additionally show $R_{\rm ZAMS}$ in gray to indicate how much the star expands during the main sequence. Note that the FGB curve is invisible when it overlaps with the CHeB curve. Shading indicates regions in which different types of mass transfer can occur due to the expansion of the donor; on the main sequence (yellow), after hydrogen exhaustion (orange) and after helium exhaustion (blue). The dashed lines indicate important transition masses in the \citet{Hurley:2000pk} fitting formulae, where $M_{\rm HT}$ is the minimum mass of a star that will complete its main sequence in a Hubble time, $M_{\rm HeF}$ is the maximum initial mass for which helium ignites degenerately in a helium flash, $M_{\rm SN}$ is the mass above which a star collapses into an \ac{NS} or \ac{BH}, possibly with a supernova explosion, $M_{\rm FGB}$ is the maximum initial mass for which helium ignites on the first giant branch and $M_{\rm WR}$ is the minimum mass at which a star will self-strip to become a Wolf--Rayet star.}
\label{fig:radius}
\end{figure*}

Figure~\ref{fig:radius} shows the maximum radial extent of a star during each evolutionary phase for a star with \ac{ZAMS} mass between $0.5$ and $150\Msun$ at solar ($Z = 0.0142$) and low ($Z =0.001$) metallicity. Phases during which a star expands are the phases when binary interactions are most likely to occur.  We discuss binary interactions further in Section~\ref{sec:binary_evolution}.

\subsection{Evolutionary Timescales}
\label{subsec:timescales}

Three key timescales in single and binary stellar evolution -- dynamical, thermal, and nuclear -- often create a very convenient timescale hierarchy.   The separation of timescales allows many approximations to be made.
 
The shortest timescale is almost always the dynamical (or freefall) timescale, defined as
\begin{equation}
    \tau_\mathrm{dyn} = \sqrt{\frac{R^{3}}{GM}} \approx 1600\,\mathrm{s} \, \left( \frac{R}{\Rsun}\right)^{3/2} \, \left( \frac{M}{\Msun} \right)^{-1/2} ,
    \label{eq:dynamical_timescale}
\end{equation}
where $R$ is the radius of a star and $M$ is its mass.
The dynamical timescale is used as the minimum timescale in \COMPAS (with a minimum cutoff of 100 s).

The Kelvin-Helmholtz (or thermal) timescale is the time required for a star's internal energy $E_\mathrm{int}$ to be radiated at its current luminosity $L$, and is given by $\tau_\mathrm{KH} = E_\mathrm{int}/{L}$.  We estimate this as 
\begin{equation}
    \tau_\mathrm{KH} \approx 3.0 \times 10^{7} \, \mathrm{yr} \,
    \left(\frac{M}{\Msun}\right) \left(\frac{M_\mathrm{s/env}}{\Msun}\right) \left(\frac{R}{\Rsun}\right)^{-1} 
    \left(\frac{L}{\Lsun}\right)^{-1} ,
    \label{eq:thermal_timescale}
\end{equation}
%
where $L$ is the luminosity of the star and $M_\mathrm{s/env}$ 
is either the total mass of the star $M$ for stellar types without a clearly defined envelope, or the mass of the envelope $M_\mathrm{env}$ for stars with a clearly defined envelope \citep[see][for more details]{Hurley:2002rf}. 
This is used, e.g., when calculating thermal timescale mass transfer (see Section \ref{subsubsec:binary_stable_mass_transfer}).

\begin{figure}
    \centering
    \includegraphics[width=\columnwidth]{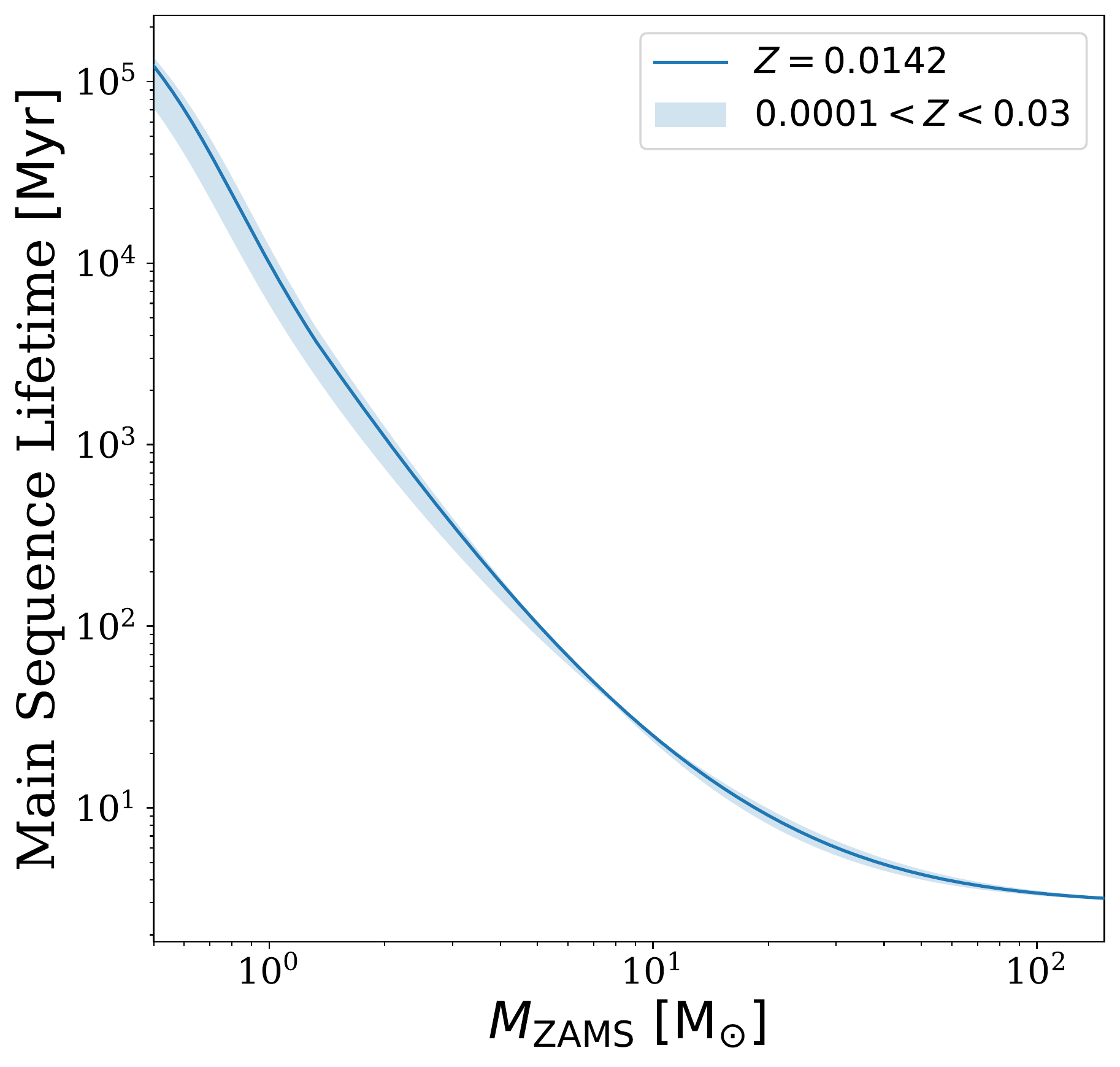}
    \caption{
        Main sequence lifetimes of stars between 0.5 and 150\Msun using \citet{Hurley:2000pk} fitting formulae across the range of metallicities at which these formulae can be used.
    }
    \label{fig:lifetimes}
\end{figure}

The nuclear timescale is relevant when nuclear fusion is setting the timescale at which the star evolves.
This applies, e.g., to main sequence stars that burn hydrogen in their centers, and stars that are undergoing central helium burning. The nuclear timescale is approximately
\begin{equation}
\begin{split}
    \tau_\mathrm{nuc} \approx \phi f_\mathrm{nuc} \frac{M c^2}{L}     
    &\approx 10^{10} \, \mathrm{yr} \,
    \left(\frac{M}{\Msun}\right)
    \left(\frac{L}{\Lsun}\right)^{-1}, 
\end{split}
\end{equation}
where $c$ is the speed of light, $\phi$ is the efficiency which the rest mass of the relevant reacting nuclei is converted into energy (for hydrogen, $\phi = 0.007$), and $f_\mathrm{nuc}$ is the fraction of the stellar mass that can serve as nuclear fuel. We do not use this timescale in any calculations; it is provided here, and optionally in \COMPAS outputs, for reference only.  Instead, we use the fitting formulae for stellar lifetimes in \citet{Hurley:2000pk}.  The main sequence lifetimes given by these fits are shown in Figure~\ref{fig:lifetimes}.  At masses of $\gtrsim 50$\Msun the \citet{Hurley:2000pk} formulae, which extrapolate the \citet{Pols:1998MNRAS} model grid in this mass range, overestimate the stellar lifetimes by approximately 30\% compared to the detailed stellar models considered in \citet{Agrawal:2020}. 

\subsection{Rotation}
\label{subsec:rotation}

Most stars rotate sufficiently slowly \citep[e.g.][]{Dufton:2013A&A,RamirezAgudelo:2013A&A} for the impact of rotation to be modest. Rotation is expected to enhance mass loss rates \citep[e.g.][see also Section~\ref{subsec:winds}]{ChiosiMaeder:1986,FriendAbbott:1986ApJ}, slightly increase main sequence luminosities and lifetimes \citep[e.g.][]{Talon:1996hc,Maeder:2000review}, and lead to increased core masses \citep[e.g.][]{Maeder:1987A&A,Langer:1992,Heger:1999ax}. 

However, very rapid rotation, especially coupled with significant tidal effects in very close binaries, can have a dramatic impact on the evolution of a star.  Sufficiently rapid rotation can lead to enhanced mixing within a star, which may lead to chemically homogeneous evolution \citep{Maeder:1987A&A}, where a star can burn almost all of its hydrogen into helium. 

The stellar tracks used in \COMPAS are based on nonrotating stellar models \citep{Pols:1998MNRAS}. We do not account for the effects of mild rotation, but we do implement \ac{CHE} following the recipes of \citet{Riley:2020}.  \COMPAS implements a metallicity-dependent rotational frequency threshold to determine whether a star is evolving chemically homogeneously. 
If a star is rotating faster than the threshold given by \citet{Riley:2020} at \ac{ZAMS}, we consider it to be evolving chemically homogeneously. 
We neglect the very limited radial evolution of a \ac{CHE} star and fix its main sequence radius equal to the \ac{ZAMS} radius of a nonrotating star of the same mass and metallicity. 
The main sequence evolution of a \ac{CHE} star thus follows the \citet{Hurley:2000pk} model of main sequence stars, albeit with a fixed radius. 
\COMPAS can be configured by the user to check, or not, the rotational frequency of the star against the \ac{CHE} threshold at every time step on the main sequence. 
In the default model, the rotational frequency check is enabled, and if the rotational frequency drops below the threshold value for \ac{CHE} (e.g., due to the orbit of the binary widening as a consequence of mass loss through winds), the star is thereafter evolved as a regular main sequence star (i.e., it immediately jumps to the track of a regular main sequence star of the same mass).  If the rotational frequency check is disabled by the user, the star evolves chemically homogeneously through its main sequence lifetime once it satisfies the \ac{CHE} threshold at \ac{ZAMS}. 
Finally, we assume that if a star evolves chemically homogeneously through the main sequence, it contracts directly into a naked helium star at the end of the main sequence, retaining its full mass at that point. 
Evolution then follows the \citet{Hurley:2000pk} models of helium stars.

\subsection{Wind Mass Loss (Single Star)}
\label{subsec:winds}

Stars lose mass through stellar winds. 
This impacts their evolution, and affects what remnants they form.
Stars lose mass throughout their lives through several mechanisms. 
Hot stars lose mass through steady-state line-driven winds \citep{CAK:1975ApJ,2001A&A...369..574V}, whilst the mechanism through which cool stars like red supergiants (RSGs) lose mass is less well understood theoretically, leading to most mass loss prescriptions for these stars being empirically derived \citep[e.g.,][]{1988A&AS...72..259D,2018MNRAS.475...55B}. 
In addition, stars close to the Humphreys--Davidson limit \citep{HumphreysDavidson:1994PASP} are known to experience eruptive mass loss. 
For a review of mass loss from massive stars, see \citet{Smith:2014ARA&A}. 
Low-mass stars generally experience weaker stellar winds than their high-mass counterparts. However, they can experience strong wind mass loss during the later stages of their evolution on the AGB \citep[see e.g.][for a recent review]{2018A&ARv..26....1H}.

Wind mass loss rates are highly uncertain, and can have a substantial impact on the evolution of a star \citep[e.g.][]{Renzo:2017,Belczynski:2019fed}. 
There is recent observational evidence that mass loss rates may be overestimated for certain evolutionary phases \citep[e.g.,][]{Smith:2014ARA&A,2018MNRAS.475...55B,Sander:2020,MillerJones:2021,Neijssel:2020CygX1}.

\COMPAS currently includes two simple analytic prescriptions for wind mass loss based on a combination of theoretical simulations and observational measurements: the original wind prescription from \citet{Hurley:2000pk} and an updated prescription from \citet{Belczynski:2010ApJ}, which is our default. 

\subsubsection{Hurley Model}
\label{subsubsec:winds_hurley}

\citet{Hurley:2000pk} define the total mass loss rate as the dominant mass loss rate during each stellar phase of the star (see Table~\ref{table:SSE_stellar_phases}), with a possible addition of LBV-like mass loss if the star is an \ac{LBV}. This can be summarized as
\begin{equation}
    \dot{M}_{H} =   \begin{cases}
                    \dot{M}_{\rm NJ}, & \text{MS} \\
                    \max \{ \dot{M}_{\rm NJ}, \dot{M}_{\rm KR}, \dot{M}_{\rm WR} \} + \dot{M}_{\rm LBV}, & \rm{if}\ \text{HG-CHeB} \\
                    \max \{ \dot{M}_{\rm NJ}, \dot{M}_{\rm KR}, \dot{M}_{\rm WR}, \dot{M}_{\rm VW} \}, & \text{AGB} \\
                    \max \{ \dot{M}_{\rm NJ}, \dot{M}_{\rm KR}, \dot{M}_{\rm WR}(\mu = 0) \}, & \text{HeMS-HeGB} \\
                    0, & \text{otherwise},
                \end{cases}
\end{equation}
where $\dot{M}_{\rm NJ}, \dot{M}_{\rm KR}, \dot{M}_{\rm WR}, \dot{M}_{\rm VW}, \dot{M}_{\rm LBV}$ are defined as follows.

For stars across the whole \ac{HR} diagram, \citet{Hurley:2000pk} apply
\begin{equation}
    \begin{split}
        \dot{M}_{\mathrm{NJ}}=9.6& \times 10^{-15} \max\left(0, \min\left(1, \frac{(L/\Lsun) - 4000 }{ 500}\right)\right) \\
        &\times\left(\frac{Z}{0.02}\right)^{1 / 2} \left(\frac{R}{\Rsun}\right)^{0.81}\\ &\times\left(\frac{L}{\Lsun}\right)^{1.24} \left(\frac{M}{\Msun}\right)^{0.16} \mathrm{M}_{\odot} \ \mathrm{yr}^{-1},
    \end{split}
\end{equation}
which is the mass loss rate from \citet{1990A&A...231..134N}, modified by the metallicity scaling $Z^{1/2}$ \citep{Kudritzki:1989AAP}. This is only non-zero for luminous massive stars with $L > 4000 \Lsun$.

For stars on the giant branch and beyond, this model adopts the results of \citet{1978A&A....70..227K} 
\begin{equation}
    \dot{M}_\mathrm{KR} = \eta \times 4 \times 10^{-13} \, \Msun \, \mathrm{yr}^{-1} \, 
    \left( \frac{L}{\Lsun} \right)
    \left( \frac{R}{\Rsun} \right)
    \left( \frac{M}{\Msun} \right)^{-1} ,
    \label{eq:mass_loss_reimers_gb}
\end{equation}
where $\eta$ is a phenomenological scaling parameter of order unity. By default, we adopt $\eta = 0.5$ following \citet{Hurley:2000pk}.

For stars on the asymptotic giant branch they use the results of \citet{VassiliadisWood:1993ApJ}
\begin{equation}
    \begin{split}
        \dot{M}_{\mathrm{VW}} = \min \Big( &10^{-11.4 + 0.0125 (P_0 - P_1)}, \\&1.36 \times 10^{-9} \frac{L}{\Lsun} \Big) \Msun \, \mathrm{yr}^{-1},
    \end{split}
    \label{eq:mass_loss_vassiliadis_wood_agb}
\end{equation}
where $P_0$, the Mira pulsation period, is given by
\begin{equation}
\begin{split}
    P_0 / \text{days} &= \text{min} \Big[1.995 \times 10^{3}, \\ 
    &8.51 \times 10^{-3} \left(\frac{M}{\Msun}\right)^{-0.9} \left(\frac{R}{\Rsun}\right)^{1.94} \Big] ,
    \label{eq:mass_loss_vassiliadis_wood_agb_P0}
\end{split}
\end{equation}
and $P_1 / \mathrm{days} = 100 \times \mathrm{max}(M/\Msun - 2.5, 0)$. 

Wolf--Rayet like wind mass loss is included for small hydrogen-envelope mass, $\mu < 1.0$, stars according to
\begin{equation}
    \dot{M}_\mathrm{WR} = 10^{-13} \left( \frac{L}{\Lsun} \right)^{1.5} (1 - \mu) \, \Msun \mathrm{yr}^{-1} .
    \label{eq:mass_loss_WR_hurley}
\end{equation}
where the parameter $\mu$ describes the ratio of the envelope mass to the total mass, so that $\mu = 1$ on the \ac{MS}, $0 < \mu < 1$ for stars with a developed core, and $\mu = 0$ for stripped stars with no hydrogen envelope. 
The full expressions for $\mu$ are given in Eq. 97 in \citet{Hurley:2000pk}.

We designate post main sequence stars with $L > 6 \times 10^{5} \Lsun$ and $\text{HD} \equiv  10^{-5} (R / \Rsun) (L / \Lsun)^{0.5} > 1.0$ as \ac{LBV}, following \citet{HumphreysDavidson:1994PASP}. 
For these \ac{LBV} stars \citet{Hurley:2000pk} add an \ac{LBV}-like mass loss, intended to account for eruptive mass loss in an averaged sense
\begin{equation}
    \dot{M}_{\mathrm{LBV}}=0.1\left(\text{HD}-1.0\right)^{3}\left(\frac{L / \Lsun}{6 \times 10^{5}}-1.0\right) \Msun \mathrm{yr}^{-1} .
\end{equation}

\subsubsection{Belczynski Model}
\label{subsubsec:winds_belczynski}

\citet{Belczynski:2010ApJ} use a model for stellar winds based on results from Monte Carlo radiative transfer simulations of \citet{2000A&A...362..295V,2001A&A...369..574V}. For stars that are not \acp{LBV} or helium stars, they define the mass loss rate as 
\begin{equation}
    \dot{M}_{\rm B} =   \begin{cases}
                            \dot{M}_{\rm H}, & \qquad\quad\ \ T/{\rm K} < 12500 \\
                            \dot{M}_{\rm Vink, BBJ}, & 12500 \le T/{\rm K} < 25000 \\
                            \dot{M}_{\rm Vink, ABJ}, & 25000 \le T/{\rm K}, \\
                        \end{cases}
\end{equation}
where $\dot{M}_{\rm H}$ is the Hurley mass loss rate defined in Section~\ref{subsubsec:winds_hurley} and the Vink mass loss rates are defined as
\begin{equation}
\begin{split}
    \log \left(\frac{\dot{M}_{\rm Vink, BBJ}}{M_\odot \, \mathrm{yr}^{-1}}\right) = -6.688 + 
    2.210 \log(L / 10^{5} \Lsun) \\
    -1.339 \log(M / 30 \Msun) 
    - 1.601 \log(V/2.0) + \\
    0.85 \log(Z / \Zsun)
    + 1.07 \log(T / 20000 \, \mathrm{K}) ,
    \label{eq:mass_loss_vink_1}
\end{split}
\end{equation}
below the bistability jump, where the ratio of the wind speed at infinity to the star's escape velocity is $V = v_\infty / v_\mathrm{esc} = 1.3$, and as
\begin{equation}
\begin{split}
    \log \left(\frac{\dot{M}_{\rm Vink, ABJ}}{M_\odot \, \mathrm{yr}^{-1}}\right)= -6.697 + 2.194 \log(L / 10^{5} \Lsun) \\ 
    -1.313 \log(M / 30 \Msun) - 1.226 \log(V/2.0) \\
    +0.85 \log(Z / \Zsun) + 0.933 \log(T / 40000 \mathrm{K}) \\
    - 10.92 \left[ \log(T / 40000 \mathrm{K}) \right]^{2} ,
    \label{eq:mass_loss_vink_2}
\end{split}
\end{equation}
above the bi-stability jump, where the ratio of the wind speed at infinity to the stars escape velocity is $V=~v_\infty / v_\mathrm{esc}=~2.6$.
The mass loss rate in equations \ref{eq:mass_loss_vink_1} and \ref{eq:mass_loss_vink_2} above scales with metallicity as
$\dot{M} \propto Z^{0.85}$, in agreement with observationally determined scaling of the mass loss rates of O and B stars with metallicity in the Milky Way and the Magellanic Clouds \citep[][]{2007A&A...473..603M}. 

For helium stars, \citet{Belczynski:2010ApJ} assume a mass loss rate
\begin{equation}
    \dot{M}_\mathrm{B, He} = f_\mathrm{WR} \times 10^{-13} 
    \left( \frac{L}{\Lsun} \right)^{1.5}
    \left( \frac{Z}{\Zsun} \right)^m
    \Msun \, \mathrm{yr}^{-1},
    \label{eq:mass_loss_helium_}
\end{equation}
from \citet{1998A&A...335.1003H}, with $m = 0.86$, as given by \citet{2005A&A...442..587V}. 
We have introduced the phenomenological scaling parameter $f_\mathrm{WR}$ to allow the strength of WR winds to be varied \citep{Barrett:2017fcw}. Our default choice is $f_\mathrm{WR} = 1$.

\ac{LBV} stars have high mass loss rates due to both line-driven winds and eruptive mass loss. The uncertainty in LBV mass loss rates is parametrized with a scaling parameter $f_\mathrm{LBV}$ with a default value of 1.5 \citep{Belczynski:2010ApJ}:
\begin{equation}
    \dot{M}_\mathrm{B, LBV} = f_\mathrm{LBV} \times 10^{-4} \, \Msun \, \mathrm{yr}^{-1} .
    \label{eq:mass_loss_flbv_belczynski}
\end{equation}

\subsection{Treatment of the Impact of Mass Loss and Gain}
\label{subsec:rejuvenation}

The stellar tracks computed by \citet{Pols:1998MNRAS} assumed no mass loss. 
When a star loses mass during a core-burning phase (either through stellar winds, as described in Section~\ref{subsec:winds}, or through mass transfer),  its luminosity will decrease.  This will extend the remaining lifetime.
Meanwhile, mass gain through mass transfer (\COMPAS does not account for accretion of winds from the companion) can \emph{rejuvenate} the star.  After mass gain such a star will appear younger than a single star of the same mass and age (a \emph{blue straggler}).

For a star losing or gaining mass through either mass transfer or winds (mass loss only) on the main sequence, we follow \citet{Hurley:2000pk} \citep[see also][]{1997MNRAS.291..732T} in modifying the lifetime according to
\begin{equation}
    t^{\prime} = f_\mathrm{rej} \frac{t_\mathrm{MS}^{\prime}}{t_\mathrm{MS}} t ,
    \label{eq:rejuvenation_hurley}
\end{equation}
where $t$ and $t_\mathrm{MS}$ are the effective age and main sequence lifetime prior to a small change in mass, and $t^{\prime}$ and $t_\mathrm{MS}^{\prime}$ are the age and main sequence lifetime of the star after a small amount of mass loss/gain. Mass changes for helium main sequence stars are treated by analogy with the above equation, with the main sequence lifetimes $t_\mathrm{MS}$ and $t_\mathrm{MS}^{\prime}$ replaced with $t_\mathrm{HeMS}$ and $t_\mathrm{HeMS}^{\prime}$ respectively \citep{Hurley:2000pk}.

The prefactor $f_\mathrm{rej}$ is unity for all mass-losing stars (and for low-mass $M \leq 0.7$\Msun main sequence stars, stellar type 0), whilst it is taken to be the ratio of the mass before/after mass gain for main sequence stars initially more massive than 0.7\Msun \citep{Belczynski:2008} and helium main sequence stars. 
The ages of Hertzsprung gap stars are updated following mass changes as described in \citet{Hurley:2000pk}. 
For giants with clearly decoupled cores and envelopes, we assume that changes to the envelope (such as mass loss/gain) do not affect the remaining lifetime. 
See \citet{2015ApJ...805...20S} and references therein for more details and possible improvements.

\subsection{Stellar Remnants}
\label{subsec:stellar_remnants}

\begin{figure}
    \centering
    \includegraphics[width=0.96\columnwidth]{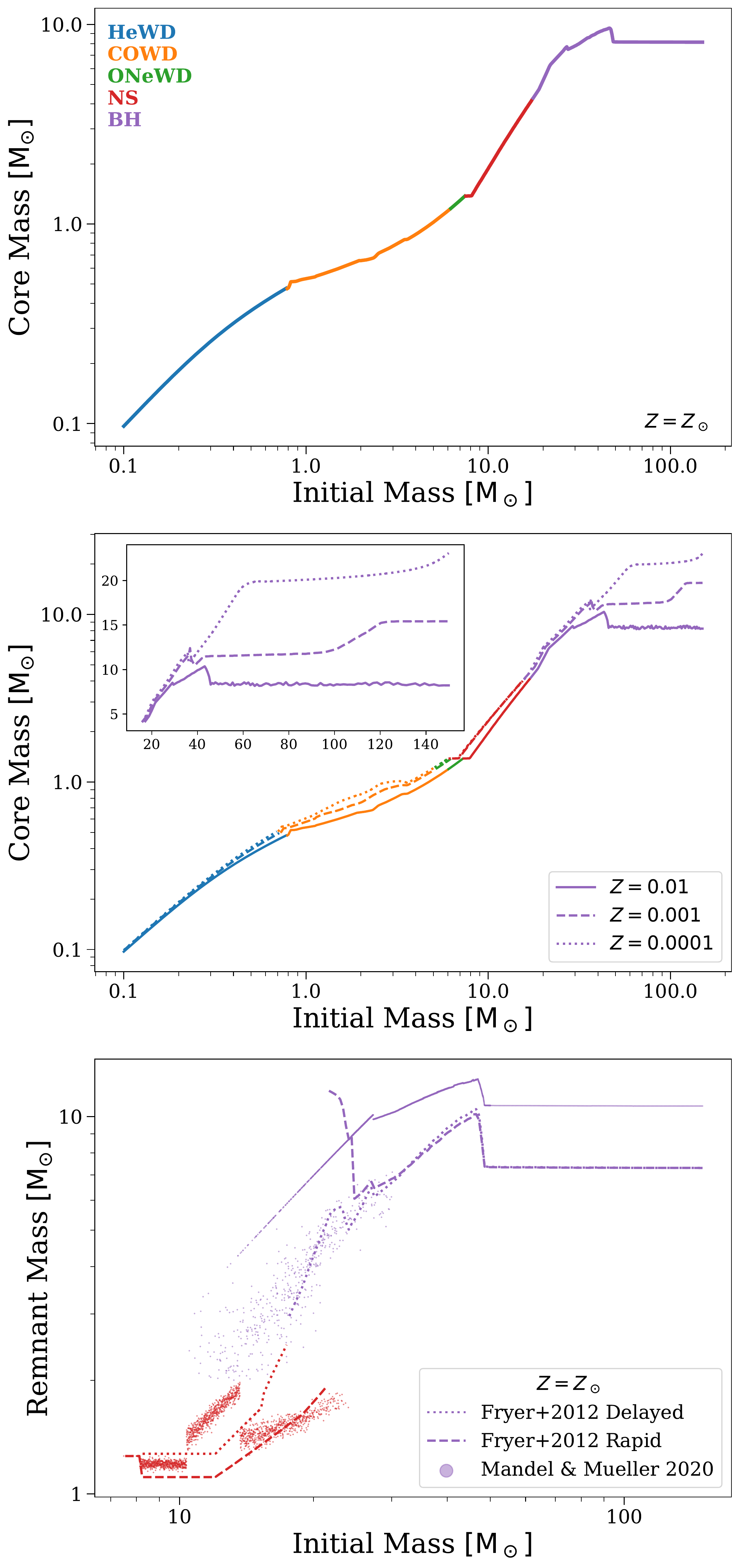}
    \caption{
    \textbf{Top:} the relation between the initial (\ac{ZAMS}) mass of single stars at solar metallicity ($Z = 0.0142$) and either the WD remnant mass or the CO core mass at the moment of compact object formation for NSs and BHs, assuming the \COMPAS default model, where colors denote the remnant type.
    \textbf{Middle:} the same relation at $Z = 0.01$ (solid), $Z = 0.001$ (dashed) and $Z = 0.0001$ (dotted) to show the metallicity dependence. The inset panel shows the same quantities as the main panel, but with linear axes, focusing on \ac{BH} masses.
    \textbf{Bottom:} the initial--compact object mass relation at $Z_\odot$ plotted for different remnant mass models (see Section \ref{subsubsec:CCSN}): the \emph{delayed} (dotted line) and \emph{rapid} (dashed line) prescriptions from \citet{Fryer:2012ApJ}, and the stochastic model from \citet{MandelMueller:2020} (points converging to a solid line).}
    \label{fig:remnant_mass}
\end{figure}

Standard stellar evolution theory predicts that low and intermediate mass stars (with initial masses $\lesssim8$\,M$_\odot$) typically end their lives as \acp{WD} (Section~\ref{subsubsec:white-dwarf-formation}), while more massive stars end their lives by collapsing into \acp{NS} and \acp{BH} (Section~\ref{subsubsec:supernovae}). We show the relation between initial masses, core masses and remnant masses of stars evolved under default \COMPAS assumptions in Figure~\ref{fig:remnant_mass}.

\subsubsection{White Dwarfs}
\label{subsubsec:white-dwarf-formation}

We distinguish between three different types of \acp{WD} based on their mass and composition, following \citet{Hurley:2000pk}. 
Stars that lose their envelopes prior to helium ignition leave behind \acp{HeWD}, while those that lose their envelopes after core-He exhaustion leave behind either \acp{COWD} or \acp{ONeWD} depending on their core masses upon reaching the base of the asymptotic giant branch. 
Larger core masses are associated with higher temperatures that allow carbon to fuse, forming oxygen-neon or oxygen-neon-magnesium cores. 
We assume that helium core masses below 1.6\Msun at the base of the asymptotic giant branch lead to \ac{COWD} formation, while core masses above that lead to \ac{ONeWD} formation.
For our default model, we find that single stars with initial masses $\lesssim 0.8$\,M$_\odot$ form \acp{HeWD} (though the evolutionary timescale for these stars to evolve is longer than a Hubble time),
while heavier stars with initial masses up to $\approx 7$ M$_\odot$ form \acp{COWD}. 
Only stars in a narrow mass range of $\approx 7$--8\,M$_\odot$ form \acp{ONeWD} (see upper panel of Figure~\ref{fig:remnant_mass}).
The boundary between stars that form WDs and those that form NSs/BHs is uncertain \citep[see, for example,][for further discussion]{Doherty:2017PASA}.
\acp{WD} have masses $\lesssim1.4$\Msun (the Chandrasekhar mass) in our model. 
We determine the radius of \acp{WD} following \citet{1997MNRAS.291..732T} and \citet{Hurley:2000pk}. 
\acp{WD} cool as they age \citep{1952MNRAS.112..583M}. 
We model the luminosity of \acp{WD} using the Mestel cooling track given in Equation 90 in \citet{Hurley:2000pk}.
By default we assume that \acp{WD} do not receive any kick during formation (but see \citealt{ElBaldry:2018MNRAS} for a discussion of evidence for small $\sim1$\,km\,s$^{-1}$ recoil velocities associated with \ac{WD} formation).

\subsubsection{Neutron Stars and Black Holes}
\label{subsubsec:supernovae}

Stars with initial masses more than $\sim 8$\,M$_\odot$ collapse into an \ac{NS} or a \ac{BH} at the end of their lives.  This collapse may be accompanied by a supernova explosion.  Asymmetry in the supernova may give the remnant a substantial momentum boost, commonly known as a \emph{natal kick} (see Section \ref{subsec:kicks}).
The mass of the supernova remnant (and by extension its stellar type), as well as the natal kick all depend on which kind of supernova it undergoes.
In \COMPAS we distinguish between several different types of supernovae, which are discussed in Section~\ref{subsec:supernova_types}. 

\COMPAS distinguishes \acp{NS} from \acp{BH} by the remnant mass; the default value of the maximum \ac{NS} mass is 2.5\Msun, following \citet{Fryer:2012ApJ}.

Figure~\ref{fig:remnant_mass} shows the core mass and remnant mass as a function of initial mass for single stars in the mass range 0.1--150\Msun in the \COMPAS default model.
The top panel of Figure~\ref{fig:remnant_mass} shows the mass ranges in which each type of stellar remnant is formed at solar metallicity. 
The middle panel of Figure~\ref{fig:remnant_mass} shows how the initial mass-core mass relation varies with metallicity, whilst the bottom panel shows how the relation depends on which remnant mass prescription is used for solar metallicity. 
This figure can be compared to results from \citet{Belczynski:2010ApJ} and \citet{Banerjee:2019jjs}.

By default, in \COMPAS we follow \citet{Hurley:2000pk} in assuming all \acp{NS} have a radius of $10$\,km. 
We also include the possibility of defining the \ac{NS} radius as a function of its mass $R_\mathrm{NS}(M_\mathrm{NS})$, as expected from the equation of state. 
Currently, the \ac{NS} equation of state from \citet{Akmal:1998cf} is implemented in \COMPAS, which gives \ac{NS} radii of 11--12\,km for \acp{NS} in the astrophysically relevant mass range 1--2.4\Msun,. The maximum \ac{NS} mass predicted by this equation of state is 2.4\Msun.

The \ac{NS} moment of inertia is assumed to follow the `universal' (equation of state insensitive) relation given by \citet{Lattimer:2004nj}. 
The luminosity of \acp{NS} is estimated following \citet{Hurley:2000pk}. 

For output purposes only, the ``radius'' of a \ac{BH} is given by its Schwarzschild radius
\begin{equation}
    R_\mathrm{BH} =  \frac{2 G M}{c^{2}} \approx 4.24 \times 10^{-6} \, R_\odot \, \frac{M}{M_\odot}
    \label{eq:radius_black_hole}
\end{equation}
and we follow \citet{Hurley:2000pk} in arbitrarily setting the \ac{BH} luminosity to $10^{-10}$\,L$_\odot$. 

Pulsar evolution is implemented in \COMPAS as described in \citet{Chattopadhyay:2020lff,Chattopadhyay:2019xye}. 
We assign a spin period and an initial magnetic field to each newly born \ac{NS}. 
We provide several options for the initial distribution of pulsar spin periods and magnetic fields, see \citet{Chattopadhyay:2019xye} for details.
The evolution of the pulsar spin period and spindown rate are followed as a function of time, assuming the canonical magnetic dipole model for a pulsar. In this model, pulsars spindown over time due to magnetic braking, but may be spun up again (or recycled) through mass accretion.
Our approach closely follows the methodology of \citet{Oslowski2011MNRAS}, \citet{Kiel:2008xw} and \citet{FaucherGiguere:2005ny} \citep[see also][]{Ye:2019luh}. 
Other pulsar properties, such as the pulsar luminosity or beaming fraction, are not directly computed by \COMPAS, but can easily be modeled in postprocessing \citep[see][for details]{Chattopadhyay:2020lff,Chattopadhyay:2019xye}.
Pulsar evolution is optional in \COMPAS, and is disabled in the default model.

\acp{BH} are assumed to be nonspinning.   However, the spins of \acp{BH} have been modeled by postprocessing \COMPAS data in a number of papers \citep[][]{Bavera:2019,Chattopadhyay:2020lff}.


\subsection{Supernova Types}
\label{subsec:supernova_types}

In \COMPAS we distinguish between several different types of supernovae. In the following, we describe the types of supernovae we model and the conditions under which each type of supernova is assumed to occur.
\COMPAS records whether the supernova progenitor star has a hydrogen-rich envelope, allowing for a crude estimation of whether it would appear observationally as a type I or II supernova.

\subsubsection{Electron-capture Supernovae }
\label{subsec:method-BPS-ECSN}
In \COMPAS we assume that a star undergoes an electron-capture supernova \citep[ECSN;][]{1980PASJ...32..303M, 1984ApJ...277..791N,1987ApJ...322..206N, 2008MNRAS.386..553I} if it has a helium core mass in the range 1.6--2.25\Msun \citep{Hurley:2002rf} at the base of the asymptotic giant branch, and the carbon-oxygen core mass reaches a threshold of 1.38\Msun. In our default model, this corresponds to a \ac{ZAMS} mass range of 7.5--8.1\Msun for single stars at $Z_\odot$.

The mass range of stars that undergo ECSNe is somewhat uncertain and model dependent. 
For example, \citet{2004ApJ...612.1044P} argued that a more realistic range of helium core masses leading to ECSNe is 1.4--2.5\Msun. 
\citet{2015ApJ...801...32A} and \citet{Vinciguerra:2020} use different core mass ranges for ECSNe (2--2.5\Msun and 1.83--2.25\Msun, respectively, where the latter is based on \citealt{Fryer:2012ApJ}, 
which could better reproduce observations of \acp{NS}). 
\citet{Willcox:2021kbg} argued that the (ZAMS) mass range of (effectively) single stars undergoing ECSNe cannot be wider than ~0.2\Msun to avoid overproducing low velocity pulsars, but ECSNe could be more common in binaries.

If a star undergoes an electron-capture supernova, we set its remnant to be a \ac{NS} with a mass of 1.26\Msun, as an approximation to the solution of Eq.~8 in \citet{1996ApJ...457..834T}, assuming a baryonic mass of $1.38\,M_\odot$. 

Massive oxygen-neon \acp{WD} (ONeWDs) close to the Chandrasekhar mass can accrete enough mass to undergo an accretion-induced collapse (AIC) to an \ac{NS} \citep{NomotoKondo:1991}.  While AIC is nominally possible in \COMPAS when the mass of a ONeWD reaches $1.38\,M_\odot$, when the ECSN prescription is followed for the remnant mass and natal kick, this is based on a very simplistic implementation of accretion onto WDs.

\subsubsection{Core-collapse Supernovae}
\label{subsubsec:CCSN}

Stars with helium core masses greater than 2.25\Msun at the base of the asymptotic giant branch undergo core-collapse supernovae (CCSNe) in \COMPAS, if/when their carbon-oxygen core mass reaches the threshold given by \citet{Hurley:2000pk}, where we replace their Chandrasekhar mass threshold with 1.38\Msun.

There is still a great deal of uncertainty regarding the mechanism of CCSNe, including which stars explode, how they explode, and what the properties of their remnants are (see \citealp{Muller:2020ard} and \citealp{Burrows:2020qrp} for recent reviews).
Some recent supernova simulations predict that whether a supernova is successful and leads to an explosion, or fails and leads to an implosion, is a nonmonotonic function of a star's initial mass \citep{Ugliano:2012ApJ,Sukhbold:2013yca,Nakamura:2014caa,Ertl:2015rga,Sukhbold:2019kzi}. 
COMPAS uses simple parameterized models (described below) to relate the properties of supernova remnants to their progenitor stars. 
We attempt to parameterize some of this uncertainty by including several different models.

By default, \COMPAS uses the \textit{delayed} supernova remnant mass prescription from \citet{Fryer:2012ApJ} to map the carbon-oxygen core masses of stars to compact object remnant masses during core-collapse supernova events. 
An alternative \textit{rapid} prescription assumes supernova explosions occur within 250 ms (compared to longer timescales assumed for the delayed model) and reproduces, by construction, a mass gap between NSs and BHs.  \citet{MandelMueller:2020} proposed a model for compact object masses and kicks which is stochastic, with both \ac{NS} and \ac{BH} formation possible in certain regions of parameter space; this model was implemented in COMPAS in \citet{Mandel:2020}.
\citet{Schneider:2020} predict that the history of mass transfer impacts the remnant mass, and we include their remnant mass prescription.  We also include the slightly older remnant mass prescriptions from \citet{Hurley:2000pk} and \citet{Belczynski:2008}, which define the remnant mass as a piecewise function or linear function, respectively, of the progenitor carbon-oxygen core mass; these models are provided for historical consistency reasons, but are somewhat outdated. 

We convert the baryonic mass of the remnant to a gravitational mass using Eq.~(13) of \citet{Fryer:2012ApJ} for \acp{NS}.  For \acp{BH}, the gravitational mass is assumed to be 0.1 M$_\odot$ less than the baryonic mass to account for mass lost in neutrinos in the default prescription \citep{Stevenson:2019rcw}.  This amount can be adjusted, or, alternatively, a fixed fraction of the mass can be lost in neutrinos during \ac{BH} formation as in Eq.~(14) of \citet{Fryer:2012ApJ}.

We show the difference in the initial-final remnant mass relation between the two models from \citet{Fryer:2012ApJ} and the stochastic model from \citet{MandelMueller:2020} in the bottom panel of Figure~\ref{fig:remnant_mass}.

\subsubsection{Ultra-stripped Supernovae}
\label{subsec:method-BPS-USSN}

Mass transfer from a helium star that re-expands after core helium burning \citep[so-called `case BB mass transfer';][]{1981A&A....96..142D,2015MNRAS.451.2123T} in short period binaries leads to severe stripping of the donor, leaving behind a helium envelope with mass $\lesssim 0.1$\Msun.  If the remaining stellar core is sufficiently massive to undergo core collapse, then we say that it undergoes an ultra-stripped supernova \citep[USSN;][]{Tauris:2013iua,2015MNRAS.451.2123T}. 
Due to the lack of envelope, USSNe remnants are \acp{NS} with characteristically lower mass, and may receive smaller natal kicks than typical \acp{NS}  
\citep{Suwa:2015saa, 2017MNRAS.466.2085M,2018MNRAS.479.3675M}. We discuss kicks further in Section~\ref{subsec:kicks}.
By default in \COMPAS we assume that case BB mass transfer is always stable \citep[cf.][]{2018MNRAS.481.4009V} and removes the entire helium envelope but none of the underlying carbon-oxygen core. 

\subsubsection{Pair-instability Supernovae}
\label{subsec:method-BPS-PISN}
Stars with helium cores in the mass range $\sim50$--$150$\Msun are believed to become unstable as a result of electron-positron pair production \citep{1964ApJS....9..201F,1967PhRvL..18..379B,Woosley:2017, 2019ApJ...887...53F}. 
This causes the radiation pressure support in the core to drop, causing the core to contract. 
As it contracts, the temperature increases, triggering explosive oxygen burning. 
This may reverse the contraction and completely unbind the star in a \ac{PISN} explosion, leaving no remnant behind  \citep{1964ApJS....9..201F,1967PhRvL..18..379B,1968Ap&SS...2...96F,2014A&A...565A..70K,2014A&A...566A.146K,2017MNRAS.464.2854K,Takahashi:2018kkb,2019ApJ...887...72L,Woosley:2019}.  \ac{BH} formation is expected again above a helium core mass of $\sim 150$\Msun \citep{Woosley2002TheStars,Woosley:2019}.
In addition, it is theoretically well established that stars with helium core masses in the range $\sim 30$--$60$\Msun lead to pulsational \acp{PISN} \citep[e.g.][]{Woosley:2017,2016MNRAS.457..351Y,2017MNRAS.470.4739S,Takahashi:2018kkb, 2019ApJ...887...53F, Marchant:2019,2020arXiv200205077R}, where material from the star is ejected in several supernova-like pulses, whilst the star returns to equilibrium between each pulse, and eventually undergoes an iron core collapse to form a \ac{BH}. 
Together, these effects lead to a dearth of \acp{BH} in the mass range $\sim45$--$130$\Msun and a possible excess in the $\sim35$--$45$\Msun mass range with the current \COMPAS default prescription.  

The implementation of \acp{PISN} in \COMPAS is discussed in detail in \citet{Stevenson:2019rcw}. 
In brief, by default stars with helium core masses in the range between 35 and 60 \Msun  lose mass through pulsational \acp{PISN} prior to collapse, while those with helium core masses between 60 and 135 \Msun explode in \acp{PISN} and leave no remnants behind.   We use fits based on results from \citet{Marchant:2019} in the \COMPAS default model for the relation between the helium core mass and the final presupernova mass in the pulsational \ac{PISN} regime.   
In addition, \COMPAS also provides the simple model from \citet{Belczynski:2016jno} to allow for comparison with StarTrack models, along with additional models based on detailed results from \citet{Woosley:2017} and \citet{2019ApJ...887...53F}.

\subsection{Supernova Natal Kicks}
\label{subsec:kicks}

Galactic pulsars are observed to have large proper motions, from which a distribution of their velocities is inferred \citep[e.g.][]{GunnOstriker:1970ApJ,LyneLorrimer:1994Nature,Hansen:1997zw,Arzoumanian:2002ApJ,2005MNRAS.360..974H,beniamini2016formation,2017A&A...608A..57V}. 
The high velocities are attributed to significant asymmetries in the supernova explosions which ``kick'' the pulsars, though the exact mechanism is uncertain and may be attributed to either hydrodynamic effects \citep[e.g.,][]{janka1994neutron,burrows1996pulsar,Wongwathanarat:2012zp} 
or neutrino emission \citep{Woosley:1987IAUS,1993A&AT....3..287B,Socrates:2005ApJ,Nagakura19}. 
See \citet{Lai:2000pk} for a broad overview of natal kicks.

By default, for \acp{NS} we draw the  the natal-kick magnitudes from a \textit{Maxwell-Boltzmann} distribution.
\begin{equation}
    p(v_\mathrm{kick} | \sigma_\mathrm{kick}) = \sqrt{\frac{2}{\pi}} \frac{v_\mathrm{kick}^{2}}{\sigma_\mathrm{kick}^{3}}
    \exp \left( -\frac{v_\mathrm{kick}^{2}}{2 \sigma_\mathrm{kick}^{2}} \right) ,
    \label{eq:maxwellian_kicks}
\end{equation}
with root-mean-square 1D velocity $\sigma_\mathrm{kick}$ (i.e., root-mean-square speed of $\sqrt{3} \sigma_\mathrm{kick}$). 
For CCSNe, we assume that $\sigma_\mathrm{kick} = \sigma_\mathrm{CCSN} = 265$\,km s$^{-1}$ \citep{2005MNRAS.360..974H}. 
ECSNe and USSNe are expected to have smaller kicks than standard iron core-collapse supernovae \citep[e.g.][]{Suwa:2015saa,Gessner:2018ekd,Muller:2018utr}. 
By default, we assume $\sigma_\mathrm{ECSN} = \sigma_\mathrm{USSN} = 30$\,km s$^{-1}$ \citep{2018MNRAS.481.4009V}.   These lower natal kicks for ECSN and USSN follow \citet{2002ApJ...571L..37P} and \citet{2004ApJ...612.1044P}, which are motivated by the subset of \acp{DNS} and NS-binary systems with low velocities and small eccentricities  \citep{2002ApJ...571..906B,2010ApJ...719..722S,beniamini2016formation,Tauris:2017omb}, as well as NS retention fractions in globular clusters \citep[e.g.][]{Pfahl:2001df}.

Several authors have proposed that \ac{NS} kicks should be proportional to the amount of ejecta, and inversely proportional to the remnant mass \citep{Bray2016,Bray:2018nzk,Giacobbo:2019fmo,MandelMueller:2020}. 
A scaling with ejecta mass would naturally account for reduced kicks in USSNe with low ejecta mass. 
In \COMPAS we have implemented the fits of this form from \citet{Bray:2018nzk} and \citet{MandelMueller:2020}, where the latter model self-consistently predicts both the remnant mass and natal kick.

COMPAS also includes a model for remnant masses and kicks based on the 1D parameterized supernova simulations of \citet{Muller:2016ujh} \citep[see][for details]{2018MNRAS.481.4009V}.

We further provide the option to use a uniform distribution of kick velocities up to some maximum $v_\mathrm{kick}^\mathrm{max}$, as well as kicks fixed at a specific value.

Whether \acp{BH} also receive natal kicks, and what their magnitudes are is an open astrophysical question. There is some evidence, both theoretical and observational, that \acp{BH} receive smaller kicks than \acp{NS} \citep[see e.g.][]{Janka:2013hfa,Mandel:2015eta,Repetto:2017gry,Atri:2019fbx}.

COMPAS currently includes four different models for \ac{BH} natal kicks: `full,' `reduced,' our default model `fallback,' and `zero.' In the `full' model, we assume that \acp{BH} receive the full kick drawn from Equation~\ref{eq:maxwellian_kicks}, where by default we assume $\sigma_\mathrm{kick} = \sigma_\mathrm{CCSN}$ for \acp{BH}. In both the `reduced' and the `fallback' models, we calculate the intensity of the kick velocity imparted to newly born \acp{BH} from the kick that a \ac{NS} would have received according to Equation~\ref{eq:maxwellian_kicks} with $\sigma_\mathrm{kick} = \sigma_\mathrm{CCSN}$. 
In the `reduced' model, we assume that \acp{NS} and \acp{BH} receive the same \emph{momentum} during the explosion, such that the kick velocity of a \ac{BH} should be rescaled according to
\begin{equation}
    v_\mathrm{BH} =  \frac{M_\mathrm{NS}}{M_\mathrm{BH}} v_\mathrm{NS} ,
    \label{eq:kick_reduced}
\end{equation}
where $M_\mathrm{NS}$ is taken to be $1.4\,\Msun$. In the `fallback' model, we scale the \ac{BH} kick by the fraction of mass falling back onto the proto-\ac{NS} $f_\mathrm{fb}$ \citep{Fryer:2012ApJ}
\begin{equation}
    v_\mathrm{BH} = v_\mathrm{NS} (1 - f_\mathrm{fb}) .
    \label{eq:kick_fallback}
\end{equation}
Note that unlike  \citet{Fryer:2012ApJ}, we apply Equation~\ref{eq:kick_fallback} even if the amount of mass falling back is less than $M_\mathrm{} = 0.2$\,M$_\odot$. 
In \COMPAS, by default we apply Equation~\ref{eq:kick_fallback} to all compact objects when using this prescription.
In the `zero' kicks model, \acp{BH} receive no kick during their collapse.

By default, we assume the supernova kick angle is drawn isotropically from the unit sphere in the rest frame of the supernova progenitor. However, there is some tentative observational evidence for spin-kick alignment in pulsars \citep[e.g.][]{Lai:2000pk,Johnston:2005ka,2012MNRAS.423.2736N,Yao:2021}, which may indicate that kicks are preferentially aligned to the spin axis of the progenitor. 
We provide various alternative models for preferential kick directionality, including along the progenitor spin axis (or in a finite cone around it), or within the progenitor spin plane (or in a wedge around it). 
We describe the impact of the supernova on the orbit of a binary in Section~\ref{subsec:binary_supernovae}.


\section{Binary Stellar Evolution}
\label{sec:binary_evolution}

Interacting binaries are the core case of study for \COMPAS.
Mass transfer is arguably the most important process in interacting binaries, modifying the component stars and the orbital properties \citep{Podsiadlowski:1992ApJ,Podsiadlowski:2010NewAR}. 
Binary evolution can lead to stellar mergers, disruption, or \ac{DCO} formation.

A binary in \COMPAS is parameterized by the orbital properties: the component masses of the primary $M_1$ and secondary $M_2$, the semi-major axis $a$ and eccentricity $e$.  The orbital angular momentum of a binary with nonrotating stars is 
\begin{equation}
    \label{eq:orbital_angular_momentum}
    J_{\rm{orb}} = \mu \sqrt{GM_{\rm{tot}}a(1-e^2)},
\end{equation}
where $M_{\rm{tot}}=M_1+M_2$ and $\mu = M_1M_2/M_{\rm{tot}}$ is the reduced mass.  
The effects of binary interactions on the stellar components, such as mass transfer and rejuvenation, are considered within our \ac{SSE} framework (Section \ref{sec:single_star}).
We currently only account for stellar rotation in the context of \ac{CHE} systems (Section \ref{subsec:rotation}).  In the absence of rotation and tides, the binary can be treated as a two-body problem in the point mass approximation.  

In this Section we present the details of our treatment of \ac{BSE}. 
In Section \ref{subsec:binary_wind_mass_loss} we discuss stellar winds in the context of binaries.
In Section \ref{subsec:binary_mass_transfer} we present our implementation of mass transfer.
In Section \ref{subsec:binary_supernovae} we describe the effect of supernovae on the orbit of the binary.
In Section \ref{subsec:gravitational_radiation} we describe our implementation of gravitational radiation.
Finally, in Section \ref{subsec:binary_caveats} we briefly discuss the main caveats in our implementation of binary evolution.

\subsection{Wind Mass Loss (Binary System)}
\label{subsec:binary_wind_mass_loss}
In Section \ref{subsec:winds} we presented the prescriptions and parameterization of stellar wind mass loss rates in \COMPAS .
Here we present the effect of wind mass loss on the orbit of the binary system.

We consider the case of gradual mass loss ($M/\dot{M}\gg P_\mathrm{orb}$) in which material rapidly leaves the system with the specific angular momentum of the mass-losing star (wind velocities are fast compared to the orbital velocity).  We assume that the fast winds are emitted spherically symmetrically from the star's surface \citep{1956AJ.....61...49H}.  This is  known as the \textit{Jeans mode} or \textit{fast winds mode} \citep{1963ApJ...138..471H}.  In the case, the orbit widens according to
\begin{equation}
    \label{eq:orbit_wind_mass_loss}
    \dfrac{\dot{a}}{a} = -\dfrac{\dot{M}_{\rm{tot}}}{M_{\rm{tot}}},
\end{equation}
which can be simplified as $aM_{\rm{tot}}=\rm{constant}$, while eccentricity is unchanged \citep{Dosopoulou:2016}.
Winds can thus be considered as an extreme case of nonconservative mass transfer (Section \ref{subsec:binary_mass_transfer}).

Currently we do not consider wind accretion \citep[e.g.,][]{Bondi1Hoyle:1944}, wind \ac{RLOF} \citep{MohamedPodsiadlowski:2007,HiraiMandel:2021}, or wind interaction with the companion \citep{BrookshawTavani:1993,2021arXiv210709675S}.

\subsection{Mass Transfer}
\label{subsec:binary_mass_transfer}
The physics and time scales involved in mass exchange are broad, complex, and their parameterizations can be convoluted.
In \COMPAS, we use a simplified approach to this complicated problem. 
Our approach is similar to that of  \cite{Belczynski:2001uc,Belczynski:2008}, \cite{Hurley:2002rf}, and \citet{postnov2014evolution}, among others.
We distinguish between a donor (subscript \textit{d}) and an accretor (subscript \textit{a}).
The donor is the star which transfers (and loses) mass, while the accretor gains mass.
The accretor can be a star or a compact object, and can fully retain the transferred mass (conservative mass transfer) or only a fraction of it (nonconservative mass transfer).
If the mass transfer is nonconservative, i.e.\ some mass is lost from the binary, there is a change in the total angular momentum of the binary.
We follow the orbital evolution of the binary through a mass transfer episode by taking the time derivative of Equation \ref{eq:orbital_angular_momentum} and rearranging it as
\begin{equation}
    \label{eq:derivative_orbital_angular_momentum}
    \dfrac{\dot{a}}{a} = 2\dfrac{\dot{J}_{\rm{orb}}}{J_{\rm{orb}}} - 2\dfrac{\dot{M}_{\rm{d}}}{M_{\rm{d}}} - 2\dfrac{\dot{M}_{\rm{a}}}{M_{\rm{a}}} + \dfrac{\dot{M}_{\rm{d}}+\dot{M}_{\rm{a}}}{M_{\rm{d}}+M_{\rm{a}}} + \dfrac{2e\dot{e}}{1-e^2} .
\end{equation}
The default \COMPAS assumption is that the binary is instantly circularized to periapsis, $r_p=a(1-e)$, at the onset of \ac{RLOF}.  \COMPAS options allow circularization with angular momentum conservation (at a separation of the semilatus rectum, $a (1-e^2)$) or mass transfer with unchanged eccentricity. 

We parameterize the fraction of mass lost by the donor which is accreted by the accretor with a factor $\beta$: 
\begin{equation}
\label{eq:mass_transfer_beta}
\dot{M}_{\rm{a}}=-\beta \dot{M}_{\rm{d}},
\end{equation}
with $0 \leq \beta \leq 1$.
We parameterize the change in angular momentum by assuming the nonaccreted matter leaves the system with $\gamma$ times the specific orbital angular momentum, i.e.\ $\dot{J}_{\rm{orb}}/\dot{M}=\gamma J_{\rm{orb}}/M$.
Following these assumptions we can rewrite Equation \ref{eq:derivative_orbital_angular_momentum} as\footnote{See the educational lecture notes on binary evolution by Onno Pols: \url{http://www.astro.ru.nl/~onnop/education/binaries_utrecht_notes/}}
\begin{equation}
    \label{eq:mass_transfer_orbital_evolution}
    \dfrac{\dot{a}}{a} = -2\dfrac{\dot{M}_{\rm{d}}}{M_{\rm{d}}} \Bigg[1-\beta\dfrac{M_{\rm{d}}}{M_{\rm{a}}}-(1-\beta)\left(\gamma + \frac12\right)\dfrac{M_{\rm{d}}}{M_{\rm{d}}+M_{\rm{a}}}  \Bigg],
\end{equation}
which is the equation we use to solve for the orbital evolution during a mass transfer episode.
Note that fully nonconservative mass transfer ($\beta=0$), where mass is lost with the specific angular momentum of the donor star ($\gamma = M_{\rm{a}}/M_{\rm{d}}$), is equivalent to fast wind mass loss from Section \ref{subsec:binary_wind_mass_loss}.

Before solving for the orbital evolution during a mass transfer phase, we first need to determine if mass transfer will occur (Section \ref{subsubsec:roche_lobe_overflow}) and, if so, whether the mass transfer episode will be dynamically stable (Sections \ref{subsubsubsec:binary_stability_criteria} and \ref{subsubsec:binary_stable_mass_transfer}). If the mass transfer episode is dynamically unstable, it will lead to a common envelope phase (Section \ref{subsubsec:binary_common_envelope}).

\subsubsection{Roche-lobe Overflow}
\label{subsubsec:roche_lobe_overflow}

A mass transfer phase can be initiated either by the radial expansion of a star as a consequence of stellar evolution or by a decrease in the binary separation.
The star overflows its Roche lobe and the surface material is transferred from the donor to the companion through the first Lagrangian point.
The first Lagrangian point is the juncture between the two Roche lobes, which are the regions that contain the gravitationally bound material around each star.
In \COMPAS we follow \cite{Eggleton:1983ApJ} and approximate the radius of the donor's Roche lobe normalized by the separation as
\begin{equation}
    \label{eq:roche_lobe_eggleton}
     r_\mathrm{RL} = R_{\rm{RL}}/a = 0.49 \frac{q_{\rm{RL}}^{2/3}}{0.6 q_{\rm{RL}}^{2/3} + \ln(1 + q_{\rm{RL}}^{1/3})},
\end{equation}
with $q_{\rm{RL}}=M_{\rm{d}}/M_{\rm{a}}$.  
Equation~\ref{eq:roche_lobe_eggleton} assumes point-like masses in a circular orbit.
The Roche lobe is generally shaped like a teardrop, but here the Roche lobe radius is defined as the radius of a sphere which has the same volume as the Roche lobe.
The condition for \ac{RLOF}, and therefore for initiating mass transfer, is when the radius of the star is larger than the Roche radius at periapsis, i.e. when $R > R_\mathrm{RL} (1-e)$.

\subsubsection{Stability Criteria}
\label{subsubsubsec:binary_stability_criteria}
When the condition for \ac{RLOF} is satisfied, we need to determine if the mass transfer episode will be dynamically stable or lead to a common envelope event.
In \COMPAS, the stability of the mass transfer phase is determined based on approximations to the mass-radius relationships $\zeta \equiv  \diff \ln R / \diff \ln M$ (see, e.g., \citealt{Soberman:1997mq}).
Namely, the response of the radius of the donor star to mass loss $\zeta_{*} \equiv  \diff \ln R_{*} / \diff \ln M$ is compared to the response of the Roche-lobe radius $\zeta_{\rm{RL}} \equiv \diff \ln R_{\rm{RL}} / \diff \ln M$  to mass transfer \citep{PaczynskiSienkiewicz1972,1987ApJ...318..794H,Soberman:1997mq}.  
If $\zeta_{*}  \geq \zeta_{\rm{RL}}$, then the mass transfer episode is assumed to be stable, otherwise, mass transfer is assumed to lead to a common envelope phase. 

We approximate the value of $\zeta_{*}$ based on stellar types (Section \ref{subsec:evolutionary_stages}).
We assume $\zeta_{*}=2$ for main sequence and HeMS stars and $\zeta_{*}=6.5$ for HG stars, as implemented in \citet{2018MNRAS.481.4009V}, based on typical values from \citet{2015ApJ...812...40G}.  These $\zeta_{*}$ values translate into critical mass ratios for stable mass transfer of $M_{\rm{d}}/M_{\rm{a}}<1.72$ ($M_{\rm{d}}/M_{\rm{a}}<2.25$) for fully conservative (fully nonconservative) mass transfer from main sequence donors and $M_{\rm{d}}/M_{\rm{a}}<3.83$ ($M_{\rm{d}}/M_{\rm{a}}<4.58$) for \ac{HG} donors.
For stellar types HG, FGB, CHeB, EAGB, and TPAGB (where we use these stellar types, defined in Section \ref{subsec:evolutionary_stages}, as loose proxies for having a convective envelope, but see, e.g., \citealt{Klencki2020bindingEnergy}) we follow \citet{Soberman:1997mq} in the form
\begin{equation}
    \label{eq:soberman}
    \begin{split}
    \zeta_\text{SPH} = \dfrac{2}{3} \bigg( \dfrac{m_\text{SPH}}{1-m_\text{SPH}} \bigg) - \dfrac{1}{3} \bigg( \dfrac{1-m_\text{SPH}}{1+2m_\text{SPH}} \bigg) \\
    -0.03m_\text{SPH} + 0.2 \bigg[ \dfrac{m_\text{SPH}}{1+(1-m_\text{SPH})^{-6}} \bigg],
    \end{split}    
\end{equation}
where $m_\text{SPH}=M_{\rm{core}}/M$ and $M_{\rm{core}}$ is the core mass as defined by \cite{Hurley:2000pk}. 
In COMPAS, we consider stripped post-helium-burning stars as a special case and assume by default that any mass transfer episode from HeHG and HeGB stellar types is always stable.
This is in agreement with the expected outcome of mass transfer episodes from stripped stars onto \acp{NS} or \acp{BH} as suggested by \citet{Tauris:2013iua,2015MNRAS.451.2123T}.
We currently do not model mass transfer from white dwarf donors (HeWD, COWD, and ONeWD).

The Roche-lobe mass-radius exponent $\zeta_{\rm{RL}}$ depends on the accreted mass fraction $\beta$ and the specific angular momentum that nonaccreted mass removes from the system (see Section \ref{subsubsec:binary_stable_mass_transfer} to see how we determine these).
We follow \citet{Soberman:1997mq} and  \citet{Woods:2012ApJ} in rewriting
\begin{equation}
    \label{eq:zeta_roche_lobe}
    \zeta_{\rm{RL}} = \dfrac{\diff \ln R_{\rm{RL}}}{\diff \ln M} = \dfrac{\partial \ln a}{\partial \ln M_{\rm{d}}} + \dfrac{\partial \ln r_\mathrm{RL}}{\partial \ln q_{\rm{RL}}}\dfrac{\partial \ln q_{\rm{RL}}}{\partial \ln M_{\rm{d}}},
\end{equation}
where the terms are taken from  Equation \ref{eq:mass_transfer_orbital_evolution} and the partial derivative of Equation \ref{eq:roche_lobe_eggleton}.

\subsubsection{Stable Mass Transfer}
\label{subsubsec:binary_stable_mass_transfer}

The amount of mass transferred during a dynamically stable mass transfer episode 
is calculated depending on the stellar type of the donor (Section \ref{subsec:evolutionary_stages}).
Broadly, we distinguish between stars that have a clear core/envelope separation (HG, FGB, CHeB, EAGB, TPAGB, HeHG, and HeGB) and those which do not (MS and HeMS).  We do not consider other donor types. 

For donor stars with a clear core/envelope separation we transfer their entire envelope.
For donor stars without a clear core/envelope separation, we calculate and remove the minimal mass necessary in order to have the donor fit within its Roche lobe.  Numerically, we accomplish this by using a root-finder with specified tolerances.  We compute the accreted mass fraction $\beta$ once, at the start of the mass transfer phase, as follows.

We compute the donor and accretor mass transfer rates.
For the donor, we assume a thermal timescale mass transfer rate
\begin{equation}
    \dot{M}_{\rm{d}} = M_{\rm{d}} / \tau_{\rm{KH,d}}.
\end{equation}
For the accreting object, the maximum accretion rate is limited depending on its stellar type \begin{equation}
\label{eq:mass_accretion_rate}
\dot{M}_{\rm{a}} = \begin{cases} 
    \dot{M}_{\rm{KH,a}} & \rm{if\ stellar\ accretor},\\
    \dot{M}_{\rm{edd}}  & \rm{if\ compact\ object\ accretor}.
       \end{cases}
\end{equation}
In Equation \ref{eq:mass_accretion_rate}, the thermal mass accretion rate is given by
\begin{equation}
    \dot{M}_{\rm{KH,a}} = C \cdot M_{\rm{a}} / \tau_{\rm{KH,a}},
\end{equation}
where the factor $C=10$ (default value) is assumed to take into account the expansion of the star due to mass transfer, following \citet{PaczynskiSienkiewicz1972, Neo:1977}, \citet{Hurley:2002rf} and \citet{2015ApJ...805...20S}.
Meanwhile, the Eddington-limited accretion rate is given by
\begin{equation}
	\dot{M}_{\rm{edd}} = 1.5 \times 10^{-8} \frac{R_*}{10\, \mathrm{km}} \frac{\mathrm{M}_\odot}{\mathrm{yr}},
\end{equation}
where for a black hole the radius $R_*$ is given by the Schwarzschild radius as in Equation \ref{eq:radius_black_hole}.
The Eddington accretion rate assumptions in \COMPAS are flexible and can be change to a user-specified function (see \citealt{vanSon:2020zbk} for more details).

By comparing the donor and accretor mass transfer rates, we can determine how conservative the mass transfer episode is.
Only for main sequence (MS and HeMS) and Hertzsprung gap (HG and HeHG) stars, mass accretion leads to stellar rejuvenation (Section \ref{subsec:rejuvenation}) as the accretor star transitions to a more massive and less evolved stellar track \citep{1997MNRAS.291..732T,Hurley:2002rf,Belczynski:2008}.
For fully conservative mass transfer, i.e. $\beta=1$, there is no angular momentum drained from the binary and Equation \ref{eq:mass_transfer_orbital_evolution} simplifies to $M_{\rm{d}}^2M_{\rm{a}}^2a=\rm{constant}$.
For nonconservative mass transfer, i.e. $\beta < 1$, the nonaccreted mass is lost, by default,  through \textit{isotropic re-emission} from the vicinity of the accreting star \citep[e.g.][]{1975MmSAI..46..217M,1991PhR...203....1B,Soberman:1997mq}.
Isotropic re-emission corresponds to $\gamma = M_{\rm{d}}/M_{\rm{a}}$.

We briefly point out some alternatives from the default mass transfer model which the user can choose in \COMPAS.
The user can specify a fixed value for $\beta$, which enforces that a fraction $\beta$ of the mass transferred by the donor is accreted, except for Eddington-limited accretion. The value of $\gamma$ can also be changed.
We have mentioned the Jeans ($\gamma = M_{\rm{a}}/M_{\rm{d}}$) mass loss mode in the context of winds, but we also include the user-defined possibility of choosing these mass loss modes in semiconservative and fully nonconservative mass transfer episodes.
\COMPAS also includes a mode of nonconservative mass transfer that represents mass loss via a circumbinary ring as an option \citep[see, e.g.,][]{Soberman:1997mq}. 
In this case, the nonaccreted mass carries a specific angular momentum $\gamma = \sqrt{a_{\rm{ring}}/a} M/\mu = \sqrt{2}M/\mu$ for a ring at radius $a_{\rm{ring}}=2a$.

\subsubsection{Common Envelope Phase (Dynamically Unstable Mass Transfer)}
\label{subsubsec:binary_common_envelope}

In \COMPAS, dynamically unstable mass transfer occurs if $\zeta_{*} < \zeta_{\rm{RL}}$ or if both stars simultaneously experience \ac{RLOF} (see Section \ref{subsubsec:binary_massive_over_contact_binaries} for an exception for \ac{CHE} binaries).
Dynamically unstable mass transfer leads to a \ac{CE} phase.
In this phase, the binary is engulfed in a shared envelope and experiences gas drag, which causes a dynamical timescale inspiral \citep{1976IAUS...73...75P,2001ASPC..229..239P,ivanova2013common}. 
\ac{CE} events are thought to be especially relevant for the formation of \acp{DCO} in tight orbits \citep{1976IAUS...73...35V,ivanova2013common}.

In \COMPAS we follow the energy formalism in the form of the $\alpha_{\rm{CE}}$-$\lambda$ prescription to estimate the post-CE orbital separation \citep{1984ApJ...277..355W,1990ApJ...358..189D}.
In the $\alpha_{\rm{CE}}$-$\lambda$ prescription the initial (pre-\ac{CE}) binding energy of the donor is equated to the orbital energy reservoir
\begin{equation}
    \label{eq:common_envelope_energy_alpha_formalism}
    E_\mathrm{bind} = \alpha_{\rm{CE}} \times \Delta E_\mathrm{orb},
\end{equation}
where $\Delta E_\mathrm{orb}$ is the difference between the binary orbital energies before and after the CE phase and $\alpha_{\rm{CE}}$ is a user-specified efficiency factor which parameterizes the fraction of the orbital energy that is used to unbind the \ac{CE}.  The default value is $\alpha_{\rm{CE}} = 1.0$.

The value of the binding energy of the envelope depends on the location of the envelope's inner boundary and the sources of energy considered.
We follow \cite{1990ApJ...358..189D} and express it in terms of a structure parameter $\lambda$:
\begin{equation}
    \label{eq:common_envelope_binding_energy}
    E_\mathrm{bind} = - \frac{G M M_{\rm{env}}}{\lambda R}.
\end{equation}
By default, \COMPAS calculates $\lambda$ using the ``Nanjing lambda'' prescription described by \citet{2010ApJ...716..114X,2010ApJ...722.1985X}, who provide fitting formulae for $\lambda$. 
Our implementation is identical to that of StarTrack \citep{2012ApJ...759...52D}, including the several improvements they have made to the $\lambda$ fits. 
This $\lambda$ was computed in two different ways; one using only the gravitational binding energy ($\lambda_\mathrm{g}$) and one that also includes the contribution of the full internal energy\footnote{Note that this includes the thermal energy, radiation energy, ionization energy, and the dissociation energy of molecular hydrogen.} ($\lambda_\mathrm{b}$).
\COMPAS allows the user to specify a linear combination of the two parameters, $\lambda = \alpha_{\rm{th}} \lambda_{\rm{b}} + (1-\alpha_{\rm{th}}) \lambda_{\rm{g}}$. 
The default is $\alpha_{\rm{th}} = 1$, i.e. including the full internal energy.

Alternative \COMPAS options to estimate $\lambda$ include prescriptions for calculating the envelope binding energy from \citet{Loveridge:2011ApJ}, fitting formulae to results from \citet{Kruckow:2016tti} as implemented in \cite{2018MNRAS.481.4009V}, and the option of using a fixed constant value.

We generalize Equations \ref{eq:common_envelope_energy_alpha_formalism} and \ref{eq:common_envelope_binding_energy} to include the potential case of a double-core \ac{CE} \citep{brown1995doubleCore} in the form

\begin{equation}
\label{eq:common_envelope_double_core}
\begin{split}    
    - \frac{G M_1 M_{\rm{1,env}}}{\lambda_1 R_1} - \frac{G M_2 M_{\rm{2,env}}}{\lambda_2 R_2} = \\ 
    = \alpha_{\rm{CE}} \Bigg ( \dfrac{G M_1 M_2}{2 a_{\rm{pre-CE}}} - \dfrac{G M_{\rm{1,core}} M_{\rm{2,core}} }{2 a_{\rm{post-CE}}} \Bigg ),
\end{split}
\end{equation}
where $a_{\rm{pre-CE}}$ and $a_{\rm{post-CE}}$ are the separation before and after the \ac{CE} phase, respectively; if one of the $M_{\rm{env}}=0$ then Equation \ref{eq:common_envelope_double_core} simplifies to the classic single-core energy formalism.
Given the simplicity of our parameterization and the short (dynamical) timescales involved in a \ac{CE} episode, we assume the phase is instantaneous. Equation~\ref{eq:common_envelope_double_core} is used to predict the post-CE orbital separation given the pre-CE binary parameters and the post-CE component masses.
The criterion for successful envelope ejection is $a_\text{post-CE} > R_1+R_2$. 

After a successful envelope ejection we always assume a circular orbit \citep[see, e.g.,][for discussion of this]{ivanova2013common}. Besides a successful envelope ejection, which leads to a close binary, the \ac{CE} phase can lead to immediate \ac{RLOF} or a stellar merger (Section \ref{subsubsec:binary_stellar_mergers}).
Immediate \ac{RLOF} implies that the stripped star or its companion is filling the respective post-\ac{CE} Roche lobe.
By default we allow these systems to engage in a mass transfer episode again, but we flag them so they can be considered as mergers in postprocessing.
 
A key uncertainty in \ac{CE} evolution is the fate of Hertzsprung gap donors \citep{Belczynski:2006zi}. 
Such stars are not expected to have developed a steep density gradient between core and envelope  \citep{2000ARA&A..38..113T,2004ApJ...601.1058I}  making it challenging to successfully eject the envelope. 
It is not clear whether Hertzsprung gap donor stars can survive \ac{CE} evolution, or whether instead, this  would lead to a merger. 
In order to account for this uncertainty, we adopt two extreme models following \citet{2012ApJ...759...52D}. In the ``optimistic" model, \ac{CE} events involving a Hertzsprung gap donor are treated in the same way as a more evolved star, determining the fate of the binary according to the energy budget (Equation~\ref{eq:common_envelope_double_core}). 
In the ``pessimistic" model, it is assumed that all \ac{CE} events involving a Hertzsprung gap donor result in a stellar merger. 
\COMPAS keeps track of systems that experience an ``optimistic" \ac{CE} event, allowing the user to remove them in postprocessing. For the results presented in this paper  we use the pessimistic model as the default assumption.

\ac{CE} events initiated by \ac{RLOF} from main sequence donors always lead to a stellar merger.  \ac{CE} events with main sequence accretors are treated as all other \ac{CE} events by default, but may optionally always lead to stellar mergers.  

There is no mass accretion onto the companion during a \ac{CE} phase in the default \COMPAS model. However, different optional \ac{CE} accretion rate prescriptions exist for the case of \ac{NS} accretors, including a user-defined fixed value, or following prescriptions from \cite{Oslowski2011MNRAS} and \cite{MacLeod2015ApJ}, as described by \citet{Chattopadhyay:2019xye}.
Mass accretion during a \ac{CE} phase involving a BH companion has been investigated with \COMPAS by \citet{vanSon:2020zbk}, who considered both accretion of a user-defined fixed fraction of the envelope and the Hoyle-Lyttleton \citep{HoyleLyttleton1939} accretion rate within a \ac{CE} following prescriptions from \cite{MacLeod2015ApJ} based on \cite{Chevalier1993}.

\subsubsection{Stellar Mergers}
\label{subsubsec:binary_stellar_mergers}

In \COMPAS, two stars are assumed to merge when $a \leq (R_1+R_2)$.  This can occur following runaway stable mass transfer, if the envelope fails to be ejected during a \ac{CE} phase, or if the direction of the supernova kick drastically shrinks the orbit.
Currently, stellar mergers are flagged and the calculation is stopped, without subsequent evolution of the merger product.

\subsubsection{Massive Overcontact Binaries}
\label{subsubsec:binary_massive_over_contact_binaries}

\COMPAS includes prescriptions for \ac{CHE} stars (Section \ref{subsec:rotation}), particularly in the context of \ac{CHE} binaries \citep{Riley:2020}.
\ac{CHE} binaries may arise from massive overcontact binaries, in which stars overflow their Roche lobes and share mass during the main sequence \cite[see, e.g.,][]{Marchant:2016wow}.
We therefore make an exception for them and relax our criteria for \ac{RLOF} and mergers \cite[see][]{Riley:2020}.
We consider that \ac{CHE} stars in massive overcontact binaries can be filling their Roche lobe throughout the main sequence as long as they don't overflow the second (outer) Lagrangian point \citep{Marchant:2016wow,Riley:2020}.
In our model, \ac{CHE} binaries filling the second (outer) Lagrangian point lead to an imminent stellar merger.

\subsection{Supernovae (Binary)}
\label{subsec:binary_supernovae}

In \COMPAS, supernovae lead to \ac{NS} or \ac{BH} formation\footnote{With the exception of \acp{PISN} as discussed in Section \ref{subsec:method-BPS-PISN}.} (the nomenclature is not necessarily associated with their observational signature).
Supernovae occur on timescales much shorter than the timescales we resolve in \COMPAS;
we therefore assume that they are instantaneous events and that they could occur with uniform probability over all orbital phases.

Supernovae affect the orbit of a binary via instantaneous mass loss and natal kicks \citep{1961BAN....15..265B,Hills:1983ApJ,Brandt:1994rr,Kalogera:1996ApJ,TaurisTakens:1998A&A,Hurley:2002rf}.
We follow Appendix B of \cite{Pfahl:2001df} to solve for the response of the orbital elements to a supernova.
This prescription accounts for the natal kick, mass ejection, interaction with the companion during the supernova, and modification of the center-of-mass velocity. 
If the postsupernova eccentricity of the binary exceeds one, we label the binary as gravitationally unbound and cease further calculations (with an option to continue stellar evolution of the noncompact-object companion, if any).

\subsection{Gravitational Radiation}
\label{subsec:gravitational_radiation}
Gravitational radiation releases energy and angular momentum from a binary, reducing both the orbital separation and the eccentricity. 
In \COMPAS we only consider \acp{GW} after \ac{DCO} formation and follow \cite{Peters:1964} to calculate the time to coalescence as
\begin{equation}
	\tau_{\rm{GW}} = \frac{15}{304} \frac{a_{\rm{DCO}}^4 c^5}{G^3 M_1 M_2 M_{\rm{tot}}} \kappa(e_{\rm{DCO}}),
	\label{eq:peters_inspiral_time}
\end{equation}
where $a_{\rm{DCO}}$ and $e_{\rm{DCO}}$ are the separation and eccentricity of the binary after the second supernova and $\kappa(e_{\rm{DCO}})$ is a function of the eccentricity given by
\begin{equation}
\begin{split}
	\kappa(e_{\rm{DCO}}) = 	 \left[  \dfrac{(1-e_{\rm{DCO}}^2)}{e_{\rm{DCO}}^{12/19}}  \left( 1+\frac{121}{304}e_{\rm{DCO}}^2  \right)^{-\frac{870}{2299}}  \right]^4\\ 
	 \times \int_0^{e_{\rm{DCO}}} \frac{e^{29/19} (1+\frac{121}{304}e^2)^{\frac{1181}{2299}}}{(1-e^2)^{\frac{3}{2}}} \, \diff e,
\end{split}
\end{equation}
where the integral is calculated as a Riemann sum over 10,000 linearly spaced eccentricity bins.
For almost circular ($e<0.01$) or very eccentric ($e>0.99$) binaries, we use the approximations as presented following Eq. 5.14 in \citet{Peters:1964}.
If $\tau_{\rm{GW}}$ is less than a Hubble time, computed with $H_0=67.8\ \rm{km\ s^{-1}\ Mpc^{-1}}$ \citep{Planck2016}, we classify the \ac{DCO} as a merger candidate (but see Section \ref{sec:cosmic_history} to see how we account for cosmological evolution).

\subsection{Caveats to Binary Evolution in \COMPAS}
\label{subsec:binary_caveats}

We currently do not include tidal evolution in \COMPAS, although tides are believed to play a non-negligible role in massive binary evolution.

We also do not include magnetic braking evolution in \COMPAS.
Most massive stars are non-magnetic \citep{Donati2009}.
Moreover, unlike for low-mass stars, the magnetic braking assumption of negligible mass loss does not hold for radiatively driven stellar winds.
We do not expect magnetic braking to play a significant role in the formation of \ac{NS} and \ac{BH} binaries.

\COMPAS has not been adequately tested in the context of systems hosting white dwarfs.

\section{Evolving a Population}
\label{sec:populations}

Typically users wish to study particular outcomes of stellar or binary system evolution: \acp{BBH} that merge within the age of the universe, \acp{DNS}, X-ray binaries, etc. The initial attributes of these systems are usually not known \textit{a priori}, so a population of systems is evolved with the expectation that some of the systems will evolve into systems of interest.

Each single star or binary system in a \COMPAS simulation is described at \ac{ZAMS} by the initial values of the salient attributes: mass and metallicity for single stars; component star masses, separation, eccentricity, and metallicity for binary systems.  Users can specify values to be used for each of the initial attributes, or allow values to be drawn from specified distributions. 

\subsection{Sampling}
\label{subsec:COMPAS_sampling}

The population of objects (stars or binary systems) synthesized by \COMPAS is intended to be a representative sample from the full population of stars or binary systems in the universe. Sampling allows us to infer information about the full population based on results from the sampled subset -- the sampled collection is used as a proxy for the actual population.

\COMPAS provides functionality that allows users to sample the initial attributes of systems outside the \COMPAS application and provide those initial attribute values to \COMPAS via input grid files (Section \ref{subsubsec:COMPAS_grid_files}). This makes it possible to interface with importance sampling via \textsc{STROOPWAFEL} \citep{Broekgaarden:2019qnw}, and in principle, allows for interfaces with other sampling tools such as Dartboard \citep{Andrews:2017ads} or emulation packages \citep[e.g.,][]{Barrett:2017tug}.

\COMPAS also includes a basic set of initial condition distributions within the main code.  This enables Monte Carlo sampling in which  users specify, for each star or binary system to be evolved, fixed values for initial attributes that should not be sampled, and \COMPAS will sample the remainder using the distributions described below. Each star or binary system is then evolved to its final state, and the results recorded in output files (see Section \ref{subsec:COMPAS_output}).

Here we briefly describe the basic initial parameter distributions available in the core \COMPAS code.  For simplicity, we assume that the overall distribution of initial parameters is an outer product over independent parameter distributions, despite evidence to the contrary \citep{1990ApJS...74..551A,2013ARA&A..51..269D, Moe:2017ApJS, 2018A&A...619A..77K}.

\subsection{Single Star/Primary Mass}
\label{subsec:COMPAS_initial_mass}

The initial mass of a single star, or the primary star (the more massive star at \ac{ZAMS}) in a binary system, \monei, is, unless specified by the user, determined by the \ac{IMF} being used. 
By default, \COMPAS uses the \citet{2001MNRAS.322..231K} \ac{IMF}, the distribution function of which is given by

\begin{equation}
    p(\monei)\propto\monei^{-2.3}
\end{equation}

\noindent above $0.5\,M_\odot$, and $\monei~\in~[5.0, 150.0]$\,M$_\odot$.  Other \ac{IMF} functions \citep[e.g.][]{1955ApJ...121..161S} are available as configurable options. For simplicity, we assume that the \ac{IMF} is the same for all metallicities.

\subsection{Mass Ratio and Secondary Mass}
\label{subsec:COMPAS_initial_mass_ratio}

The mass of the secondary star (less massive at \ac{ZAMS}), \mtwoi, in a binary system being evolved by \COMPAS is, unless specified by the user, determined by the mass ratio 

\begin{equation}
\qi~\equiv~\mtwoi~/~\monei,
\end{equation}

\noindent where $\qi~\in~[0.01, 1.0]$.

In the default \COMPAS model, the mass ratio is drawn from a flat distribution \citep{Sana:2012Sci, kobulnicky2014toward}. Other distributions for mass ratio provided in \COMPAS as options are the distributions described by \citet{1991A&A...248..485D} and \citet{Sana:2012Sci} (Table S3).  
A minimum value for \mtwoi can be specified by the user. 

\subsection{Metallicity}
\label{subsec:COMPAS_initial_metallicity}

The metallicity, \Zi, of a single star, or both component stars of a binary system, is, unless otherwise specified by the user, given by $Z_i=Z_\odot=0.0142$ \citep{Asplund}.

Users can specify that metallicity be sampled for each single or binary star (for binary stars, both component stars use the same, sampled, value for metallicity), using a log-uniform distribution:

\begin{equation}
    p(Z_i)\propto\frac{1}{Z_i},
\end{equation}

\noindent where $\Zi~\in~[0.0001, 0.03]$.

\subsection{Semimajor Axis}
\label{subsec:COMPAS_initial_semi_major_axis}

The initial semi-major axis of a binary star, $a_i$, is, unless specified by the user, sampled independently of the masses using a log-uniform distribution:

\begin{equation}
    p(a_i)\propto\frac{1}{a_i},
\end{equation}

\noindent where $a_i~\in~[0.01, 1000]$\AU~\citep{1924PTarO..25f...1O, 1983ARA&A..21..343A}.

Other distributions for semi-major axis provided in \COMPAS as options are those described by \citet{DuquennoyMayor:1991A&A} and \citet{Sana:2012Sci}.  A custom method allows the user to specify parameters of the distribution.

\subsection{Orbital Period}
\label{subsec:COMPAS_initial_orbital_period}

The orbital period can be specified for binary stars instead of the semi-major axis. The user can either specify a value for the orbital period, $P_i$, or that the value be sampled.  If the value is sampled, a log-uniform distribution is used:

\begin{equation}
    p(P_i)\propto\frac{1}{P_i},
\end{equation}

\noindent where $P_i~\in~[1.1, 1000]$~days.

\subsection{Orbital Eccentricity}
\label{subsec:COMPAS_initial_eccentricity}

Unless otherwise specified by the user, the \COMPAS default model assumes all binary stars are circular at birth (i.e.,~initial eccentricity $\ei=0$).  Other distributions for eccentricity provided in \COMPAS as options are:

\begin{itemize}
    \item {a flat distribution, $p(\ei)=1$,}
    \item {a thermal eccentricity distribution $p(\ei)=2\ei$ \citep{1975MNRAS.173..729H},}
    \item {the M35 distribution described by \citet[][]{2013AJ....145....8G},}
    \item {the distribution described by \citet{1991A&A...248..485D}, and}
    \item {the distribution described by \citet[][Table S3]{Sana:2012Sci}.}
\end{itemize}

\subsection{Supernova Kicks}
\label{subsec:COMPAS_initial_kicks}

Stars that experience a supernova event may experience a momentum boost as a result of the explosion (see Section~\ref{subsec:stellar_remnants}). Unless otherwise specified by the user, the attributes of these so-called natal kicks (i.e. magnitude and direction) are drawn from default distributions, depending upon the supernova type and expected remnant.  The \COMPAS model for natal kicks is described in detail in Section~\ref{subsec:kicks}.

\subsection{Stellar Rotation}
\label{subsec:COMPAS_initial_rotation}

Unless otherwise specified by the user, the COMPAS default model assumes that all stars are nonrotating at birth (i.e., the initial rotational velocity $v_\mathrm{rot} = 0$\,km/s). Other rotation distributions provided in \COMPAS as options are:

\begin{itemize}
    \item {the initial, individual rotational velocities for the two stars can be provided separately,}
    \item {the rotational velocity distribution from \citet{Hurley:2000pk};}
    \item {a rotational velocity distribution for O and B stars based on results from the VLT-FLAMES survey \citep{RamirezAgudelo:2013A&A}.}
\end{itemize}

We describe the limited aspects of rotation currently modeled by \COMPAS in Section~\ref{subsec:rotation}.
For stars in binaries, if CHE is enabled (default), \COMPAS overwrites the drawn or zero rotational velocities of the binary components with the orbital velocity of the binary (i.e. tidal locking is assumed).

\section{Postprocessing}\label{sec:postprocessing}
Using binary population synthesis simulations to make predictions and calculations for astrophysical populations requires converting the data from the simulation into  meaningful astrophysical quantities. The foremost example, for compact objects, is calculating \ac{DCO} formation and merger rates as a function of redshift and/or component masses.  The strengths of the \COMPAS suite include its publicly available postprocessing scripts where these calculations are performed. This section describes the methodology behind the main postprocessing scripts that are publicly available in \COMPAS{} at \url{https://github.com/TeamCOMPAS/COMPAS}.

\subsection{Recording Properties}
\label{subsec:PostProcessing-recording-Properties}
As described in Section \ref{subsec:COMPAS_output}, during the simulation, \COMPAS calculates and records properties of the stars and/or  binary systems such as the ages, masses, stellar radii, effective temperatures, velocities, eccentricities,  and separations. 
The user can specify which properties are recorded and when during the simulation they are reported.
Examples include the option to save the properties at every time step (\textit{detailed output}, an example is given in Section~\ref{subsec:Detailed-evolution-of-a-binary}) or only printing the properties of the binary at important evolutionary stages of the binary such as \ac{CE} episodes and \acp{SN}. 
A detailed description of how the output can be specified is given in the code documentation; \docCOMPAS.

\subsection{Selecting Binary Systems of Interest}\label{subsec:Sub-selecting-binary-systems-of-interest}
During a simulation, \COMPAS returns the recorded properties of all of the simulated  binary systems. Further subselection of binaries of specific interest, such as \acp{BBH} that merge in a Hubble time or systems that experienced a \ac{CE} event, is done by means of postprocessing. The ``optimistic'' and ``pessimistic'' \ac{CE} selection (as mentioned in Section~\ref{subsubsec:binary_common_envelope}) is also performed in postprocessing. The subselection of systems of interest  is typically done in \COMPAS by \textit{``slicing''} or \textit{``masking''} the data, which is described in the publicly available jupyter notebooks.

\subsection{Converting to Yields per Star-forming Mass}
\label{subsec:method-BPS-merger-rate-per-M-SFR}
\begin{figure*}
		\includegraphics[width=\textwidth]{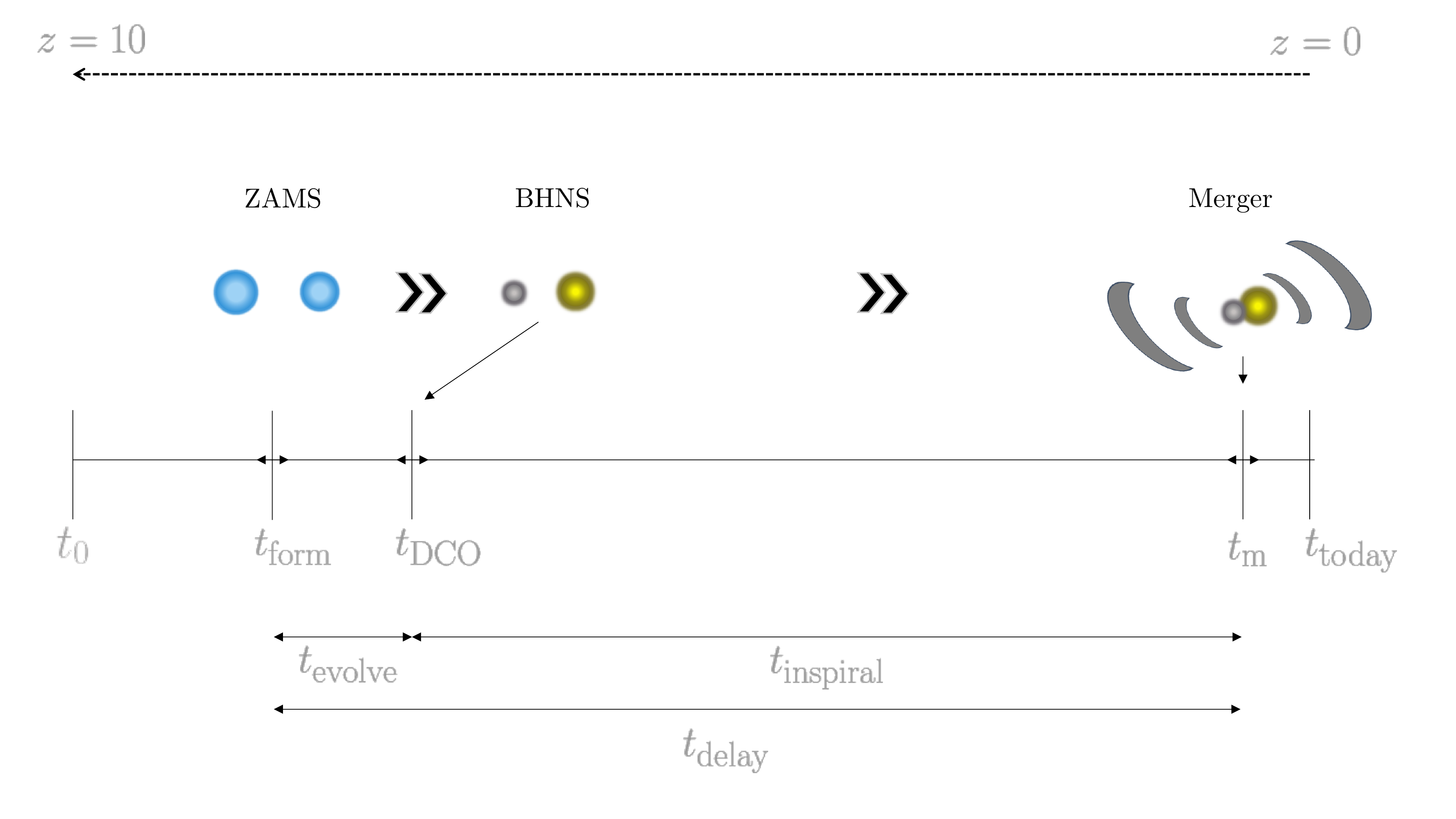}
    \caption{Schematic display of the different times in the formation and evolution of a binary system that impact the time \tmerger at which a  \ac{DCO} (here a BHNS) system will merge in the history of the universe. The relevant timescales are:  the moment the binary is formed at \ac{ZAMS} from a gas cloud, \tform, the moment of \ac{DCO} formation, \tDCO, the time at which the merger takes place, \tmerger, the time it takes the binary to evolve from \ac{ZAMS} to the \ac{DCO} system, $\tevolve = \tDCO - \tform$, the inspiral time, $\tinspiral= \tmerger -\tDCO$ and the time between binary formation at \ac{ZAMS} and merger, $\tdelay = \tmerger - \tform$.  All mergers that occur within the horizon distance of \ac{GW} detectors are potentially detectable today. Arrows indicate that those times that vary per binary system. Figure from \citet{Broekgaarden:2021} available at \url{https://github.com/FloorBroekgaarden/BlackHole-NeutronStar/tree/main/plottingCode/Fig_1/variations}. }
   \label{fig:timescalesEvolution}
\end{figure*}

A {\COMPAS} simulation is typically performed by modeling only a fraction of the underlying stellar population by, for example, not simulating single stars and/or not drawing the simulated binaries from their full initial birth distributions (e.g., by only simulating stars with masses $\geq 5\Msun$).  To obtain meaningful  estimates for formation rates of the binaries of interest, the population synthesis simulation is typically re-normalized to a formation yield per unit star forming mass. 
In this section we will often write this \ac{DCO} formation yield as a function of: birth metallicity \Zi\footnote{Before we used $Z$ instead of \Zi for metallicity. In the remaining section we use \Zi to be consistent with the notation for other birth parameters.}, delay time \tdelay (i.e., the time between the formation and the merger of a binary, see Figure~\ref{fig:timescalesEvolution}), and compact object masses \monef, \mtwof. 
Doing so, the \ac{DCO} formation yield for a binary with metallicity \Zi, delay time \tdelay and  final compact object masses \monef and \mtwof can be written as  
$$\frac{\diff^4 \Nform}{\diff \MSFR \diff \tdelay \diff \monef \diff \mtwof} (\Zi, \tdelay, \monef, \mtwof),$$ 
where \Nform is the number of systems of interest that form and \MSFR is a unit of star forming mass. To get the total yield $\diff \Nform / \diff \MSFR$ this is marginalized over \tdelay, \monef and \mtwof. This yield is typically computed with a Monte Carlo approach through \COMPAS simulations.  The subsequent conversion to merger and detection rates is then done in postprocessing, as described below.

\subsection{Calculating Astrophysical Rates over the Cosmic History of Our Universe: The Case of Double Compact Object Mergers}
\label{sec:cosmic_history}
These \COMPAS suite scripts to calculate cosmological formation and merger rates of astrophysical events, often referred to as ``cosmological integration'', are based on the work presented in \citet{2019MNRAS.490.3740N}. 
We describe the method behind these postprocessing scripts for the example of calculating \ac{DCO} merger rates over redshift below, similarly to \citet{2019MNRAS.490.3740N} and \citet{Broekgaarden:2021}, but the idea can be easily generalized to other phenomena.

\subsubsection{Double Compact Object Merger Rates for Ground-based GW Detectors}

The \ac{DCO} merger rate measured by a comoving observer in the source frame of the merger at a given merger time \tmerger (measured since the Big Bang) is given by
\begin{align}
&\rate_{\rm{m}} (\tmerger,\monef, \mtwof) \equiv \frac{\diff^4 \Nmerger }{\diff \tmerger \diff \Vc  \diff \monef \diff \mtwof} (\tmerger,\monef, \mtwof)  \notag \\
		&= \int \diff \Zi  \int_0^{\tmerger} \diff \tdelay \, {\rm{SFRD}}(\Zi, z(\tform=\tmerger-\tdelay)) \, \times \notag \\ 
		&\, \frac{\diff^4 \Nform}{\diff \MSFR \diff \tdelay \diff \monef \diff \mtwof} (\Zi, \tdelay, \monef, \mtwof),  
\label{eq:MSSFR-merger-rate}
\end{align}
where we convolve the yield with the the metallicity-specific star formation rate density ${\rm{SFRD}}(\Zi, z(\tform))$, which is a function of birth metallicty \Zi and redshift $z$\footnote{We use SFRD for the star formation rate density, and \SFRD for the metallicity-specific star formation rate density.}.  In Equation \ref{eq:MSSFR-merger-rate} \Vc is the comoving volume and the relevant star formation rate is computed at a redshift corresponding to the formation time $\tform = \tmerger - \tdelay$.  Delay times and metallicities are integrated over.  We describe the merger rate in Equation~\ref{eq:MSSFR-merger-rate} as an explicit function of \monef and \mtwof because the probability of a \ac{DCO} merger detection depends on their values (see below), but the total merger rate is often computed by marginalizing over the \ac{DCO} component masses.

\begin{figure*}
\includegraphics[width=1\textwidth, trim=0 12cm 0 2.5cm, clip]{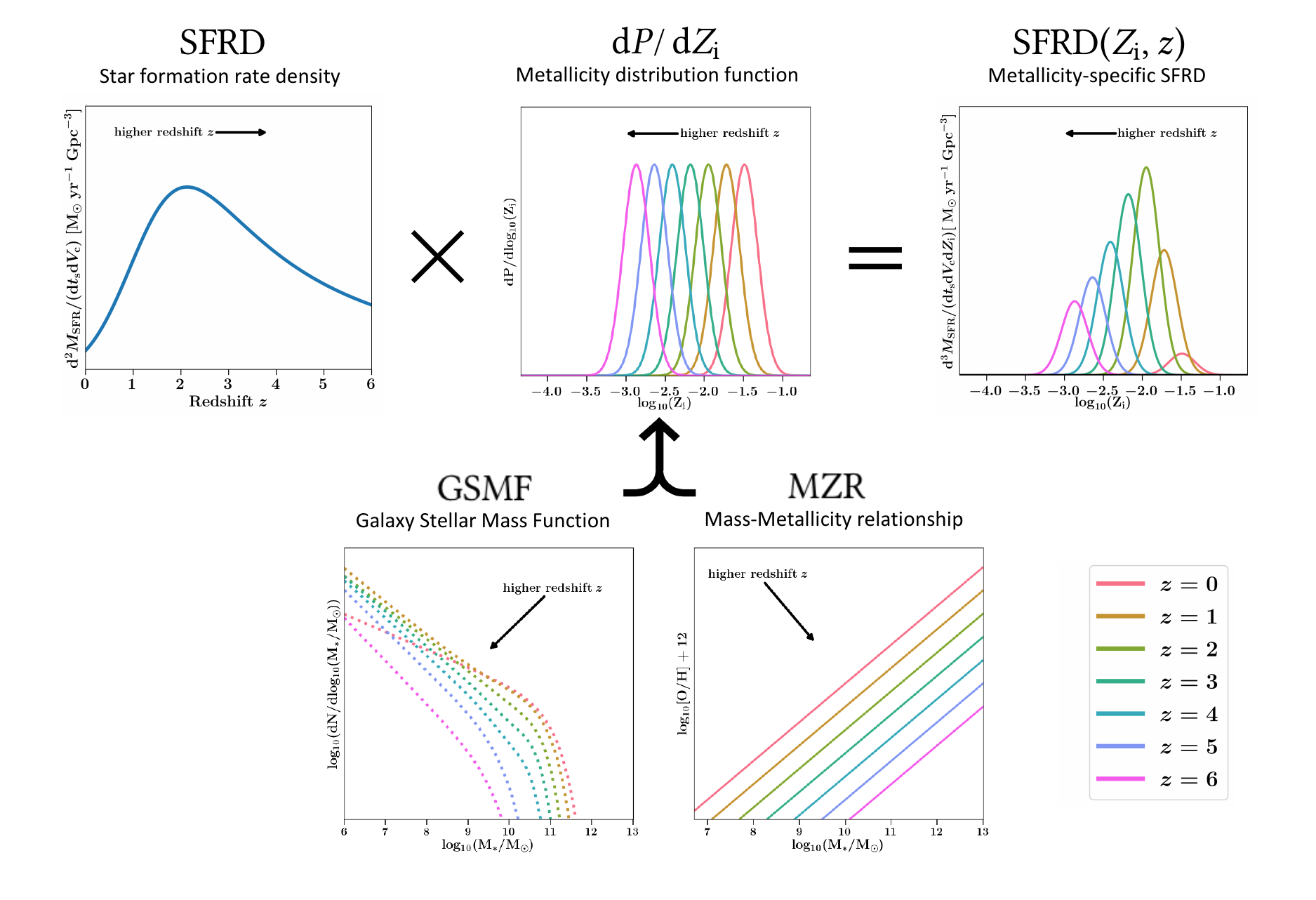}
   \caption{Schematic depiction of how models for the metallicity-specific star formation rate density, \SFRD,  can be created in \COMPAS by multiplying a star formation rate density with a metallicity probability distribution function, $\diff P / \diff \Zi$. 
   The metallicity distribution function is typically constructed in the available \COMPAS postprocessing scripts by convolving a galaxy stellar mass function with a mass-metallicity relationship. An exception is the `preferred' model from \citet{2019MNRAS.490.3740N},  which uses a phenomenological model directly for the metallicity distribution function. 
   The arrows in each subfigure indicate in which direction the distribution moves as redshift increases. Original figure from \citet{Broekgaarden:2021} {available at \url{https://github.com/FloorBroekgaarden/BlackHole-NeutronStar/tree/main/plottingCode/Fig_2}.} }
    \label{fig:MSSFR-sketch}
\end{figure*}

We obtain the \SFRD by 
multiplying the total \ac{SFRD} with a metallicity probability density function
\begin{align}
	&{\rm{SFRD}}(\Zi, z_{\rm{form}}) \equiv 
	\frac{\diff^3 \MSFR}{\diff \ts \diff \Vc \diff \Zi}(z_{\rm{form}})  
	 \notag \\
	 &= \underbrace{\frac{\diff^2 \MSFR}{\diff \ts \diff \Vc }(z_{\rm{form}})}_\text{SFRD} 
	 \times 
	 \underbrace{\frac{\diff P }{\diff \Zi}(z_{\rm{form}})}_\text{GSMF $+$ MZR}, 
	 \label{eq:MSSFR-equation}
\end{align}
where we used the short hand notation $z_{\rm{form}} = z(\tform)$.  
In the available \COMPAS postprocessing scripts, the metallicity distribution function, ${\diff P }/{\diff \Zi}$, is typically described as a convolution between a \ac{GSMF} and the \ac{MZR}.  
This is discussed in more detail  in the following sections and schematically shown in Figure~\ref{fig:MSSFR-sketch}.

In practice, the integral in Equation~\ref{eq:MSSFR-merger-rate} is approximated by a Monte Carlo estimate, integrating over simulated metallicities. The metallicities in the integral limit in Equation~\ref{eq:MSSFR-merger-rate} that fall outside of the simulated metallicity range can be either included in the edge bins (see, e.g., \citet{Broekgaarden:2021} for more details) or conservatively ignored (effectively curtailing ${\diff P }/{\diff \Zi}$ to zero outside the range of simulated metallicities).

The \ac{DCO}  merger rate in Equation~\ref{eq:MSSFR-merger-rate} is then converted to a local detection rate $\rate_{\rm{det}}$ by integrating over the comoving volume and taking into account the probability \Pdet of detecting a gravitational-wave source  (Section~\ref{subsec:detection-probability}) using
\begin{align}
	&\rate_{\rm{det}}(\tdet, \monef, \mtwof) \equiv 
	\frac{\diff^3 \Ndet}{\diff \tdet  \diff \monef \diff \mtwof} 
	\notag  \\
	&= \int 
	\diff z \frac{\diff \Vc}{\diff z}  \,
	 \frac{\diff \tmerger}{\diff \tdet}  \,  
	 \rate_{\rm{m}}(\tmerger) \, 
	  \Pdet (\monef, \mtwof, z(\tmerger)),
\label{eq:CI-detector-rate-mergers}
\end{align}
where \tdet is the time in the detector (i.e. the observer) frame,  $\Pdet (\monef, \mtwof, z)$ is the probability of detecting a gravitational-wave signal from a binary with component masses \monef, \mtwof merging at redshift $z$, and $\frac{\diff \tmerger}{\diff \tdet} = \frac{1}{(1+z)}$ and $\frac{\diff \Vc}{\diff z}$ are given by, e.g., \citet{1999astro.ph..5116H}.  In practice, this integral over redshift can be computed as a Riemann sum over discrete redshift bins; see, e.g., previous work by \citet{2013ApJ...779...72D, 2015ApJ...806..263D,2016ApJ...819..108B,2016MNRAS.458.2634M, 2019MNRAS.482..870E, 2019PhRvD.100f4060B,Bavera:2019, 2019MNRAS.482.5012C}. The total detection rate per unit time can be obtained by further integrating over \ac{DCO} masses.

\begin{table*}[tbh]
\resizebox{\textwidth}{!}{%
\centering
\begin{tabular}{llll}
\hline
\hline
{xyz} index &  {SFRD} [x]                    & {GSMF}  [y]     & {MZR} [z]                       \\ \hline \hline
0 (default)   				& \multicolumn{3}{c}{`preferred' phenomenological  model from   \citet{2019MNRAS.490.3740N} }          \\
\hline
1              			       	& {\citet{2014ARA&A..52..415M}}  &  \citet{2004MNRAS.355..764P} & \citet{2006ApJ...638L..63L}   \\
2              					&  \citet{2004ApJ...613..200S}							& \citet{2015MNRAS.450.4486F} single Schechter   & \citet{2006ApJ...638L..63L} $+$ offset    \\
3              			      	& \citet{2017ApJ...840...39M}     		& \citet{2015MNRAS.450.4486F} double Schechter          &  \citet{2016MNRAS.456.2140M}             \\ \hline
\end{tabular}%
}
%
\caption{Examples of the by-default available options to construct a  metallicity-specific star formation density model \SFRD in the publicly available cosmic integration postprocessing scripts of \COMPAS. A \SFRD model can be obtained by combining a star formation rate (SFR) with a galaxy stellar mass function (GSMF) and mass-metallicity relation (MZR). See  Sections~\ref{sec:cosmic_history} for more details. The labels 0, 1, 2, 3 are solely used in the figures to refer to these models.
The code is flexible to easily adopt a user specified prescription for the \ac{SFRD}, \ac{GSMF} or \ac{MZR}. 
\label{tab:MSSFR-variations-labels}
}
\end{table*}

\subsection{Metallicity-specific Star Formation Rate Density Prescriptions}
\label{subsec:MSSFR-variations}
In {\COMPAS}, the current publicly available postprocessing scripts use convolutions between analytical prescriptions for the \ac{SFRD}, \ac{GSMF} and \ac{MZR} (Equation~\ref{eq:MSSFR-equation}). The analytical equations are based on observations and simulations and can be flexibly adapted to the user's preferences. We present below the prescriptions existing in \COMPAS postprocessing based on the work from \citet{2019MNRAS.490.3740N}. 
An overview of the default available options for those prescriptions is given in Table~\ref{tab:MSSFR-variations-labels}.

\subsubsection{Star Formation Rate Density Prescriptions}
\label{sec:MSSFR-SFR}
\begin{figure}
\centering
\includegraphics[width=\columnwidth]{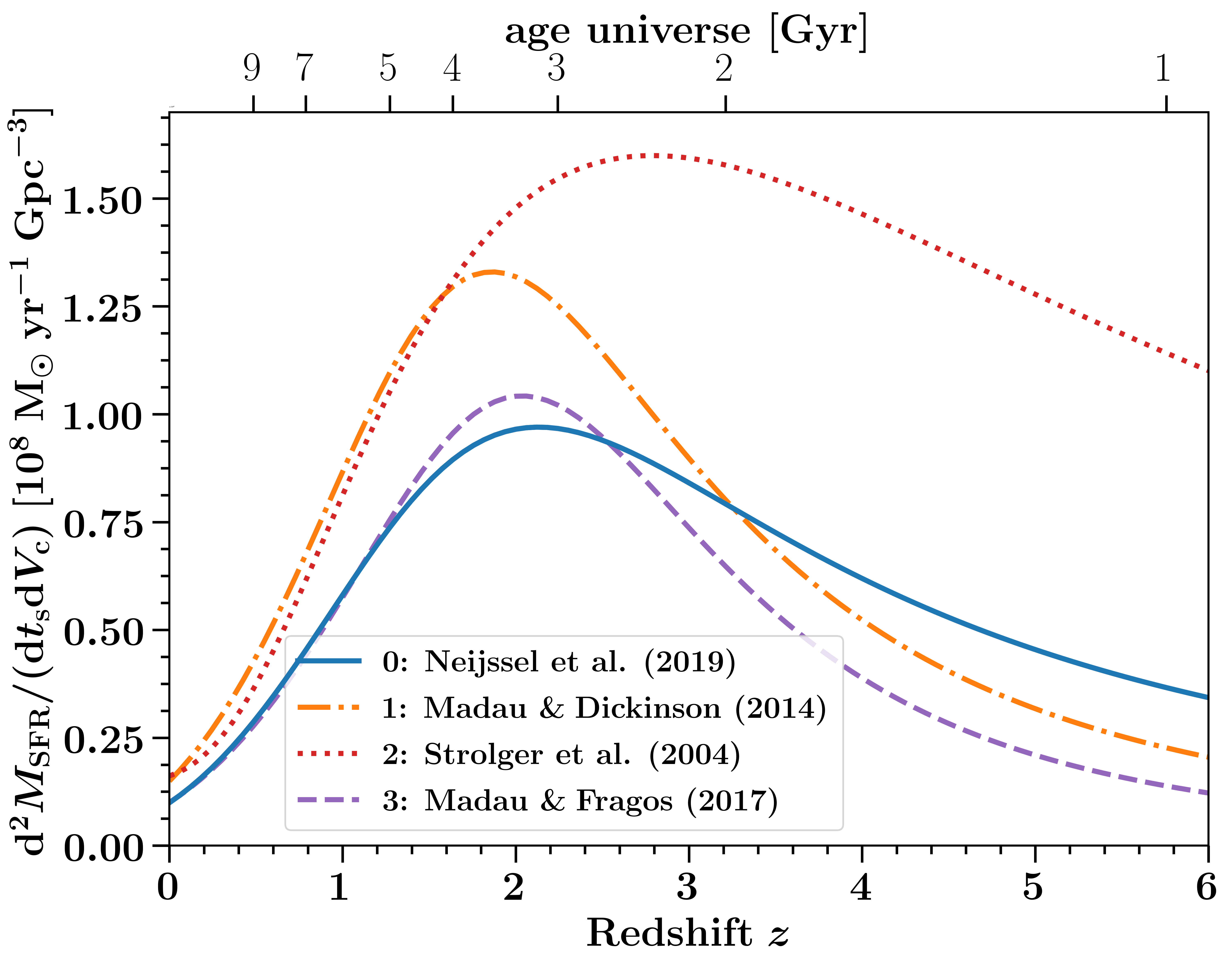}
\caption{The star formation rate density as a function of redshift for the four default available  options in the cosmic integration postprocessing scripts of \COMPAS. }
\label{fig:SFR-options-in-COMPAS}
\end{figure}

 The \COMPAS cosmic integration postprocessing scripts have  different options for the \ac{SFRD}.  Several examples are given in the second column of Table~\ref{tab:MSSFR-variations-labels} and shown in Figure~\ref{fig:SFR-options-in-COMPAS}. 
One of the \ac{SFRD} prescriptions is based on the `preferred' model from \citet{2019MNRAS.490.3740N}, which is
calibrated to match  
the \ac{GW} detections from the first two observing runs of LIGO and Virgo.  
Two other options are the \citet{2014ARA&A..52..415M} and  \citet{2017ApJ...840...39M} \acp{SFRD}. All three \acp{SFRD} are given by the functional form
\begin{equation}
    \frac{d^2 \MSFR}{d \ts d \Vc }(z) = b_{\rm{1}} \frac{(1+z)^{b_{2}}}{1+[(1+z)/b_{3}]^{b_{4}}} \Msun \yearmin \MpcminThree.
    \label{eq:star_formation_rate}
\end{equation}
where $z$ is the redshift. 
The parameters $b_{\rm{1}}$, $b_{\rm{2}}$, $b_{\rm{3}}$ and $b_{\rm{4}}$ in the equation for the \ac{SFRD} are defined by:
$b_{\rm{1}} =0.01,\, b_{\rm{2}}=2.77,\, b_{\rm{3}}=2.9$ and $b_{\rm{4}}=4.7$ for the preferred model in \citet{2019MNRAS.490.3740N}; $b_{\rm{1}}=0.015,\,  b_{2}=2.7,\, b_{\rm{3}}=2.9$ and $b_{\rm{4}}=5.6$ for the \citet{2014ARA&A..52..415M} prescription (see their Equation 15) and $b_{\rm{1}}=0.01,\, b_{\rm{2}}=2.6,\, b_{\rm{3}}=3.2$ and $b_{\rm{4}}=6.2$ for the \citet{2017ApJ...840...39M} prescription (see their Equation 1). 
Another available \ac{SFRD} is the one from 
\citet[][see their Equation~(1)]{2004ApJ...613..200S}, which uses an extinction-corrected model for \ac{SFRD} described as a function of the universe's age: 
\begin{align}
    &\frac{d^2 \MSFR}{d \ts d \Vc }(\tform) = \\
    &c_1\left[\tform^{c_2} \exp\left({-\frac{\tform}{c_3}}\right) + c_4\, \exp\left({c_4 \frac{\tform-t_0}{c_3}}\right)\right]\frac{ \Msun}{\mathrm{yr}\,\mathrm{Mpc^3}},\nonumber 
\label{eq:star_formation_rate_strolger}
\end{align}
with $c_1=0.182,\, c_2=1.26,\, c_3=1.865,\, c_4=0.071$ and $t_0 = 13.47$\Gyr.

\subsubsection{Metallicity distribution function over redshift (GSMF + MZR)}
\label{subsubsec:GSMF_MZR}
The chemical evolution of star formation in our universe is described in the \COMPAS postprocessing scripts by the metallicity probability density, $\diff P / \diff \Zi $, which is a function of redshift. 
The cosmic integration postprocessing scripts offer several options for the metallicity density function.

\begin{figure}
\centering
\includegraphics[width=\columnwidth]{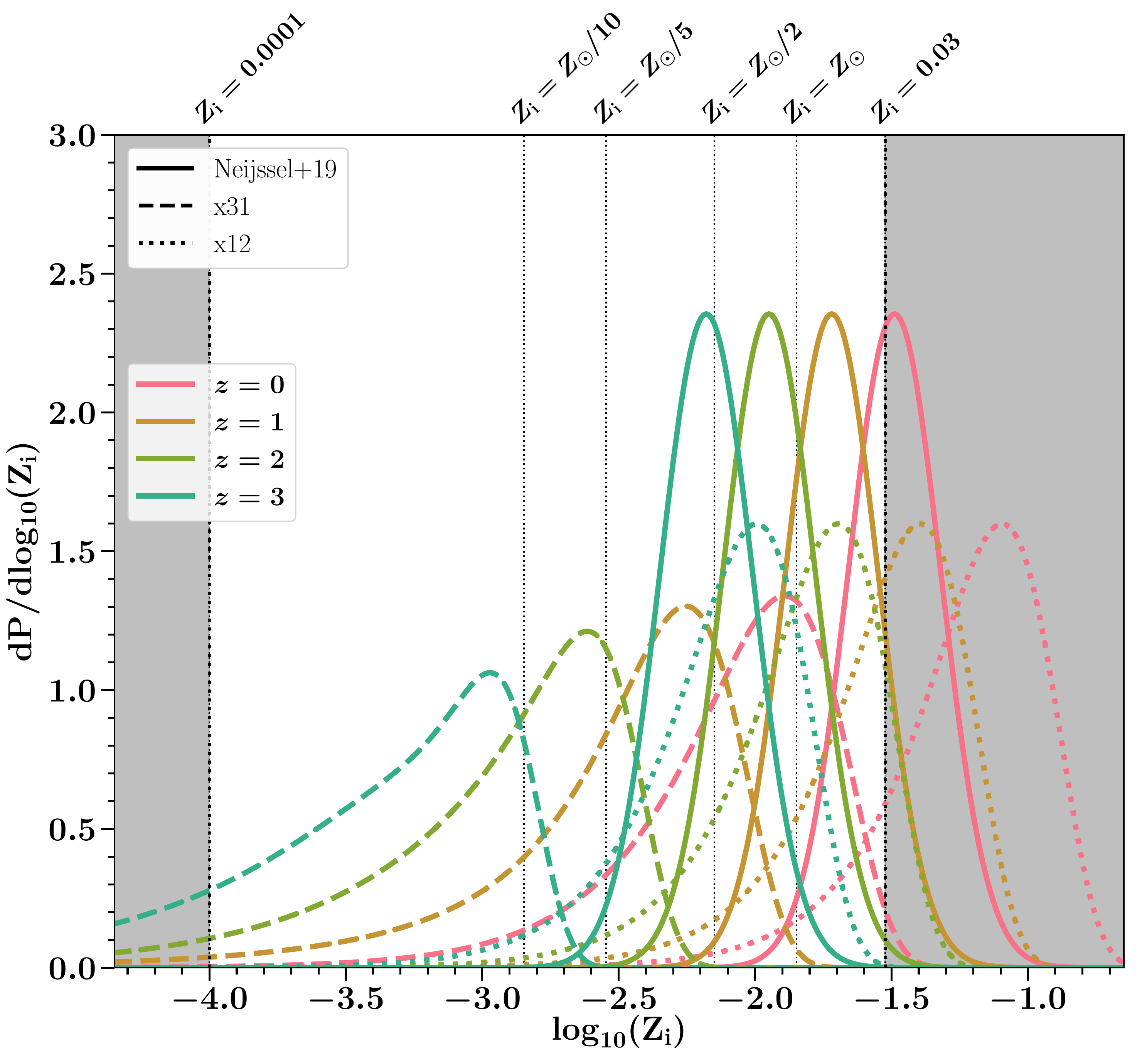}
\caption{Three examples of metallicity probability distribution functions  ($\diff P / \diff \Zi $) available in the  postprocessing scripts of \COMPAS.  For each of the prescriptions the metallicity distribution is shown for four different redshifts. Gray areas show \Zi values that fall outside of the metallicity range of SSE tracks  from \citet{Hurley:2000pk}. }
\label{fig:MSSFR-Z-PDFs}
\end{figure}

The preferred model from  \citet{2019MNRAS.490.3740N} ($\rm{yz}=00$ in Table~\ref{tab:MSSFR-variations-labels}) uses a %
phenomenological model, which defines the metallicity density function as a symmetric log-normal distribution:
\begin{align}
\frac{\diff P }{\diff \Zi}(z) = \frac{1}{\Zi \sigma_1 \sqrt{2 \pi}} \exp\left({-\frac{(\ln(\Zi) - \mu_1(z))^2}{2 \sigma_1^2}}\right),
\end{align}
where  $\sigma_1$ is the standard deviation in $\ln(Z)$-space and $\mu_1(z)$ is the redshift-dependent mean in $\ln(Z)$-space. \citet{2019MNRAS.490.3740N} use a redshift independent $\sigma_1 =0.39$ and mean $\mu_1= \left<\ln(Z)\right>$ defined by 
\begin{align}
\left<Z(z)\right>  =  \exp{(\mu_1 + \sigma_1^2/2)} = Z_{0} 10^{e_1 z}, 
\label{eq:MSSFR-mu}
\end{align}
where $Z_{0}$ is the mean metallicity at redshift 0.  In the preferred phenomenological model, $Z_{0}=0.035$ and $ e_1 = -0.23$. 

This parameterization of the mean (Eq.~\ref{eq:MSSFR-mu}) follows the work by \citet{2006ApJ...638L..63L}.  
Observational evidence suggests the metallicity distributions are likely not symmetric in log-metallicity \citep[e.g.,][]{2006ApJ...638L..63L,2019MNRAS.482.5012C,Boco:2020pgp}. 
The other prescriptions for the metallicity distribution function, which are available in the \COMPAS postprocessing, are therefore asymmetric convolutions of a \ac{GSMF} and an \ac{MZR}.

\subsubsection{Galaxy Stellar-Mass Function Prescriptions}
\label{sec:MSSFR-GSMF}
The available \acp{GSMF} in the \COMPAS postprocessing scripts  are all based on observations of luminosity distributions of galaxies, which are converted to galaxy mass distributions based on a luminosity-mass relation. The options are listed in the third column of Table~\ref{tab:MSSFR-variations-labels} and in Figure~\ref{fig:GSMF-options-in-COMPAS}. More details are provided in Appendix A3 of \citet{2019MNRAS.490.3740N}.

The \acp{GSMF} 1 and 2 use a functional form of a single Schechter function given by 
\begin{align}
 \Phi_{M_{*}}(z) \diff M_{*} = \phi_1(z) \left(\frac{M_{*}}{M_\mathrm{c} (z)}\right)^{-\psi_1(z)}  \exp\left({\frac{-M_{*}}{M_\mathrm{c} (z)}}\right) \diff M_{*},  
\end{align}
where $M_{*}$ is the galaxy stellar mass,  $\phi_1$ is the overall normalization,  $\psi_1$  is the parameter for the slope of the \ac{GSMF} for $M_{*} \leq M_\mathrm{c}$ and $M_\mathrm{c}$ is the cutoff where the \ac{GSMF} moves from a power-law into an exponential drop off (see Figure~\ref{fig:GSMF-options-in-COMPAS}). 
\citet[][Equation~1]{2004MNRAS.355..764P} use a z-independent single Schechter function with $\phi_1 =7.8\cdot 10^{-3}$\,Mpc$^{-3}$, $\psi_1=+1.16$ and $M_{\rm{c}} =  7.64 \cdot 10^{10}$\Msun. 
The \ac{GSMF} option ``\citet{2015MNRAS.450.4486F} single'' ($\rm{y}=2$ in Table~\ref{tab:MSSFR-variations-labels}), uses a linear fit by \citet{2019MNRAS.490.3740N} to the tabulated redshift-dependent values for $\phi_1, \psi_1, $ and $M_{\rm{c}}$.  The \ac{GSMF} option ``\citet{2015MNRAS.450.4486F} double'' ($\rm{y}=3$ in Table~\ref{tab:MSSFR-variations-labels}) is instead based on a double Schechter function given by
\begin{align}
 &\Phi_{M_{*}}(z) \diff M_{*} = \exp\left({\frac{-M_{*}}{M_c(z)}}\right)  \times\\
 &\left[ \phi_1(z) \left(\frac{M_{*}}{M_{c}(z)}\right)^{-\psi_1(z)}  +  \phi_2(z) \left(\frac{M_{*}}{M_{c}(z)}\right)^{-\psi_2(z)}  \right]  \diff M_{*}, \nonumber
\end{align}
which is fitted in a similar way based on tabulated data.

\begin{figure}
\centering
\includegraphics[width=\columnwidth]{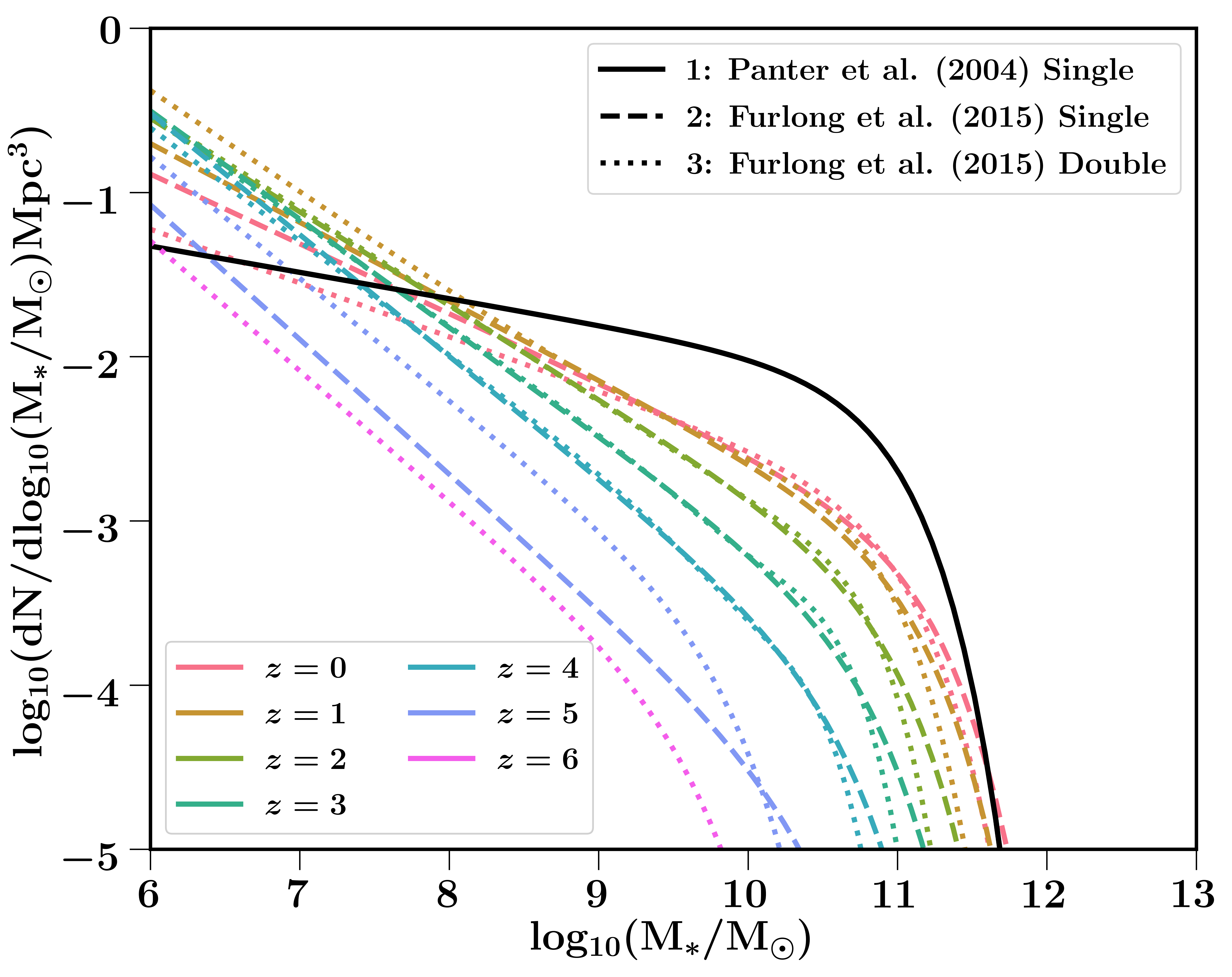}
\caption{The galaxy stellar mass functions (\acp{GSMF}) for the three default available options in the cosmic integration postprocessing scripts of \COMPAS (see also Table~\ref{tab:MSSFR-variations-labels}). The \acp{GSMF} are shown at seven different redshift values, except for the Panter et al.\ \ac{GSMF}, which is not redshift-dependent.}
\label{fig:GSMF-options-in-COMPAS}
\end{figure}

\subsubsection{Mass-Metallicity Relation Prescriptions}
\label{sec:MSSFR-MZR}
The default \ac{MZR} in the \COMPAS postprocessing scripts are analytical formulas for the mapping between galaxy stellar mass and the metallicity of star formation.
Outside of the phenomenological model by \citet{2019MNRAS.490.3740N}, there are three different  \ac{MZR} available by default in \COMPAS. 

The \ac{MZR} 1 and 2 in Table~\ref{tab:MSSFR-variations-labels} are based on \citet{2006ApJ...638L..63L}, who derive a \ac{MZR} based on \citet{2005ApJ...635..260S} observations of 56 galaxies in the Gemini Deep Deep  and Canada-France Redshift Survey. 
They obtain a bisector fit, which is then simplified by \citet{2006ApJ...638L..63L} resulting in the \ac{MZR} 
%
\begin{align}
     \frac{M}{M_{\rm{*}}} =
     \left( 
        \frac{\Zi}{Z_{\odot}} 
    \right)^2,
\end{align}
with $M_{\rm{*}}= 7.64 \cdot 10^{10} \Msun$ from  \citet{2004MNRAS.355..764P}, and a mean metallicity that scales with redshift as Equation~\ref{eq:MSSFR-mu} with the \COMPAS default values of $Z_0 = 0.035$, $\alpha=0.3$. 
Combining these two relations results in the \ac{MZR} by \citet{2006ApJ...638L..63L} given in Table~\ref{tab:MSSFR-variations-labels} with the label $\rm{z}=1$. 
However, since this \ac{MZR} has an offset from the \citet{2005ApJ...635..260S} bisector fit, \citet{2019MNRAS.490.3740N}
also added a second default \ac{MZR} prescription based on \citet{2006ApJ...638L..63L} with a fixed offset in metallicity to match the results by \citet{2005ApJ...635..260S}.

The \ac{MZR} from  \citet{2016MNRAS.456.2140M} is based on theoretical models of cosmological simulations combined with population synthesis simulations. 
Their \ac{MZR} is given by
\begin{align}
    &\log_{10} \left(Z_{\rm{gas}} / Z_{\odot}  \right) 
    = \\
    &0.35 
    \left[
    \log_{10} 
        \left(  
            \frac{M_*}{M_{\odot}} 
        \right)
    -10
    \right]
    +0.93 \exp\left({-0.43z}\right) - 1.05,\nonumber
\end{align}
with $Z_{\rm{gas}}$ the metallicity of the star forming gas (here \Zi). 

\begin{figure}
\centering
\includegraphics[width=\columnwidth]{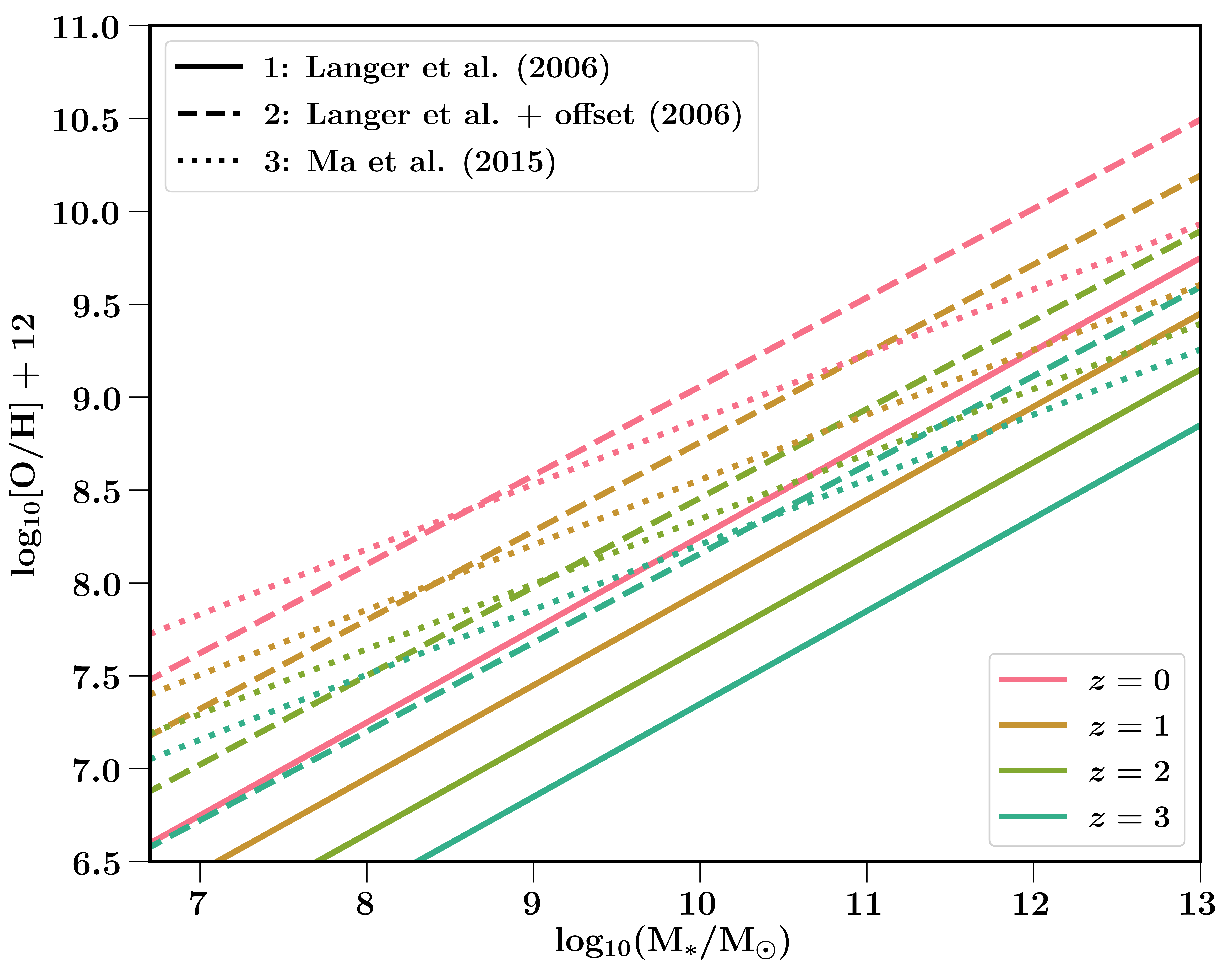}
\caption{The mass-metallicity relations for the three default available options in the cosmic integration postprocessing scripts of \COMPAS. The \acp{MZR} are shown at different redshift values. }
\label{fig:MZR-options-in-COMPAS}
\end{figure}

The three \ac{MZR} options available by default in \COMPAS are given in Figure~\ref{fig:MZR-options-in-COMPAS}.

\subsection{Gravitational-wave selection effects}
\label{subsec:detection-probability}
Whether a \ac{DCO} merger is detectable by a \ac{GW} interferometer depends on its distance (i.e. redshift), orientation, inclination, and source component masses \monef and \mtwof.  
The detectability by a network is approximated by checking whether the source signal-to-noise ratio (S/N) in a single detector surpasses a predefined threshold. 
\COMPAS contains publicly available postprocessing scripts that calculate the detection probability of a gravitational-wave source,
based on the method described in \citet{Barrett:2017fcw}.  Source component spins and any residual in-band eccentricity are not currently included in these detection probability estimates.

Typically a user sets in the \COMPAS postprocessing scripts  a S/N threshold of ${\rm{S/N}} = 8$ for a single ground-based \ac{GW} detector \citep{1993PhRvD..47.2198F}, such that sources with a higher S/N are detectable, but this can be changed to the user-specified value. 
The S/Ns of the DCO mergers are calculated by computing the source waveforms using a user-chosen waveform model included in the LAL software suite \citep{lalsuite}, such as {\sc{IMRPhenomPv2}} \citep{2014PhRvL.113o1101H,2016PhRvD..93d4006H,2016PhRvD..93d4007K} and {\sc{SEOBNRv3}} \citep{2014PhRvD..89h4006P,2017PhRvD..95b4010B}. 
We marginalize over the sky localization and source orientation of the binary using the antenna pattern function from \citet{1993PhRvD..47.2198F}. 
The detector sensitivity can be chosen by the user. 
Available options include the sensitivity of a LIGO instrument at design sensitivity and O1, O2, and O3 configurations \citep{2015CQGra..32g4001L,2016LRR....19....1A}, as well as the third-generation Einstein Telescope detector \cite{ET:2011}.


\section{Usage Examples}
\label{sec:results}

In this section, we show a few practical examples of \COMPAS usage.  Although \COMPAS has mainly been applied to large population studies (see examples in Section \ref{sec:intro}), it is often useful to visualize the full evolutionary path of a given isolated binary, especially when trying to reproduce a specific system.  We show an example in Section \ref{subsec:Detailed-evolution-of-a-binary}, with the caveat that the approximate treatment of rapid population synthesis cannot match the precision of detailed stellar codes for individual systems.  In Section \ref{subsec:chirpmassdist} we show a more typical application of \COMPAS postprocessing tools to predict a distribution over a population, in this case, the chirp mass distribution of detectable \ac{BBH} mergers.

\subsection{Detailed Evolution of a Binary}\label{subsec:Detailed-evolution-of-a-binary}
\begin{figure*}
\centering
\includegraphics[width=\textwidth]{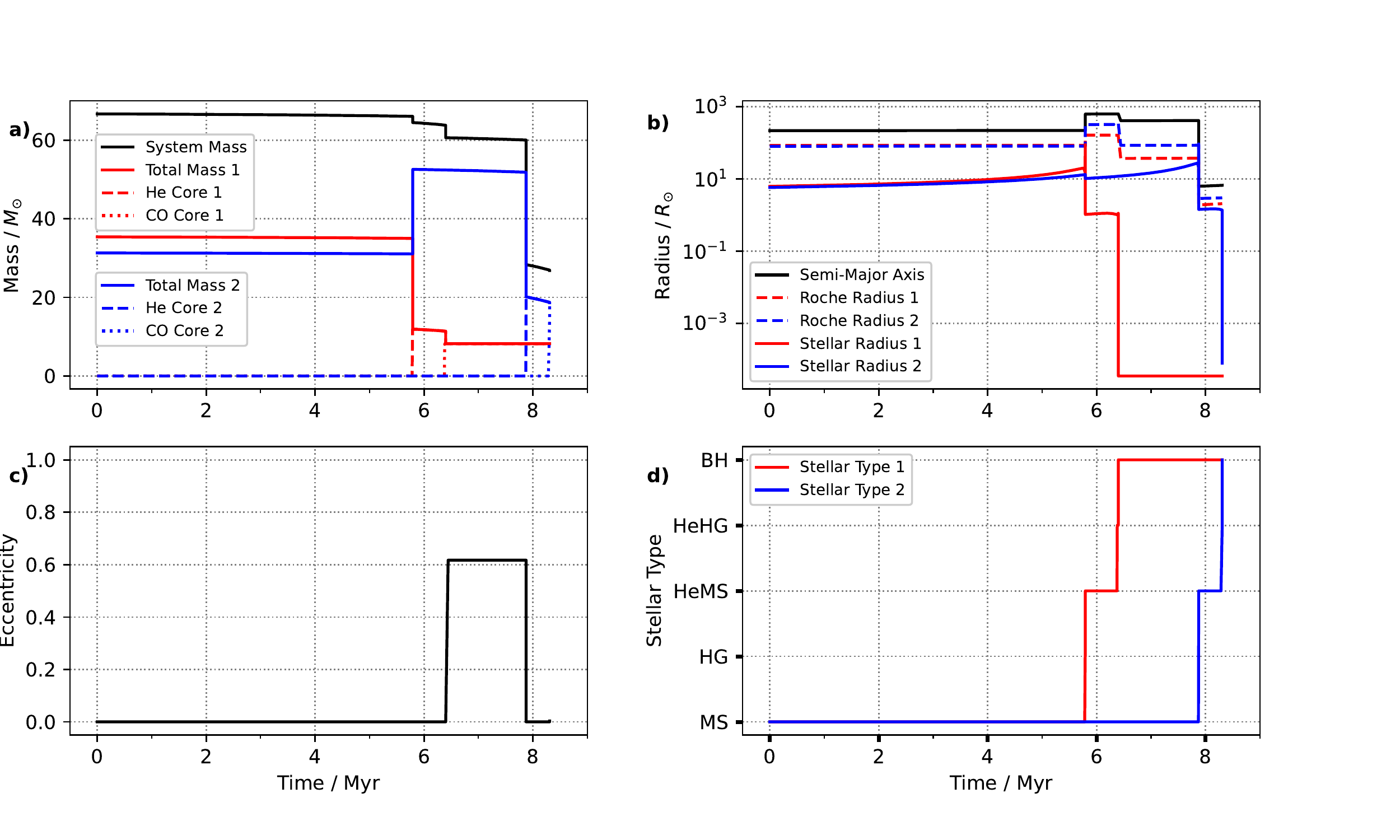}
\caption{Plots produced using detailed \COMPAS output of the evolution of a with initial masses $M_{\textrm{ZAMS},1} = 35.4 \Msun$, $M_{\textrm{ZAMS},2} = 29.3 \Msun$ on a circular orbit with semi-major axis $a=1.02$AU at metallicity $Z = 0.001$. These initial parameters produce a GW151226-like \ac{BBH} with component masses 15.9 and 8.2 \Msun and a coalescence time of 280 Myr (GW151226 had measured component masses of $14.2^{+8.3}_{-3.7}$ and $7.5^{+2.3}_{-2.3} \Msun$, \citealt{Abbott2016gw151226}). Panels (a)--(d) plot respectively the time evolution of the mass parameters (masses of both stars and their He and CO cores, and the total mass of the binary), the radius parameters (stellar radii and Roche radii for both stars, and the semi-major axis), eccentricity, and stellar types (as described in Table \ref{table:SSE_stellar_phases}). Red, blue, and black lines denote the primary star, secondary star, and the binary, respectively.)}  
\label{fig:demo-detailed-evolution}
\end{figure*}

We provide an example plot of the \COMPAS detailed output in Figure \ref{fig:demo-detailed-evolution}, which records the detailed evolution of the progenitor to a GW151226-like \ac{BBH} \citep{Abbott2016gw151226,Stevenson2017FormationEvolution}. 
The code to reproduce this binary and the detailed output of any binary is available at \url{https://github.com/TeamCOMPAS/COMPAS}.
can be used to plot the detailed output of any binary.
We now describe the evolution of this example binary system.
Descriptions of the stellar types referenced below can be found in Table \ref{table:SSE_stellar_phases}.

\begin{enumerate}[label=(\roman*)]
    \item Figure \ref{fig:demo-detailed-evolution}(b) shows that the primary star (red line) exceeds its Roche lobe at 5.8 Myr and initiates mass transfer as it expands rapidly once it evolves off the main sequence.
    This dynamically stable mass transfer episode is nonconservative, as reflected by the concurrent dip in total mass (black curve in Figure \ref{fig:demo-detailed-evolution}(a)).
    This causes the semi-major axis to nearly triple, despite the larger initial mass of the donor (see Equation \ref{eq:mass_transfer_orbital_evolution}). Figure \ref{fig:demo-detailed-evolution}(a) shows that the primary loses 23 \Msun (corresponding to its hydrogen envelope), of which 21 \Msun is accreted by the secondary. The primary emerges as a stripped Helium star (HeMS), as shown in Figure \ref{fig:demo-detailed-evolution}(d), with total mass equal to its He-core mass (Figure \ref{fig:demo-detailed-evolution}(a)).
    
    \item After another 0.6 Myr, the primary collapses into a \ac{BH} in a \ac{CCSN}, ejecting 3 \Msun in the process (Figure \ref{fig:demo-detailed-evolution}(a)). Its natal kick induces an orbital eccentricity of 0.62 (Figure \ref{fig:demo-detailed-evolution}(c)), but due to a fortuitous combination of kick magnitude and direction, the orbital semi-major axis shrinks by ${\sim}1/3$ (Figure \ref{fig:demo-detailed-evolution}(b)).
    
    \item The secondary evolves off the main sequence at 7.8 Myr (Figure \ref{fig:demo-detailed-evolution}(d)). It too expands and exceeds its Roche lobe shortly thereafter, triggering dynamically unstable mass transfer back onto the primary (now a \ac{BH}, Figure \ref{fig:demo-detailed-evolution}(b)).  The binary enters a \ac{CE}, characterised by orbital tightening by several orders of magnitude (Figure \ref{fig:demo-detailed-evolution}(b)) on the dynamical timescale of the donor. 
    In the default \COMPAS model, \ac{CE} phases are assumed to completely circularise the binary, removing the eccentricity imparted by the first \ac{SN} (Figure \ref{fig:demo-detailed-evolution}(c)).
    
    \item The secondary is stripped by the \ac{CE} episode and continues its evolution as a stripped helium star (HeMS) (Figure \ref{fig:demo-detailed-evolution}(d)), collapsing into a \ac{BH} at 8.3 Myr. The complete fallback in the second supernova results in no natal kick and a consequently small post-\ac{SN} eccentricity (Figure \ref{fig:demo-detailed-evolution}(c)).
    
\end{enumerate}

\subsection{Chirp Mass Distribution of LIGO BBHs}\label{subsec:chirpmassdist}
We show an example application of \COMPAS population postprocessing tools to predict the chirp mass distribution of \acp{BBH} detected by gravitational-wave detectors.

\begin{figure}
\centering
\includegraphics[width=\columnwidth]{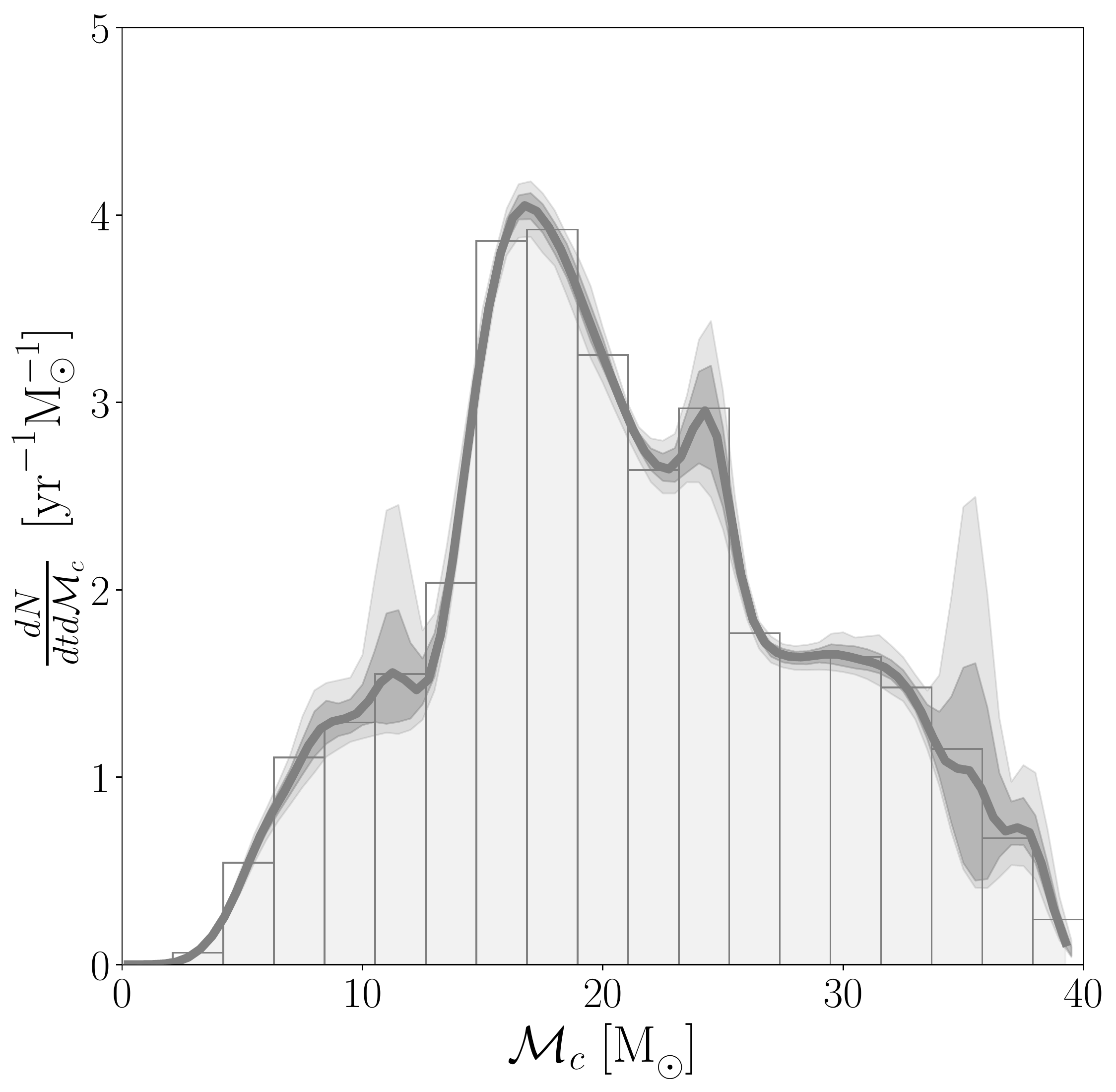}
\caption{Distribution of the detection rate of \ac{BBH} mergers over chirp mass, assuming LIGO O3 sensitivity, an S/N threshold of 8, and the default metallicity-specific star formation history model from \citet{2019MNRAS.490.3740N}. The gray bars show a binned histogram. The thick gray smooth solid line is a kernel density estimate of the distribution, and the shaded region shows the one and two $\sigma$ sampling uncertainties estimated through bootstrapping.}
\label{fig:chirpmassdist}
\end{figure}

We evolve ten million binaries with \COMPAS assuming the default model. Metallicities are sampled log-uniformly in the range $\log(Z)\in [0.0001, 0.03]$.  A Monte Carlo integral over a smooth distribution of metallicities (see Section \ref{sec:cosmic_history}) avoids the metallicity binning artefacts encountered when using a discrete metallicity grid, as discussed, e.g., by \citet{2015ApJ...806..263D}. We compute the merger rate distribution under the assumption of a phenomenological, metallicity-specific star-formation history (see Section \ref{subsec:MSSFR-variations}) detailed in \cite{2019MNRAS.490.3740N}.  We apply gravitational-wave selection effects (see Section \ref{subsec:detection-probability}) for a LIGO detector network operating at O3 sensitivity and an S/N detection threshold of 8. Figure \ref{fig:chirpmassdist} shows the chirp mass distribution of detectable \acp{BBH}.

The script used to make this plot, which makes use of \COMPAS's postprocessing classes, can be found at \url{https://github.com/TeamCOMPAS/COMPAS}.

\section{Conclusions}
\label{sec:conclusions}

We described \COMPAS, a public rapid binary population synthesis code.  \COMPAS \vCOMPAS evolves $\sim 100$ binaries per second on a modern laptop computer.  Given its parallel structure (jobs can be readily split across multiple cores with no need for communication until results are re-combined), a population of a billion binaries can be evolved in 24 hours on a modest 128 core cluster.  \COMPAS is designed to make it easy to specify desired parameterized prescriptions or  introduce new models for various stages of stellar and binary evolution.  Together with its postprocessing toolbox, \COMPAS is thus well suited for inference studies on observed stellar binary populations \citep[e.g.,][]{Barrett:2017fcw}.

The following are some of the planned enhancements to \COMPAS that we hope to include in future versions:

\begin{itemize}

\item \COMPAS currently relies on \citet{Hurley:2000pk} models for single stellar evolution.  In order to evaluate the impact of uncertainties in these models, we plan to incorporate single stellar evolution tracks interpolated from other stellar evolution codes with METISSE \citep[Method of Interpolation for Single Star Evolution,][]{Agrawal:2020}.

\item Extend \COMPAS to more accurately treat low-mass stars.  While the code can evolve binaries with low-mass components, a number of features, ranging from white dwarf novae to magnetic braking, are either not included or insufficiently tested.

\item Include a proper treatment of stellar mergers, allowing the future evolution of merger products to be tracked.

\item Update and re-activate the treatment of tidal interactions, including tidal synchronization and circularization, to include the latest models \citep[e.g.,][]{Vick:2020}.

\end{itemize}

COMPAS is a public code, and we encourage the community to use it and, should they wish, to become involved in its development.  In particular, any defects or enhancement requests can be brought to our attention via the \href{https://github.com/TeamCOMPAS/COMPAS/issues}{github issue tracker}\footnote{\url{https://github.com/TeamCOMPAS/COMPAS/issues}} or by e-mail, \href{mailto:compas-user@googlegroups.com}{compas-user@googlegroups.com}.  COMPAS is published in the Journal of Open Source Software under: Compas et al., (2021). COMPAS: A rapid binary population synthesis suite. Journal of Open Source Software, 6(68), 3838. \url{https://doi.org/ 10.21105/joss.03838}

   
\section{Acknowledgements}
The authors thank Ben Bradnick, Isobel Romero-Shaw and Rajath Sathyaprakash for past contributions to the code, and Simone Bavera, Chris Belczynski, Christopher Berry, Jan Eldridge, Tassos Fragos, David Hendriks, Jarrod Hurley, Vicky Kalogera, Morgan MacLeod, Pablo Marchant, Javier Mor\'{a}n Fraile, Philipp Podsiadlowski, Carl Rodriguez, Dorottya Sz\'{e}csi and Michael Zevin for discussions and advice.  
Multiple authors are supported by the Australian Research Council Centre of Excellence for Gravitational Wave Discovery (OzGrav), through project number CE170100004.
Multiple authors were funded in part by the National Science Foundation under grant No. (NSF grant No. 2009131),  the Netherlands Organization for Scientific Research (NWO) as part of the Vidi research program BinWaves with project number 639.042.728 and by the European Union's Horizon 2020 research and innovation program from the European Research Council (ERC, grant agreement No. 715063).  F.S.B. is supported in part by the Prins Bernard Cultuurfonds studiebeurs.  I.M. is a recipient of an Australian Research Council Future Fellowship (FT190100574).  A.V.G. acknowledges funding support by the Danish National Research Foundation (DNRF132).\\
This research has made use of NASA’s \href{http://adsabs.harvard.edu/}{Astrophysics Data System Bibliographic Services}\footnote{\url{http://adsabs.harvard.edu/}}. 
   
\software{COMPAS is written in C++ and  we acknowledge the use of the GNU C++ compiler, GNU scientific library (gsl), the BOOST C++ library, and the HDF5 C++ library from \url{http://www.gnu.org/software/gsl/} \citep{galassi2002gnu}. The COMPAS suite makes use of Python from the Python Software Foundation. Python Language Reference Available at \url{http://www.python.org} \citep{CS-R9526}. In addition, the COMPAS suite makes use of the python packages \href{http://www.astropy.org}{Astropy} \citep{2013A&A...558A..33A,2018AJ....156..123A}, \href{https://docs.h5py.org/en/stable/}{hdf5}\footnote{\url{https://docs.h5py.org/en/stable/}} \citep{collette_python_hdf5_2014}, the 
\href{http://ipython.org}{IPython}\footnote{\url{http://ipython.org}} and \href{https://jupyter.org/}{Jupyter notebook package}\footnote{\url{https://jupyter.org/}} \citep{PER-GRA:2007,kluyver2016jupyter}, \href{http://www.matplotlib.org}{Matplotlib}\footnote{\url{http://www.matplotlib.org}}  \citep{2007CSE.....9...90H},  \href{http://www.NumPy.org/}{NumPy}\footnote{\url{http://www.NumPy.org/}} \citep{harris2020array}, \href{https://www.scipy.org}{SciPy}\footnote{\url{https://www.scipy.org}} \citep{2020SciPy-NMeth}, Seaborn \citep{waskom2020seaborn}.
The COMPAS post-processing code for detection probability currently makes use of precomputed results from the LALSuite toolkit \citep{lalsuite}, such as the {\sc{IMRPhenomPv2}} waveform \citep{2014PhRvL.113o1101H,2016PhRvD..93d4006H,2016PhRvD..93d4007K}. }

\facilities{Some of the results in this manuscript were obtained using the following computing facilities: FAS Research Computing, Harvard University, and the OzSTAR national facility at Swinburne University of Technology. The OzSTAR program receives funding in part from the Astronomy National Collaborative Research Infrastructure Strategy (NCRIS) allocation provided by the Australian Government.}

\section*{Data availability}
We encourage the community to make results obtained with COMPAS publicly available at  \url{https://zenodo.org/communities/compas/}. The scripts to reproduce the data and plots for all figures in this manuscript using \vCOMPAS of COMPAS are provided in the corresponding directories at \url{https://github.com/TeamCOMPAS/COMPAS}.

\newpage 

\bibliographystyle{aasjournal}
\bibliography{bib}



\end{document}